\documentclass[12pt]{iopart}
\usepackage{epsfig}
\newcommand{\slp}{/\!\!\!p}

\newcommand{\bvec}{\mathbf}

\newcommand{\gapproxeq}
{\lower .7ex\hbox{$\;\stackrel{\textstyle >}{\sim}\;$}}
\newcommand{\lapproxeq}
{\lower .7ex\hbox{$\;\stackrel{\textstyle <}{\sim}\;$}}
\renewcommand\theequation{\thesection.\arabic{equation}}
\newcommand{\be}{\begin{equation}}
\newcommand{\ee}{\end{equation}}
\newcommand{\bea}{\begin{eqnarray}}
\newcommand{\eea}{\end{eqnarray}}
\newcommand{\bt}{\begin{tabular}}
\newcommand{\et}{\end{tabular}}
\newcommand{\eqn}[1]{(\ref{#1})}
\newcommand{\pp}{~~~.}
\newcommand{\vv}{~~~,}
\newcommand{\ds}{\displaystyle}
\newcommand{\nn}{\nonumber}
\newcommand{\ov}{\overline}
\newcommand{\rt}{\rightarrow}
\newcommand{\lrt}{\leftrightarrow}
\newcommand{\pe}{{\phi_e}}
\newcommand{\p}{{\rm p}}

\newcommand{\hnb}{\hat{n}_B}
\newcommand{\hH}{\widehat{H}}
\newcommand{\hmu}{\widehat{M}_u}
\newcommand{\hdmi}{\Delta \widehat{M}_i}
\newcommand{\hdmj}{\Delta \widehat{M}_j}
\newcommand{\hgi}{\widehat{\Gamma}_i}

\newcommand{\hrho}{\hat{\rho}}
\newcommand{\hp}{\hat{\rm p}}

\def\neb{\hbox{$\ov{\nu}_e \!$ }}

\def\ca{{C_{\scriptscriptstyle A}}}
\def\cv{{C_{\scriptscriptstyle V}}}

\newcommand{\tre}{\left( \cv^2 + 3 \ca^2 \right)}
\renewcommand\theequation{\thesection.\arabic{equation}}
\def\caq{{C_{\scriptscriptstyle A}^2}}
\def\cvq{{C_{\scriptscriptstyle V}^2}}
\newcommand{\Nature}{{\it Nature\,}}

\newcommand{\ApJ}{{\it Astrophys. J.\,}}
\newcommand{\ApJS}{{\it Astrophys. J. Suppl.\,}}
\newcommand{\ARAA}{{\it Annu. Rev. Astron. Astrophys.\,}}

\begin{document}
\hfill MPP-2004-82

\hfill DSF 19/2004

\title[Nuclear Reaction Network for Primordial Nucleosynthesis]{Nuclear Reaction Network for Primordial
Nucleosynthesis: a detailed analysis of rates, uncertainties and
light nuclei yields.}

\author{P.D. Serpico\dag, S. Esposito\ddag, F. Iocco\ddag, G. Mangano\ddag,\\ G.
Miele\ddag, and O. Pisanti\ddag\footnote[3]{serpico@mppmu.mpg.de\\
sesposit@na.infn.it\\ iocco@na.infn.it\\ mangano@na.infn.it\\
miele@na.infn.it\\ pisanti@na.infn.it}}

\address{\dag\ Max Planck Institut f\"{u}r Physik, Werner Heisenberg
Institut, \\F\"{o}hringer Ring 6, 80805, M\"{u}nchen, Germany}

\address{\ddag\ Dipartimento di Scienze Fisiche, Universit\`{a} di
Napoli {\it Federico II}, \\and INFN, Sezione di Napoli,
\\Complesso Universitario di Monte Sant'Angelo, Via Cintia,
I-80126 Napoli, Italy}

\begin{abstract}
We analyze in details the standard Primordial Nucleosynthesis scenario. In
particular we discuss the key theoretical issues which are involved in a
detailed prediction of light nuclide abundances, as the weak reaction
rates, neutrino decoupling and nuclear rate modeling. We also perform a
new analysis of available data on the main nuclear processes entering the
nucleosynthesis reaction network, with particular stress on their
uncertainties as well as on their role in determining the corresponding
uncertainties on light nuclide theoretical estimates. The current status of
theoretical versus experimental results for $^2$H, $^3$He, $^4$He and
$^7$Li is then discussed using the determination of the baryon density as
obtained from Cosmic Microwave Background anisotropies.
\end{abstract}

\pacs{98.80.Ft, 26.35.+c, 98.80.-k}

\maketitle

\eqnobysec

\section{Introduction}

The huge amount of data coming from astronomical observations in the recent
years represents the basic impulse and the {\it a priori} condition for the
development of what has been fairly called {\it precision cosmology}.
Results on Cosmic Microwave Background (CMB) anisotropies by the WMAP
Collaboration \cite{wmap03} and the second data release of the Sloan
Digital Sky Survey \cite{sdss} are perhaps the most striking examples of
such a massive effort. On one hand these experimental results give a
beautiful confirmation of our present understanding of the evolution of the
Universe and provide accurate information on several cosmological
parameters. On the other hand they stimulated further efforts in increasing
the level of accuracy of theoretical studies. This general trend is clearly
recognized in the analysis of several cosmological observables, such as the
CMB anisotropies and polarization, the Large Scale Structure formation and,
for what matters for the present study, Big Bang Nucleosynthesis.

Big Bang Nucleosynthesis (BBN) is one of the pillars of the
cosmological model, and it also represents one of the most
powerful tools to test fundamental physics. Presently, this aspect
is even more relevant than in the past. In fact since the baryon
density is now independently measured with very high precision by
CMB anisotropies, the theory of BBN is basically parameter free in
its standard formulation so that comparing experimental
determination of light nuclides with corresponding theoretical
estimate can severely constraint any exotic physics
(for a general overview on the subject, see
the review by B.D. Fields and S. Sarkar
in the Particle Data Book~\cite{PDG} and references therein).

In the last decade the accuracy of BBN theory has been increased
by a careful analysis of many of its key aspects. The accuracy of
the weak reactions which enter the neutron/proton chemical
equilibrium has been pushed up to less than percent level
\cite{Lopez,EMMP1}. Similarly, the neutrino decoupling has been
carefully studied by several authors by explicitly solving the
corresponding kinetic equations, see {\it e.g.}
\cite{dolgovrep,Mangano:2001iu} and References therein. These two
issues are mainly affecting the prediction of $^4$He mass
fraction, which presently has a very small uncertainty, of the
order of 0.1 \%, due to the experimental uncertainty on neutron
lifetime.

One of the most relevant aspects to get an accurate determination
of light nuclide abundances is the evaluation of the several
nuclear reaction rates which enter the BBN reaction network, as
well as the corresponding uncertainties. This task involves a
careful study of the available data or predictions on each
reaction, the choice of a reasonable protocol to combine them in
order to obtain a best estimate and an error and finally, the
calculation of the corresponding thermal rates entering the BBN
network.

After the first analysis performed in \cite{fh64}-\cite{ca88}, and
later studies \cite{KrRom90,SKM93}, which have been the status of the art
results for decades, an important step has been the
publication of a large nuclear rate catalogue by NACRE
Collaboration \cite{NACRE}. Despite of the fact that this
catalogue by its own nature is an {\it all-purpose} compilation
and not only
devoted to BBN studies, and that in particular not all BBN
reactions are in fact covered, nevertheless it
is an extremely important reference. Other nuclear databases, such
as \cite{Exfor} or \cite{TUNL} collect relevant information on
cross sections, nuclear parameters and beta-unstable nuclide
lifetimes.

Recently, new and more precise experimental data have been obtained for
relevant nuclear reactions in the energy range of interest for the BBN, as
for example the LUNA Collaboration results on $^2$H + p $\lrt$ $\gamma$ +
$^3$He process \cite{Casella02}. In view of this new developments and of
the fact that as stressed already there is presently a demand for an
increased precision of BBN predictions, it is important to undertake a
critical review of the whole BBN nuclear network, and in particular to
quantify the uncertainties which affect the nuclide abundances as obtained
by propagating the error on the several nuclear rates entering this
network. This study is the main goal of this paper. Similar analysis has
been also performed in \cite{Cyburt:2004cq}, \cite{cuocoetal} and
\cite{Descouvemont04}.

In the following Sections we cover all the main theoretical aspects of BBN
with the aim of providing a self contained description of the several
issues which enter an accurate determination of nuclide abundances. To help
the reader to distinguish the review parts collecting the results of
previous studies from the ones devoted to original new results first
presented in the present paper, the former have been tagged by a ``$*$''.

In Section 2 we review in details the general BBN framework by
recalling the corresponding set of equations. We also describe the
neutrino decoupling as obtained via a numerical solution of the
corresponding kinetic equation, including as in
\cite{Mangano:2001iu} the effect of QED radiative corrections and
in particular the non thermal distortion in neutrino distribution
functions due to $e^+-e^-$ entropy release.

Section 3 is devoted to the analysis of all main rates entering
the BBN network. We first review the calculation of $n \lrt p$
weak rates, evaluated up to order $\alpha$ QED radiative
corrections, and also taking into account the effect of finite
nucleon mass and thermal radiative effects. We also discuss the
effect on these rates of neutrino decoupling. After introducing
the nuclear astrophysics formalism used in our analysis, we
describe our method for the data reduction and rate estimate and
finally describe in details the data and the results obtained for
all leading reactions entering the BBN network. We also analyze
some sub-leading reaction which, in view of the present
uncertainties on the corresponding rates, may still play a role in
determining the eventual light nuclide yields.

In Section 4 we first summarize the present status of observations on
$^2$H, $^4$He, and $^7$Li, as well as what is known on other light
nuclides, such as $^3$He or $^6$Li. We then give our results for main
nuclide abundances, as obtained by a FORTRAN code that we have developed in
the recent years modifying the original public code of \cite{KawCode92} in
order to consistently introduce all the issues described in this paper
(neutrino decoupling, radiative corrections to $n \lrt p$ rates, etc.) and
employing a different numerical resolution method, as explained in Section
2.3.

All results are given for a standard BBN scenario, with three non
degenerate neutrinos, and the baryon density as fixed by the WMAP
result combined with results of CBI and ACBAR experiments on CMB
anisotropies, as well as with 2dFGRS data on power spectrum,
$\omega_b \equiv \Omega_b h^2= 0.023 {\pm} 0.001$ \cite{wmap03},
where $\Omega_b$ is the ratio of the baryonic matter density with
respect to the critical one, and $h$ is the Hubble constant in
units of 100 Km $s^{-1}$ Mpc$^{-1}$. A careful study of this case
seems to us of particular relevance since there are no free
parameters left. Comparing experimental results with theoretical
expectations is therefore a clean consistency test of the simplest
BBN dynamics. It may also give useful hints to point out possible
systematics in the experimental results for nuclide abundances. A
particular stress is given to the present theoretical uncertainty
on each nuclide yield, as well as on the role of the several
nuclear rates in building up this uncertainty. We also comment on
the impact of different values adopted for $\omega_b$ as obtained
by WMAP with different choices of priors or combining different
observations. Finally, we discuss how the baryon density is
constrained by BBN as compared to the very precise measurement by
CMB anisotropies.

Our Conclusions are reported in Section 5.

\section{The BBN general framework $*$}

\subsection{Generalities $*$}
\label{s:generalities}

We consider $N_{nuc}$ species of nuclides, whose number densities, $n_i$,
are normalized with respect to the total number density of baryons, $n_B$,
\be X_i=\frac{n_i}{n_B} \quad\quad\quad i=n,\,p,\,^2{\rm H}... \pp \ee A
list of all nuclides which are typically included in a BBN analysis is
reported in Table 1. To quantify $^2$H, $^4$He and $^7$Li abundances, we
also use in the following the parameters \be X_{^2{\rm
H}}/X_p,\,\,\,\,\, X_{^3{\rm
He}}/X_p,\,\,\,\,\,Y_p = 4 X_{^4{\rm He}}, \,\,\,\,\,X_{^7{\rm Li}}/X_p \vv
\ee i.e. the $^2$H, $^3$He and $^7$Li number density normalized to hydrogen, 
and the $^4$He mass fraction\footnote{Though this definition is widely used,
and we will use it as well, yet it is only approximately related to the
real mass fraction, since the $^4$He mass is not given by 4 times the
atomic mass unit. The difference is quite small, of the order of $0.5 \%$.
However, in view of the present precision of theoretical analysis on $^4$He
yield, this difference cannot be neglected and we think it is worth to
stress this point in order to avoid possible misinterpretations.}. For the
sake of clarity we stress that our notation $X_i$ for the nuclide number
density normalized to hydrogen is not adopted by all authors. The symbols
$Y_i$ can be also frequently found in the literature.
\begin{table*}
\begin{center}
\begin{tabular}{|cc|cc|cc|cc|cc|}
\hline 1) &n   &   7) & $^6$Li  &  13) & $^{10}$B  &  19) & $^{13}$C  &  25) & $^{15}$O \\
\hline 2) &p   &   8) & $^7$Li  &  14) & $^{11}$B  &  20) & $^{13}$N  &  26) & $^{16}$O \\
\hline 3) &$^2$H  &   9) & $^7$Be  &  15) & $^{11}$C  &  21) & $^{14}$C  &   &     \\
\hline 4) &$^3$H  &  10) & $^8$Li  &  16) & $^{12}$B  &  22) & $^{14}$N  &   &       \\
\hline 5) &$^3$He &  11) & $^8$B   &  17) & $^{12}$C  &  23) & $^{14}$O  &   &     \\
\hline 6) &$^4$He &  12) & $^9$Be  &  18) & $^{12}$N  &  24) & $^{15}$N  &   & \\
\hline
\end{tabular}
\end{center}
\caption{Nuclides considered in the BBN analysis} \label{nuclnumb}
\end{table*}

The neutrino/antineutrino distributions are denoted by
$f_{\nu_e}(\left|\vec{p}\right|,t)$,
$f_{\bar{\nu}_e}(\left|\vec{p}\right|,t)$ and
\be f_{\nu_\mu}=f_{\nu_\tau} \equiv f_{\nu_x}(\left|\vec{p}\right|,t) \vv
~~~ f_{{\bar \nu}_\mu}=f_{{\bar \nu}_\tau} \equiv f_{{\bar
\nu}_x}(\left|\vec{p}\right|,t) \vv
\ee
In the temperature range we are interested in, $10 \,{\rm MeV} > T > 0.01\,
{\rm MeV}$, electrons and positrons are kept in thermodynamical equilibrium
with photons by fast electromagnetic interactions. Thus, they are
distributed according to a Fermi-Dirac function $f_{e^{\pm}}$, with chemical
potential $\mu_e$.

The set of differential equations ruling primordial nucleosynthesis is the
following (see for example \cite{Wagoner,Esposito:1999sz,Esposito:2000hh})
\bea &&\frac{\dot{R}}{R}  = H = \sqrt{\frac{8\, \pi G_N}{3}~ \rho}
\vv
\label{e:drdt} \\
&&\frac{\dot{n}_B}{n_B} = -\, 3\, H \vv
\label{e:dnbdt} \\
&&\dot{\rho} = -\, 3 \, H~ (\rho + \p) \vv
\label{e:drhodt} \\
&&\dot{X}_i = \sum_{j,k,l}\, N_i \left(  \Gamma_{kl \rt ij}\,
\frac{X_l^{N_l}\, X_k^{N_k}}{N_l!\, N_k !}  \; - \; \Gamma_{ij \rt kl}\,
\frac{X_i^{N_i}\, X_j^{N_j}}{N_i !\, N_j  !} \right) \equiv \Gamma_i(X_j)
\vv
\label{e:dXdt} \\
&&L \left(\frac{m_e}{T}, \pe\right) = \frac{n_B}{T^3}~ \sum_j Z_j\, X_j \vv
\label{e:charneut}\\
&&\left(\frac{\partial}{\partial t} - H \, \left|\vec{p}\right|
\frac{\partial}{\partial \left|\vec{p}\right|} \right) \,
f_{\nu_\alpha}(\left|\vec{p}\right|,t) = I_{\nu_\alpha} \left[
f_{\nu_e},f_{\bar{\nu}_e},f_{\nu_x},f_{\bar{\nu}_x},f_{e^-},f_{e^+}
\right] \vv \label{nuboltz} \eea with $\nu_\alpha=\nu_e$,
$\bar{\nu}_e$, $\nu_x$, $\bar{\nu}_x$, where $\rho$ and $\p$
denote the total energy density and pressure, respectively, \bea
\rho &=& \rho_\gamma + \rho_e + \rho_\nu + \rho_B  = \rho_{NB} + \rho_B\vv \\
\p &=& \p_\gamma + \p_e + \p_\nu + \p_B = \p_{NB} + \p_B\vv \eea
where $i,j,k,l$ denote nuclides, $\rho_B$ and $\rho_{NB}$ are the
baryon and non baryonic energy density respectively, $Z_i$ is the
charge number of the $i-$th nuclide, and the function $L(\xi,y)$
is defined as \be L(\xi,\omega) \equiv \frac{1}{\pi^2}
\int_\xi^\infty d\zeta~\zeta\, \sqrt{\zeta^2-\xi^2}~ \left(
\frac{1}{e^{\zeta-\omega}+1} - \frac{1}{e^{\zeta+\omega}+1}
\right) \pp \label{lfunc} \ee Equation (\ref{e:drdt}) is the
definition of the Hubble parameter, $H$, with $G_N$ the
gravitational constant, whereas Equations (\ref{e:dnbdt}) and
(\ref{e:drhodt}) state the total baryon number and entropy
conservation in the comoving volume, respectively. The set of
$N_{nuc}$ Boltzmann equations (\ref{e:dXdt}) describes the density
evolution of each nuclide specie, with $\Gamma_{kl \rt ij}$ the
rate per incoming particles averaged over kinetic equilibrium
distribution functions. While in fact chemical equilibrium among
nuclides cannot be assumed, as BBN strongly violates Nuclear
Statistical Equilibrium, it is perfectly justified to assume
kinetic equilibrium, which is maintained  by fast strong and
electromagnetic processes. Equation (\ref{e:charneut}) states the
Universe charge neutrality in terms of the electron chemical
potential, with $\phi_e \equiv \mu_e/T$ and $T$ the temperature of
$e^{\pm},\gamma$ plasma, and finally Equations (\ref{nuboltz}) are
the Boltzmann equations for neutrino species, with $I_{\nu_\alpha}
\left[ f_{\nu_e},f_{\nu_x} \right]$ standing for the collisional
integral which contains all microscopic processes creating or
destroying the specie $\nu_\alpha$. In the following we do not
consider neutrino oscillations, whose effect has been proved to be
sub-leading and affect $^4{\rm He}$ mass fraction for $\sim 1
{\cdot} 10^{-4}$ \cite{hannestad}.

In the standard scenario of no extra relativistic degrees of
freedom at BBN epoch apart from photons and neutrinos, the
neutrino chemical potential is bound to be a small fraction of
neutrino temperature, smaller than approximately
$|\xi_\nu|\equiv|\mu_\nu/T_\nu| \leq 0.1$
\cite{cuocoetal,Dolgov:2002ab,melch1,melch2}. This bound applies
to all neutrino flavors, whose distribution functions are
homogenized via flavor oscillations~\cite{Dolgov:2002ab,Abazajian:2002qx}.
In view of this, as far as
this analysis is concerned, we will focus on non degenerate
neutrinos, so that $f_{\nu_e}=f_{\bar{\nu}_e}$ and
$f_{\nu_x}=f_{\bar{\nu}_x}$.

The neutrino energy density and pressure are defined in terms of their
distributions as
\be
\rho_\nu = 3 \, \p_\nu =  2 \, \int
\frac{d^3 p}{(2 \pi)^3} \, \left|\vec{p}\right|\, \left[
f_{\nu_e}+ 2 \, f_{\nu_x} \right]
\vv
\ee
whereas for baryons we have
\bea
\rho_B &=& \left[M_u  + \sum_i \left(
\Delta M_i + \frac{3}{2} \, T
\right)~ X_i \right] n_B\,\,\, ,
\label{neupress}
\nonumber\\
\p_B & = & T \, n_B \, \sum_i X_i \,\,\, . \label{barpress} \eea
with the $\Delta M_i$ and $M_u$ the i-th nuclide mass excess and the atomic
mass unit respectively. Finally, pressure and energy density for the
electromagnetic plasma ($e^{\pm}$ and $\gamma$) is calculated by improving the
incoherent scattering limit result (particles correspond to poles of the
propagators at, respectively, $p^2=0$ (photons) and $p^2 = m_e^2$
(electrons/positrons)) and considering the effect of finite temperature QED
corrections. It has been shown in fact that at the time of BBN, these
corrections affect Equations
\eqn{e:drdt}-\eqn{nuboltz} at some extent and slightly influence
the $^4$He abundance \cite{Lopez}. Indeed finite temperature QED
corrections modify the electromagnetic plasma equation of state and thus
influence Equation (\ref{e:drhodt}), the expression of the expansion rate
$H$ and, as it will be discussed in the following, the conversion rates $n
\leftrightarrow p$.

The change in the electromagnetic plasma equation of state can be
evaluated by considering the corrections induced on the $e^{\pm}$
and photon masses, using the tools of Real Time Finite Temperature
Field Theory \cite{rtftft}. For the electron/positron mass, up to
order $\alpha \equiv e^2/\left( 4 \,\pi \right)$, we find the
additional finite temperature contribution
\cite{Mangano:2001iu,heckler}
\begin{eqnarray} \fl
\delta m_e^2\left( |\bvec{p}|, \,T \right) &=&
\frac{2\,\pi\,\alpha\,T^2}{3} + \frac{4\,\alpha}{\pi} \, \int_0^\infty
d |\bvec{k}| \, \frac{|\bvec{k}|^2}{E_k} \, \frac{1}{{\rm e}^{E_{k}/T}+1}
\nonumber\\ &-&
\frac{2 \, m_e^2 \,\alpha}{\pi \,|\bvec{p}|} \, \int_0^\infty d |\bvec{k}| \,
\frac{|\bvec{k}|}{E_{k}} \, \log\left| \frac{|\bvec{p}|+|\bvec{k}|}
{|\bvec{p}|-|\bvec{k}|} \right| \frac{1}{{\rm e}^{E_{k}/T}+1} \vv
\label{dme}
\end{eqnarray}
where $E_{k} \equiv \sqrt{|\bvec{k}|^2 + m_e^2}$. We note that in
this equation the term depending on the $e^{\pm}$ momentum
$|\bvec{p}|$ contributes for less than $10\,\%$ to $\delta
m_e^2\,$ and thus it can be safely neglected. This highly
simplifies the analysis and is fully legitimate at the level of
precision we are interested in (corrections to $Y_p$ at the level
of $0.0001$).

The renormalized photon mass in the electromagnetic plasma is instead, up
to order $\alpha$, given by (see for example \cite{nicolao})
\begin{equation}
\delta m_\gamma^2 \left( T \right) = \frac{8\,\alpha}{\pi} \,
\int_0^\infty d |\bvec{k}| \, \frac{|\bvec{k}|^2}{E_k} \,
\frac{1}{{\rm e}^{E_{k}/T}+1}\pp
\label{dmg}
\end{equation}
The corrections~(\ref{dme}) and (\ref{dmg}) modify the
corresponding dispersion relations as $E_i^2 =|\bvec{k}|^2 + m_i^2
+ \delta m_i^2 \left( T \right)\,$ ($i = e,\, \gamma$). Thus the
total pressure and energy density of the electromagnetic plasma
result to be
\begin{eqnarray}
\p_\gamma+\p_e &=& \frac{T}{\pi^2} \, \int_0^\infty d |\bvec{k}| \, |\bvec{k}|^2 \, \,
\log\left[\frac{\left( 1 + {\rm e}^{-E_e/T} \right)^2}{ \left( 1 -
{\rm e}^{-E_\gamma/T} \right)} \right]\vv \label{press}\\
\rho_\gamma + \rho_e &=& \left( T \,
\frac{d}{d T} - 1\right) (\p_\gamma+\p_e ) \pp
\label{energ}
\end{eqnarray}
Expanding these expressions with respect to $\delta m_e^2$ and
$\delta m_\gamma^2$, one obtains the first order correction
\begin{eqnarray}
\delta \p_\gamma + \delta \p_e &=& - \int_0^\infty \frac{|\bvec{k}| \, d|\bvec{k}|}
{2 \, \pi^2} \left[
\frac{|\bvec{k}|}{E_k} \, \frac{\delta m_e^2 \left( T \right)}{{\rm
e}^{E_{k}/T}+1} + \frac{1}{2} \, \frac{\delta m_\gamma^2 \left( T
\right)}{{\rm e}^{k/T}-1} \right] \,\,.
\label{pint}
\end{eqnarray}
The energy density is then obtained by using the expression for
$\p_\gamma+\p_e$ in Equation (\ref{energ}).

The equations (\ref{e:drdt}), (\ref{e:drhodt}) and (\ref{nuboltz})
can be solved separately since they are essentially determined by
relativistic particles, and only negligibly affected by the
baryons contribution\footnote{The effect of baryons on neutrino
decoupling is due to nucleon--neutrino scattering processes, as
well as to the baryon contribution to Hubble parameter $\rho_B$.
Both effects are expected to produce very small corrections of
order $\eta$ or $M_u \eta /T$.}. This allows to solve the
evolution of the neutrino species first, and then to substitute
the result into the remaining equations.

\subsection{Neutrino decoupling $*$}
\label{s:neudec}

The evolution of neutrino distribution function can be highly
simplified by using the scale factor as evolution variable $x
\equiv m_e\, a$, and the comoving momentum $y \equiv \left|\vec{p}
\right |\, a$. We also define the {\it rescaled} photon
temperature as $\bar{z} \equiv T\,  a$. By using these
definitions, the set of equations (\ref{e:drhodt}) and
(\ref{nuboltz}), once baryonic contribution is neglected, becomes
\bea && \frac{d}{dx} \bar{\rho}(x) = \frac 1x\,
\left(\bar{\rho}-3\bar{\p}
\right) \vv \label{energy2} \\
&& \frac{d}{d x}\, f_{\nu_\alpha}(x,y) = \frac{1}{x H}\, I_{\nu_\alpha}
\left[ f_{\nu_e},f_{\nu_x} \right] \vv ~~~~
\mbox{with} ~~~ \nu_\alpha=\nu_e,  \nu_x \pp \label{boltz} \eea In
Equation \eqn{energy2}, which states the conservation of the total energy
momentum, $\bar{\rho}$ and $\bar{\p}$ are the dimensionless energy density
and pressure of the primordial plasma, respectively, \be \bar{\rho} = \rho
\left (\frac{x}{m_e} \right)^4 \sim \rho_{NB}
\left (\frac{x}{m_e} \right)^4
,\,\, \qquad \bar{\p} = \p \left (\frac{x}{m_e} \right)^4 \sim \p_{NB}
\left (\frac{x}{m_e} \right)^4 \pp \ee As
in Reference \cite{Mangano:2001iu,Esposito:2000hi} the unknown neutrino
distributions are parameterized as
\begin{equation}
f_{\nu_\alpha} \left( x,y \right) = \frac{1}{e^{y }+1} \left( 1 + \delta
f_{\nu_\alpha}(x,y) \right) = \frac{1}{{\rm e}^y +1} \, \left[ 1 +
\sum_{i=0}^\infty\, a_{i}^\alpha(x)\, P_i \left( y \right) \right]  \,,
\label{expan}
\end{equation}
where $P_i(y)$ are orthonormal polynomials with respect to the Fermi
function weight
\begin{equation}
\int_0^\infty \frac{dy}{{\rm e}^{y}+1} \, P_i \left( y \right) \, P_j
\left( y \right) = \delta_{ij} \,\,.
\label{ortho}
\end{equation}
With these definitions the problem of finding the (momentum
dependent) distortion in neutrino distribution function is then
reduced to determine the time evolution of suitable linear
combinations of the lower momenta of these distributions. This
method has been shown to provide results in quite a good agreement
with different approaches, which instead use discretized comoving
variables and solve Boltzmann equations on a grid in the $y$
variable \cite{DHS,replyDHS}.

By substituting Equation (\ref{expan}) into Equations
(\ref{boltz}), including the QED corrections discussed previously,
and using the covariant conservation of the energy momentum tensor
(\ref{energy2}) as an evolution equation for $\bar{z}$, one gets
\begin{eqnarray}
&\frac{d \bar{z}}{dx} &= \frac{\frac{x}{\bar{z}}J(x/\bar{z})-
\frac{1}{2 \pi^2
\bar{z}^3}\int_0^\infty~dy\,y^3\left(\frac{df_{\nu_e}}{dx}+
2\frac{df_{\nu_x}}{dx}\right)+G_1(x/\bar{z})}
{\frac{x^2}{\bar{z}^2}J(x/\bar{z})+Y(x/\bar{z})+\frac{2\pi^2}{15}+G_2(x/\bar{z})}
\vv
\label{dzdx}\\
&\frac{d}{d x}&\, a_{i}^\alpha(x) = \frac{1}{x H}\, \int_0^\infty
dy \, P_i \left( y \right) \, I_{\nu_\alpha}\left[
f_{\nu_e},f_{\nu_x} \right] \vv \label{eqc}
\end{eqnarray}
where
\begin{eqnarray} \fl
&&G_1(\omega) = 2 \pi \alpha \left[ \frac{1}{\omega}
\left(\frac{K(\omega)}{3} + 2 K(\omega)^2-
\frac{J(\omega)}{6}- K(\omega)~J(\omega) \right) \right. \nonumber \\
\fl &&+ \left.  \left(
\frac{K'(\omega)}{6}-K(\omega)~K'(\omega)+\frac{J'(\omega)}{6} +
J'(\omega)~K(\omega) + J(\omega)~K'(\omega)\right ) \right]
\\ \fl &&G_2(\omega) =
-8 \pi \alpha \left( \frac{K(\omega)}{6} + \frac{J(\omega)}{6} -
\frac{1}{2} K(\omega)^2 + K(\omega)~J(\omega)\right) \nonumber\\
\fl &&+2 \pi \alpha \omega
\left(\frac{K'(\omega)}{6}-K(\omega)~K'(\omega)
+\frac{J'(\omega)}{6} + J'(\omega)~K(\omega) +
J(\omega)~K'(\omega)\right) \vv
\end{eqnarray}
with
\begin{eqnarray}
K(\omega) & = & \frac{1}{\pi^2}~\int_0^\infty
du~\frac{u^2}{\sqrt{u^2+\omega^2}}~
\frac{1}{\exp\left(\sqrt{u^2+\omega^2}\right)+1} \vv \\
J(\omega) & = &  \frac{1}{\pi^2}~\int_0^\infty
du~u^2~\frac{\exp\left(\sqrt{u^2+\omega^2}\right)}
{\left(\exp\left(\sqrt{u^2+\omega^2}\right)+1\right)^2} \vv \\
Y(\omega) & = &  \frac{1}{\pi^2}~\int_0^\infty
du~u^4~\frac{\exp\left(\sqrt{u^2+\omega^2}\right)}
{\left(\exp\left(\sqrt{u^2+\omega^2}\right)+1\right)^2} \pp
\end{eqnarray}
The functions $K'(\omega)$ and $J'(\omega)$ stand for the first
derivative of $K(\omega)$ and $J(\omega)$ with respect to their
argument.
\begin{figure}
\begin{center}
\includegraphics[scale=0.5]{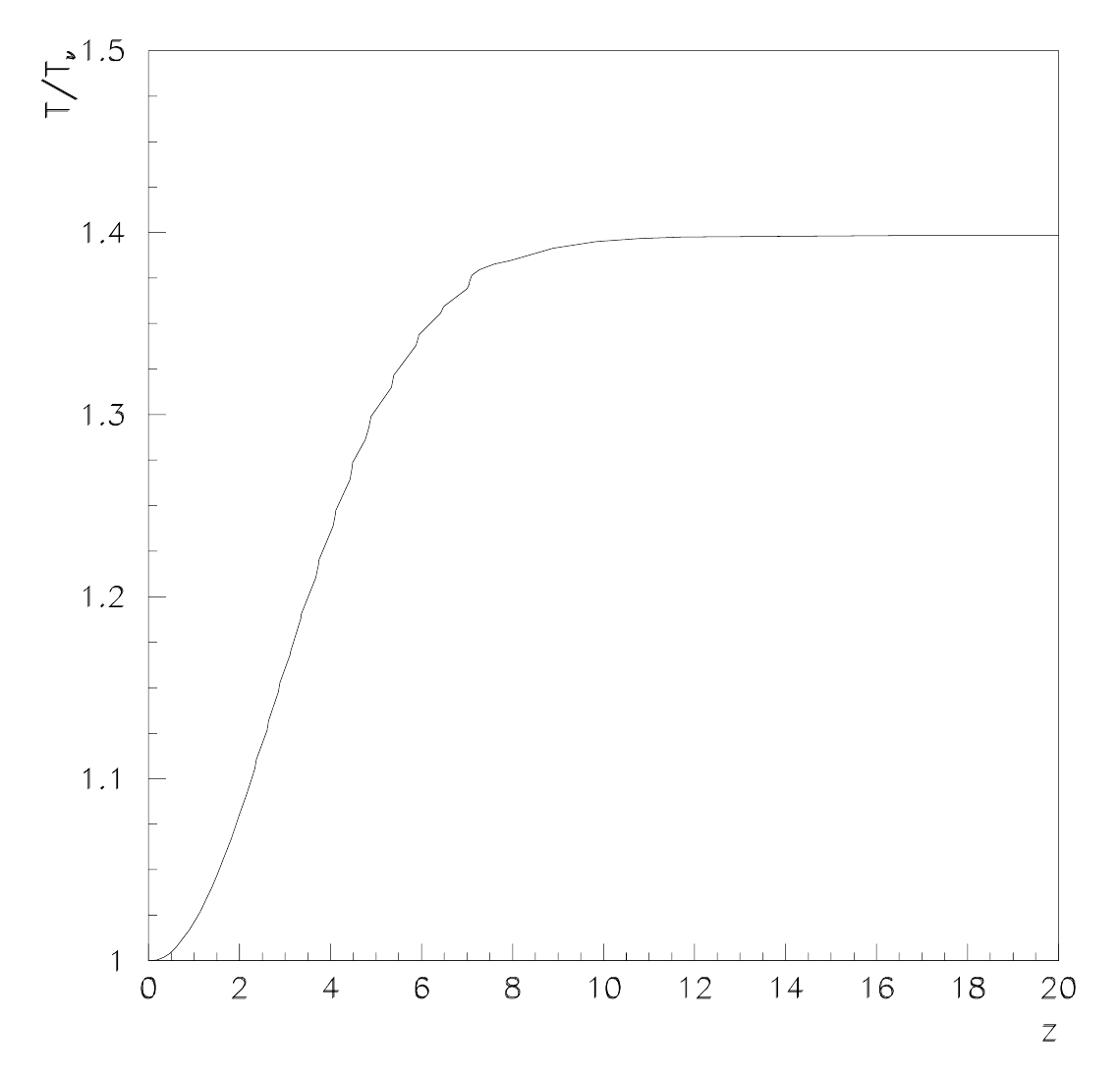}
 \caption{The evolution of
$\bar{z}=T/T_\nu$ versus $z= m_e/T$. The asymptotic value at small
temperatures is $\bar{z}$=1.3984 }\label{zbarevol}
\end{center}
\end{figure}

Since at high temperature, $T \sim 10\, {\rm MeV}$, the neutrinos
are in thermal equilibrium with the electromagnetic plasma, the
initial condition for the coefficients is $a_{i}^\alpha=0$ and we
can always choose the initial value for the scale factor such that
we have, at $10\, {\rm MeV}$, $\bar{z}=1$. From Equation
\eqn{dzdx}, neglecting the terms proportional to the derivative of
neutrino distributions, as well as QED corrections, one gets the
asymptotic value $\bar{z}_{eq}^D=(11/4)^{1/3}=1.4010$ which
represents the ratio between the photon and neutrino temperatures
after the complete annihilation of $e^+e^-$ pairs into photons
only. This limit is often referred to as the instantaneous
decoupling value. A fully numerical treatment of the decoupling
shows indeed that the asymptotic value of $T/T_\nu$,
$\bar{z}=1.3984$, is lower than $\bar{z}_{eq}^D$, since neutrino
plasma is slightly heated by the $e^+e^-$ annihilations. We report
in Figure \ref{zbarevol} the actual behavior of $\bar{z}$. In
\cite{Mangano:2001iu,Esposito:2000hi} it was shown that a very
good approximation, giving results with accuracy of $1 \%$, is to
consider polynomials in (\ref{expan}) up to third order. The
neutrino distribution function can be also written as
\begin{equation}
f_{\nu_\alpha} \left( x,y \right) \simeq \frac{1}{e^{y }+1} \left(
1 + \sum_{i=0}^3\, c_{i}^\alpha(x)\, y^i\right) \, \pp
\label{expan2}
\end{equation}
\begin{figure}
\begin{center}
\includegraphics[scale=0.5]{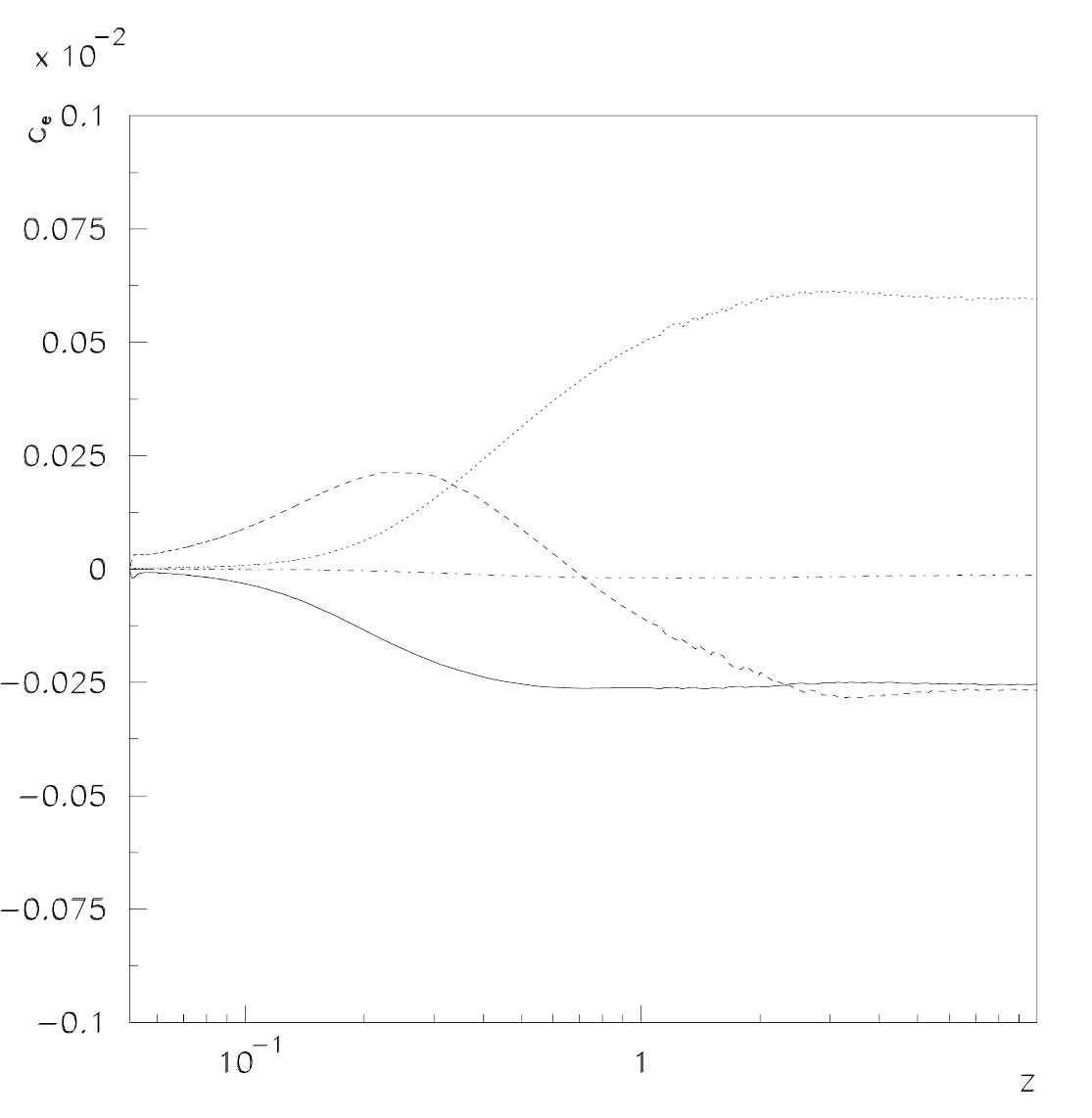}
\caption{The evolution
of the electron neutrino distortion coefficients $c_0^e$ (solid),
$c_1^e$ (dashed), $c_2^e$ (dotted) and $c_3^e$ (dot-dashed) versus
$z= m_e/T$}\label{f:ce}
\end{center}
\end{figure}
\begin{figure}
\begin{center}
\includegraphics[scale=0.5]{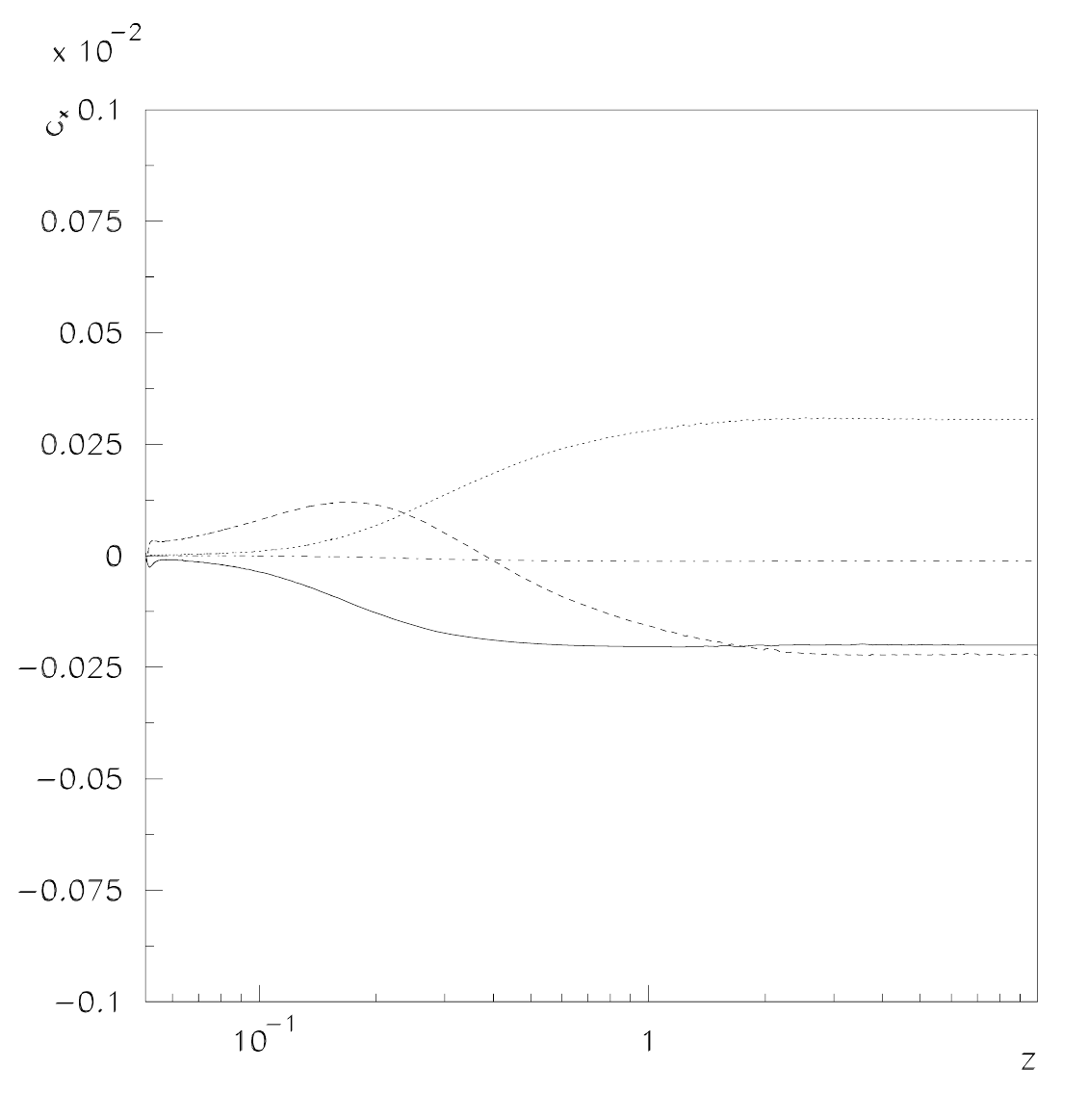}
\caption{The evolution
of the $\mu,\tau$ neutrino distortion coefficients $c_0^x$
(solid), $c_1^x$ (dashed), $c_2^x$ (dotted) and $c_3^x$
(dot-dashed) versus $z= m_e/T$}\label{f:cx}
\end{center}
\end{figure}
In Table 2 we report the asymptotic values of the $c_{i}^\alpha$
coefficients, as found in \cite{Mangano:2001iu}, while their
evolution versus $z$ is shown in Figures \ref{f:ce} and
\ref{f:cx}, respectively.
\begin{table*}
\begin{center}
\begin{tabular}{ccccc}
\hline Flavor ($\alpha$)& $c^\alpha_0$ & $c^\alpha_1$ &
$c^\alpha_2$ & $c^\alpha_3$\\ \hline $e$& -2.507& -2.731& 6.010&
-0.1419\\ $\mu,\tau$& -2.003& -2.196& 3.061& -0.1091\\ \hline
\end{tabular}
\end{center}
\caption{Values of the coefficients of Equation (\ref{expan2}) in
unit of $10^{-4}$} \label{coeff2}
\end{table*}

\subsection{Numerical solution of the BBN set of equations $*$}
\label{s:numsol}

Once neutrino distribution functions are determined, the BBN set
of equations is reduced to Equations (2.4)-(2.8) of Section 2.1.
This set can be further reduced. Actually it is more convenient to
follow the evolution of the $N_{nuc}+1$ unknown functions $(\pe,~
X_j)$ in terms of the dimensionless variable $z=m_e/T$, and to use
Equation \eqn{e:charneut} to get $n_B$ as a function of $\pe$. The
new set of differential equations may be cast in the form \be
\frac{d \pe}{dz}\, =\, \frac{1}{z}~ \frac{L~ E~ F + (z\, L_z - 3\,
L)~ G}{L~ E~ \frac{\ds \partial \hrho_e}{\ds
\partial \pe} - L_\pe~ G} \vv
\label{e:basic1a}
\ee
\be
\frac{dX_i}{dz}\, =\, - \frac{\hgi}{z}~ \frac{L_\pe~ F + (z\, L_z
- 3\, L)~ \frac{\ds
\partial \hrho_e}{\ds \partial \pe}}{L~ E~ \frac{\ds \partial
\hrho_e}{\ds \partial \pe} - L_\pe~ G} \vv \label{e:basic2a}
\ee
where the functions $E$, $F$ and $G$ are given by
\bea
E(z,\,
\pe,\, X_j) &=& 3\, \hH - \frac{ \ds \sum_i~ Z_i\, \hgi}{\ds
\sum_j~ Z_j\, X_j} \vv
\label{e:funce} \\
F(z,\, \pe,\, X_j) &=& 4\, \hrho_{e,\gamma} + \frac32~ \hp_B - z\,
\frac{\partial \hrho_e}{\partial z} -z\frac{\partial
\hat{\rho}_\gamma}{\partial z} \vv
\label{e:funcg} \\
G(z,\, \pe,\, X_j) &=& 3\, \hH \left(\hrho_{e,\gamma} +
\hp_{e,\gamma}+\hp_B + \frac{N(z)}{3}\right) \nonumber \\
&+& \frac{z\, L}{\sum_j Z_j\, X_j}\sum_i \left( \hdmi +
\frac{3}{2\, z} \right)\hgi \vv \eea with \be N(z) = \left.
\frac{1}{\bar{z}^4} \left( x \frac{d}{dx} \bar{\rho}_\nu \right)
\right|_{x=x(z)} \vv \label{e:funcf} \ee and $H \equiv m_e\, \hH$,
$n_B \equiv m_e^3~ \hnb$, $\Gamma_i \equiv m_e\, \hgi$, $\rho
\equiv T^4\,\hrho$, $\p \equiv T^4\, \hp$, and finally
$\bar{\rho}_\nu = \rho_\nu (x/m_e)^4$. Note that by using the
previous definitions, it is possible to express $\hrho_B$ and
$\hp_B$ as functions of $z$, $\pe$, and $X_i$ only \bea \hrho_B
&=& \frac{z\, L (z, \pe)}{\sum_j Z_j\, X_j}~ \left[ \hmu + \sum_j
\left( \hdmj\, +\, \frac{3}{2\, z} \right)\, X_j \right] \vv \\
\hp_B &=& \frac{L (z, \pe)}{\sum_j Z_j\, X_j}~ \sum_j X_j \pp \eea
With $\hdmi$ and $\hmu$ we denote the i-th nuclide mass excess and
the atomic mass unit, respectively, normalized to $m_e$. The
values of the partial derivative of $L$ with respect to $z$ and
$\pe$, denoted with $L_z$ and $L_{\pe}$, and the quantities
$\hrho_e$, $\partial \hrho_e/\partial z$ and $\partial
\hrho_e/\partial \pe$ in a form which is suitable for a BBN code
implementation can be find in Appendix A of
\cite{Esposito:2000hh}.

The neutrino contribution to the previous equations, via the
function $N(z)$, can be obtained by the solution of Equations
(\ref{dzdx}), (\ref{eqc}), written as function of the variable
$z$. To this end it is necessary to invert (numerically) the
relation \be \frac{x}{\bar{z}(x)} = z \longrightarrow x = x(z) \pp
\ee It is interesting to notice that neutrinos only contribute to
the $z$ evolution of $\phi_e$ and $X_i$ via the non equilibrium
terms in their distribution functions, since \be x \frac{d}{dx}
\bar{\rho}_\nu=  \sum_{i=0}^3\,  \left( \frac{1}{\pi^2}\int dy\,
y^{3+i} \frac{1}{e^y+1} \right) \,x \frac{d}{dx}\, c_{i}^\alpha(x)
\pp \ee This expression can be numerically calculated as a
function of $x$, and then it should be evaluated at $x(z)$. This
result is quite expected. In fact neutrinos well before decoupling
share the same temperature of the electromagnetic plasma, so
$\bar{z}$ is a constant in this limit. On the other hand, once
they are fully decoupled at low temperatures, they satisfy an
independent entropy conservation condition, and so they do not
further contribute to the time evolution of $z$. Only during the
electron/positron annihilation phase they do affect this
evolution, via the non equilibrium terms $\delta
f_{\nu_\alpha}(x,y)$, which in fact are the genuine effects of
their residual coupling to the electromagnetic plasma.

\begin{figure}[!hbtp]
\begin{center}
\includegraphics[scale=0.5]{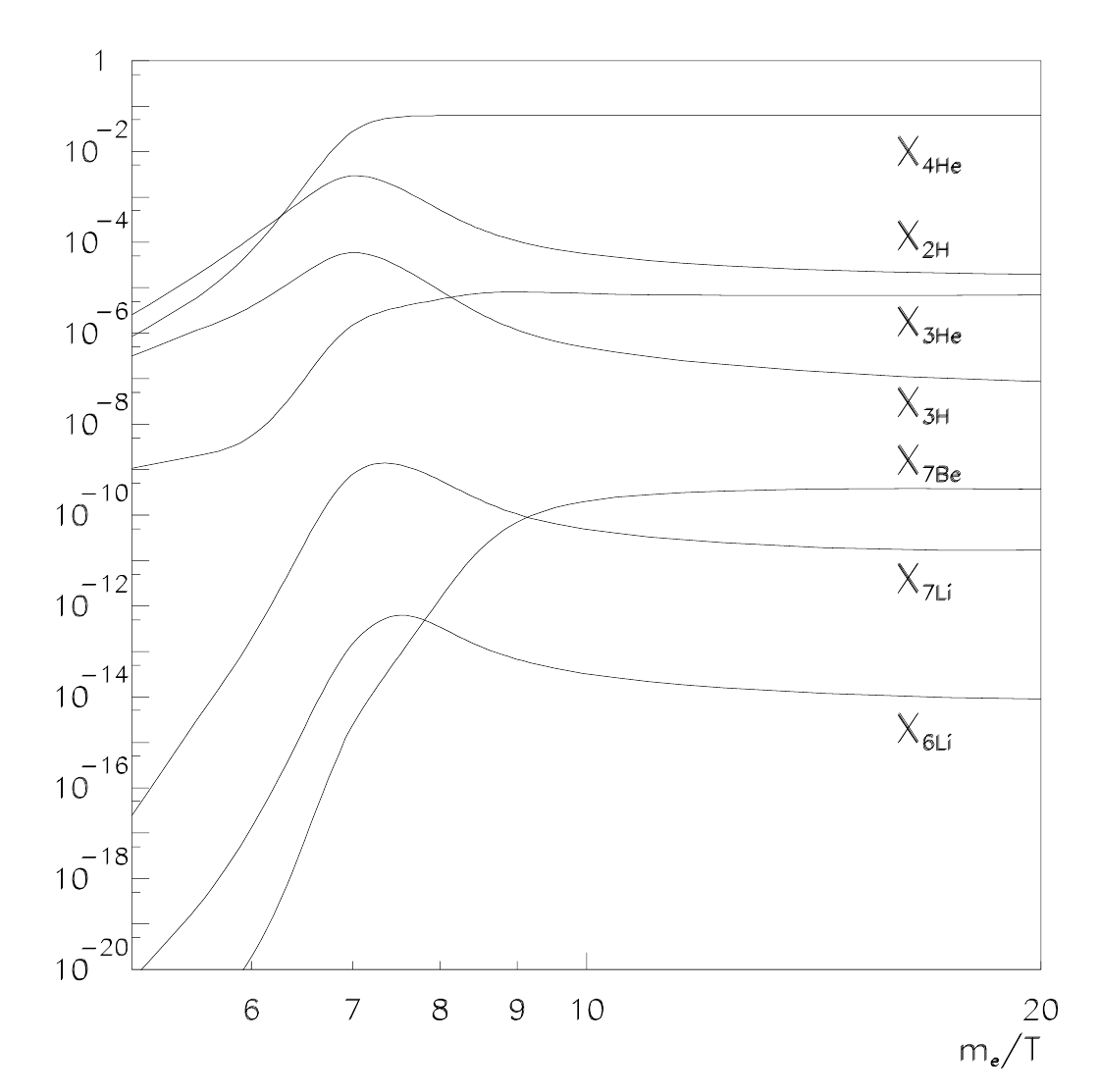}
\caption{Nuclear abundances depart from their equilibrium values, undergo an
out-of-equilibrium phase, and eventually reach their final
values}\label{BBNevol}
\end{center}
\end{figure}

Equations \eqn{e:basic1a}-\eqn{e:basic2a} are solved by imposing the
following initial conditions at $z_{in}= m_e/(10\, {\rm MeV})$
\bea
\pe (z_{in}) &=& \pe^0 \vv \\
X_n (z_{in}) &=& \left(\exp\{\hat{q}\, z_{in}\}+1\right)^{-1} \vv \\
X_p(z_{in}) &=& \left(\exp\{-\hat{q}\, z_{in}\}+1\right)^{-1} \vv \\ X_i
(z_{in}) &=& \frac{g_i}{2}~ \left( \zeta(3) \sqrt{\frac{8}{\pi}}
\right)^{A_i-1} ~ A_i^\frac{3}{2}\, \left( \frac{m_e}{M_N z_{in}}
\right)^{\frac{3}{2} (A_i-1)} \eta_i^{A_i-1}\, X_p^{Z_i}\, X_n^{A_i-Z_i}  \nn
\\
&{\times}& \exp \left\{ \hat{B}_i \, z_{in} \right\}
\quad\quad\quad\quad\quad i= \,^2{\rm H}, \,^3{\rm H}, ... \pp
\eea In the previous equations $\hat{q}=(M_n- M_p)/m_e$, and the
quantities $A_i$ and $\hat{B}_i$ denote the atomic number and the binding
energy of the $i-$th nuclide normalized to electron mass, respectively.
Finally $\eta_i$ is, the initial value of the baryon to photon number
density ratio at $T=10\,{\rm MeV}$ (we will discuss the final to initial
value ratio of $\eta$ in the following), and $\pe^0$ the solution of the
implicit equation \be L (z_{in},\,
\pe^0) = \frac{2\, \zeta(3)}{\pi^2}~ \eta_i~ \sum_i Z_i\, X_i
(z_{in}) \pp \ee The method of resolution of the BBN equations
\eqn{e:basic1a}, \eqn{e:basic2a} is the same applied in
\cite{Esposito:1999sz}. It uses a class of Backward Differentiation
Formulas with Newton's method, implemented in a NAG routine with adaptive
step-size (see \cite{Esposito:1999sz} for more details). As an example of
the evolution of nuclei abundance we report in Figure \ref{BBNevol} the
result of our numerical code for $\omega_b =0.023$ and three neutrinos
versus $m_e/T$.

\section{The Nuclear rates}\label{seteq}

\subsection{Weak Reactions $*$}

The weak reactions transforming $n \leftrightarrow p$, namely
\bea
(a)~~~ \nu_e + n \rightarrow e^- + p &~~~,~~~~~~& (d)~~~ \neb + p
\rightarrow e^+ + n \vv \nonumber \\
(b)~~~e^- + p \rightarrow \nu_e + n &~~~,~~~~~~& (e) ~~~n \rightarrow e^- +
\neb + p \vv \nonumber \\
(c)~~~ e^+ + n \rightarrow \neb + p &~~~,~~~~~~& (f)~~~ e^- + \neb
+ p \rightarrow n\pp \label{e:reaction} \eea are the leading
processes in fixing the neutron abundance at the onset of BBN and
thus a key quantity in determining the $^4$He mass fraction. In
view of this, much effort has been devoted to refining the
theoretical accuracy in evaluating these processes, which
presently is at the order of $0.1 \%$. We here summarize the main
results obtained in the literature, while more details can be
found in \cite{Lopez,EMMP1,Esposito:1999sz}.

\begin{figure}[!hbtp]
\begin{center}
\includegraphics[scale=0.5]{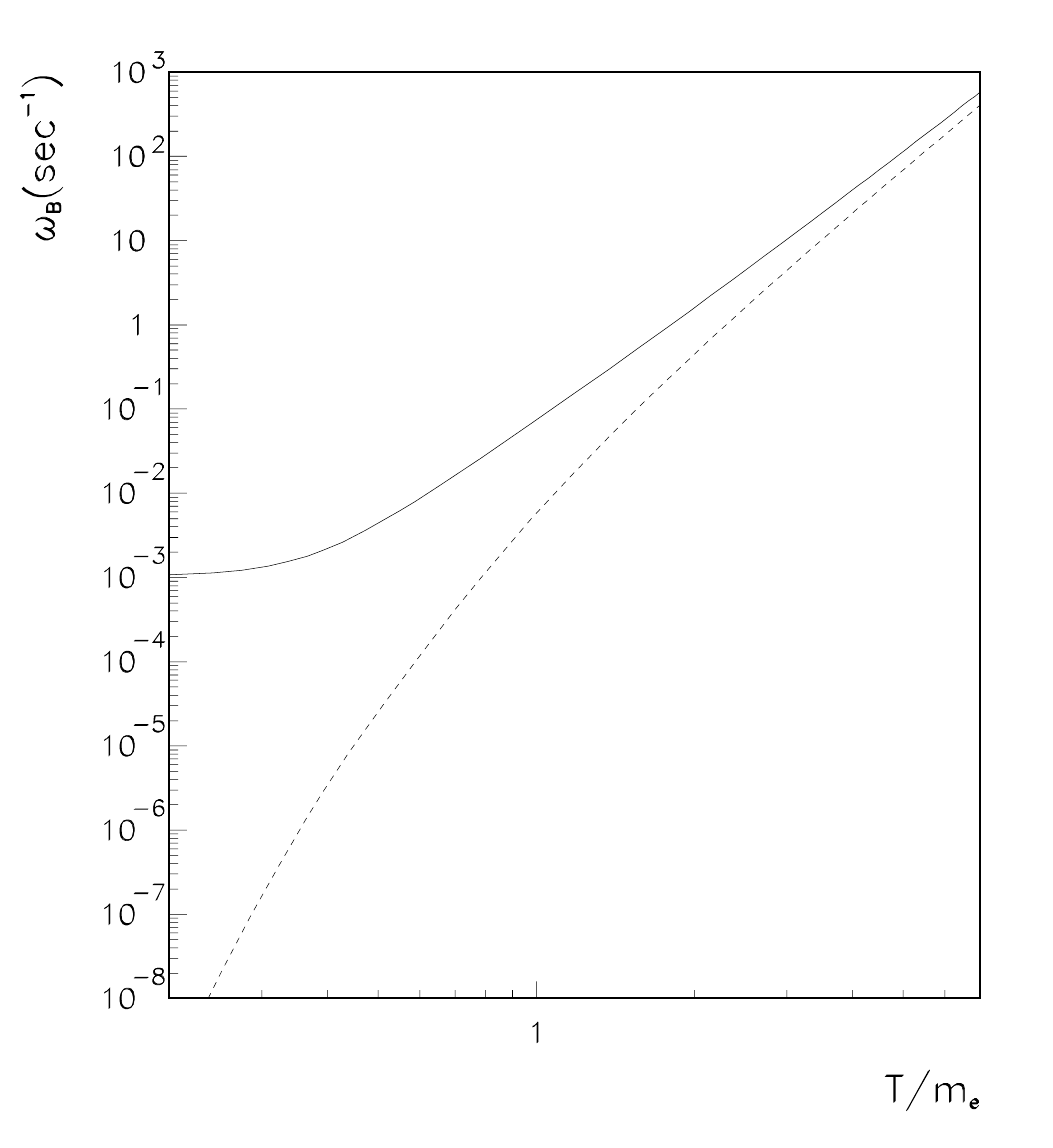}
\end{center}
\caption{The total Born rates, $\omega_B$, for $n \rightarrow p$
(solid line) and $p \rightarrow n$ transitions (dashed line).}
\label{f:born}
\end{figure}

The Born rates are the tree level estimates obtained with $V-A$ theory and
with infinite nucleon mass. As an example, for the neutron decay process
$(e)$ we have \bea  \fl \omega_B ( n \rightarrow e^- + \neb + p ) \, &=& \,
\frac{G_F^2 \tre}{2 \pi^3} \,
\int_0^\infty d {|\bvec{p}'| \,|\bvec{p}'|^2} \,  q_0^2 \,
\Theta(q_0) {\times} \nonumber \\
&{\times}& \left[ 1 - f_{\bar{\nu}_e} (q_0)\right]  \left[ 1 -
f_e(p_0') \right] \vv \label{e:cb15} \eea where $G_F$ is the Fermi
coupling constant, $\cv$ and $\ca$ the nucleon vector and axial
coupling. According to our notation $\bvec{p}'$ and $p_0'$ are the
electron momentum and energy, and $q_0= M_n - M_p - p_0' \equiv
\Delta - p_0'$ the neutrino energy. The rates for all other
processes $(a)-(d), (f)$ can be simply obtained from
(\ref{e:cb15}) properly changing the statistical factors and the
expression for $q_0$ (see for example \cite{EMMP1}). Average is
performed at this level of approximation over equilibrium
Fermi-Dirac distribution for leptons, i.e. neglecting the effects
of distortion in neutrino/antineutrino distribution functions. In
Figure \ref{f:born} we report the Born rates $\omega_B$ for $n
\leftrightarrow p$ processes,
The accuracy of Born approximation results to be, at best, of the
order of $7\%$. This can be estimated by comparing the prediction
of Equation \eqn{e:cb15} for the neutron lifetime at very low
temperatures, with the experimental value $\tau_n^{ex} = (885.7
{\pm} 0.8)~s$ \cite{PDG}.

A sensible improvement is obtained by considering four classes of
effects
\begin{itemize}
\item[{1)}] electromagnetic radiative corrections, which largely
contribute to the rates of the fundamental processes; \item[{2)}]
finite nucleon mass corrections, which are of the order of $T/M_N$
or $m_e/M_N$, $M_N$ being the nucleon mass; \item[{3)}]
thermal/radiative effects, proportional to the surrounding plasma
temperature; \item[{4)}] non instantaneous neutrino decoupling
effects on the Born rates. Distortion in electron neutrino
distribution function directly enters the thermal average of weak
rates, as well as the detailed evolution of the temperature ratio
$T/T_\nu=\bar{z}$.
\end{itemize}

\subsubsection{Electromagnetic radiative corrections.}
\label{s:rad}

Electromagnetic radiative corrections to the Born amplitudes for processes
(\ref{e:reaction}) are typically split into $outer$ and $inner$ terms (for
a review see e.g. \cite{Wilkinson}). The first ones involve the nucleon as
a whole and consist of a multiplicative factor to the squared modulus of
transition amplitude of the form \be 1+
\frac{\alpha}{2 \pi} g(p_0',q_0)~~~. \label{e:outer} \ee The function
$g(p_0',q_0)$, can be found in Reference \cite{Sirlin}, and depends on both
electron and neutrino energies. On the other hand, the inner corrections
are deeply related to the nucleon structure. They have been estimated in
Reference \cite{Marciano}, and result in the additional multiplicative
factor
\be 1 + \frac{\alpha}{2 \pi} \left( 4 \, \ln
\frac{M_Z}{M_p} \, + \, \ln \frac{M_p}{M_A} \, + \, 2 C \, + A_g
\, \right)~~~, \label{e:inner} \ee where the first term is the
short--distance contribution and $A_g=-0.34$ is a perturbative QCD
correction. The other two terms are related to the axial--induced
contributions, with $M_A= (400 \div 1600) {\rm MeV}$ a low energy cut-off
in the short-distance part of the $\gamma W$ box diagram, and $C$ is
related to the remaining long distance term.

Summing up these two kind of corrections, and resumming all
leading logarithmic corrections $\alpha^n ln^n(M_Z)$
\cite{Marciano2}, one gets the global multiplicative factor \be
\fl {\cal G}(p_0',q_0) = \left[ 1 \, + \,\frac{\alpha}{2 \pi}
\left( \ln \frac{M_p}{M_A} \, + \, 2 C  \right) \, + \,
\frac{\alpha (M_p)}{2 \pi} \, \left[ g(p_0',q_0) \, + \, A_g
\right] \right] S(M_p , M_Z)~~~, \label{e:inout} \ee where $\alpha
(\mu)$ is the QED running coupling constant defined in the
$\ov{MS}$ scheme and $S(M_p,M_Z)$ a short distance rescaling
factor, defined in \cite{EMMP1}.

Another electromagnetic effect is present when both electron and
proton are either in the initial or final states (namely,
processes $(a), (b), (e)$ and $(f)$). It is the {\it Coulomb
correction} due to rescattering of the electron in the field of
the proton and leading to the Fermi function  \be {\cal F}(p_0')
\, \simeq \, \left( 1 \, + \, \alpha \pi \,
\frac{p_0'}{{|\bvec{p}'|}} \right)\pp \label{e:coulomb} \ee
Summing all these corrections and using the most recent estimates
for $G_F=(1.16637{\pm} 0.00001){\cdot} 10^{-5}\, GeV^{-2}$,
$C_V=0.9725 {\pm} 0.0013$ and the ratio $C_A/C_V=-1.2720 {\pm}
0.0018$, \cite{PDG}, and including the (small) finite nucleon mass
corrections discussed in the next Section, the theoretical
prediction for neutron lifetime, which as we said can be used as a
check of accuracy, is now $\tau_n^{th} = 886.5~s$. The inclusion
of electromagnetic radiative corrections therefore gives quite a
satisfactory result for $\tau_n$, with accuracy of the order of
$0.1 \%$. As in \cite{Esposito:1999sz,EMMP1} we deal with this
tiny residual discrepancy by assuming the presence of a further
multiplicative factor which properly renormalizes the theoretical
prediction. Such a factor $1+ \delta_{\tau}= \tau_n^{th}
/\tau_n^{ex}= 1.001$, is then applied as a rescaling factor to all
processes (\ref{e:reaction}). Notice how this value is remarkably
closer to unity with respect to what estimated in
\cite{Esposito:1999sz,EMMP1}. This is simply because of the
different experimental value of $\tau_n^{ex}$ quoted in \cite{PDG}
with respect to previous estimates. In closing we have to mention
that there may be still a systematic effect induced by adopting
this procedure, since it assumes that all residual discrepancy
parameterized by $\delta_\tau$ is independent on energies of the
outgoing particles. However it is worth saying that with present
data on $^4$He abundance, this effect is hardly playing a role in
a comparison of theoretical prediction for BBN with data.

\subsubsection{Finite nucleon mass corrections.$*$}
\label{s:finitemass}

There are relevant contributions to the $n \leftrightarrow p$
rates which appear by relaxing the infinite nucleon mass
approximation. These effects are proportional to $m_e/M_N$ or
$T/M_N$, and in the temperature range relevant for BBN, can be as
large as the radiative corrections. This has been first pointed
out in \cite{Seckel} and then also numerically evaluated in
\cite{Lopez,EMMP1}.

At order $1/M_N$, the weak hadronic current receives a
contribution from the weak magnetic moment coupling \be J_\mu^{wm}
= i \frac{G_F}{\sqrt{2}} \frac{f_2}{M_N} \, \ov{u}_p(p) \,
\sigma_{\mu \nu} \, (p-q')^\nu u_n(q')~~~, \ee where, from CVC,
$f_2 = V_{ud} (\mu_p - \mu_n)/2= 1.81 V_{ud}$. Both scalar and
pseudoscalar contributions can be shown to be much smaller and
negligible for the accuracy we are interested in. At the same
order $1/M_N$ the allowed phase space for the relevant scattering
and decay processes get changed, due to nucleon recoil. Finally,
one also has to consider the additional contribution due to the
initial nucleon thermal distribution. In fact in the infinite
nucleon mass limit the average of weak rates over nucleon
distribution is trivial, since the nucleon is at rest in any
frame. For finite $M_N$, by considering only $1/M_N$ terms, the
effect of the thermal average over the thermal spreading of the
nucleon velocity produces a purely {\it kinetic} correction
$\Delta \omega_K$, whose expression can be reduced to a
one--dimensional integral over electron momentum and then
numerically evaluated. The explicit expression, which we do not
report for brevity, can be found in Section 4.2 and Appendix C of
\cite{EMMP1}.

\subsubsection{Thermal-Radiative corrections.}
\label{s:thermalrad}

\begin{figure}[!hbtp]
\begin{center}
\includegraphics[scale=0.5]{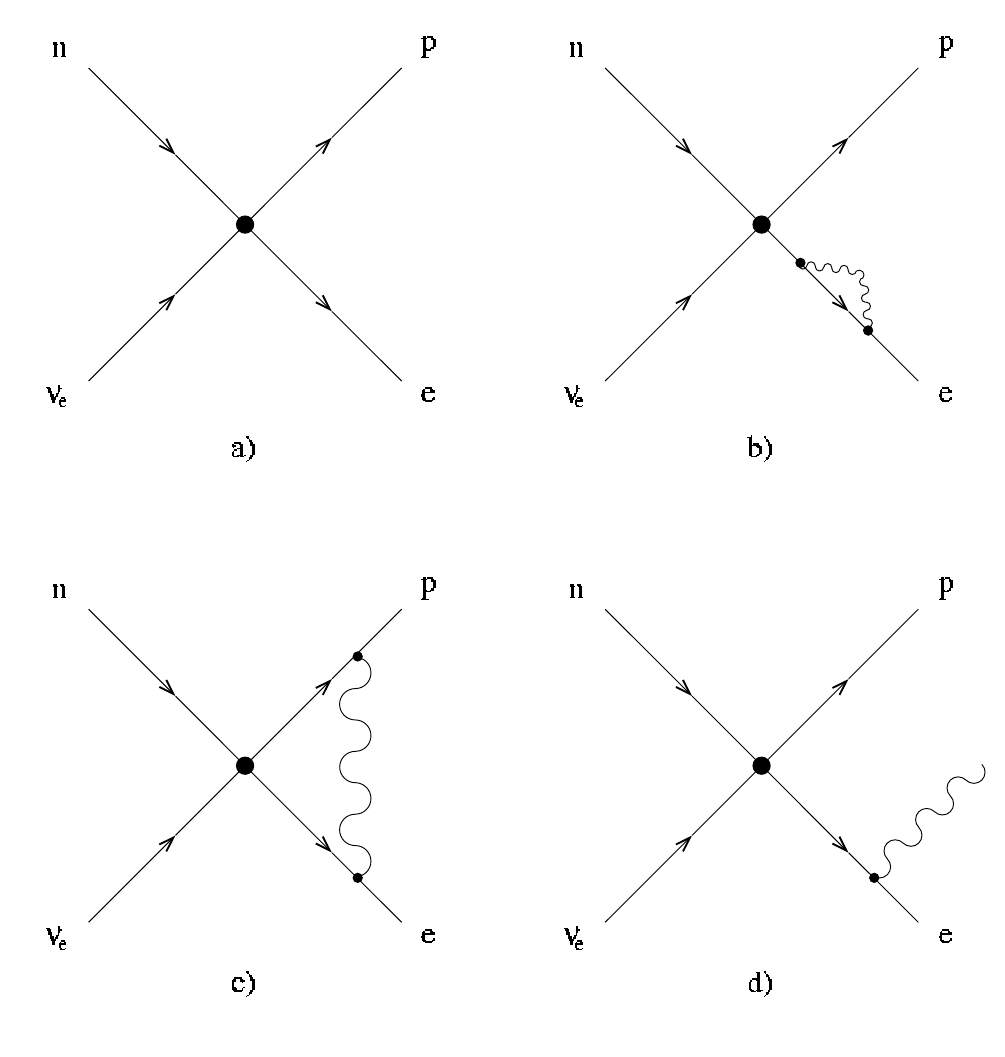}
\end{center}
\caption{The tree level Born (a), the one-loop (b),(c), and the
photon emission/absorption diagrams (d) for $n \rightarrow p$
processes.} \label{f:grafici}
\end{figure}

The $n \leftrightarrow p$ rates get slight corrections from the
presence of the surrounding electromagnetic plasma. To compute
these corrections once again one may use the standard Real Time
formalism for Finite Temperature Field Theory \cite{rtftft} to
evaluate the finite temperature contribution of the graphs
reported in Figure \ref{f:grafici}, for the $n \rightarrow p$
processes. Inverse processes $p \rightarrow n$ are obtained by
inverting the momentum flow in the hadronic line. The first order
in $\alpha$ is given by interference of one-loop amplitudes of
Figure \ref{f:grafici} b) and c) with the Born result (Figure
\ref{f:grafici} a)) . As usual, photon emission and absorption
processes (Figure \ref{f:grafici} d)), which also give an order
$\alpha$ correction, should be included to cancel infrared
divergences.
Notice that photon emission (absorption) amplitudes by the proton
line are suppressed as $M_p^{-1}$.

All field propagators get additional on shell contributions
proportional to the number density of that particular specie in
the surrounding medium. For $\gamma$ and $e^{\pm}$, neglecting in
this case the small electron chemical potential, we have \bea  i
\Delta_\gamma^{\mu \nu}(k) = - \left[\frac{i}{k^2} + 2 \pi \,
\delta(k^2)~f_\gamma(k_0) \right] g^{\mu \nu}~~~,
\label{e:a1} \\
i \, S_e(p')= \frac{i}{\slp' - m_e} - 2 \pi~\delta({p'}^2 -
m_e^2)~ f_e(p_0')~(\slp' + m_e)~~~, \label{e:a2} \eea with
$f_\gamma$ the photon distribution function. The entire set of
thermal/radiative corrections, at first order in its typical scale
factor, i.e. $\alpha T/m_e$, have been computed by several authors
\cite{FTQFT1}-\cite{FTQFT13} with quite different results.

\begin{figure}[!hbtp]
\begin{center}
\includegraphics[scale=0.5]{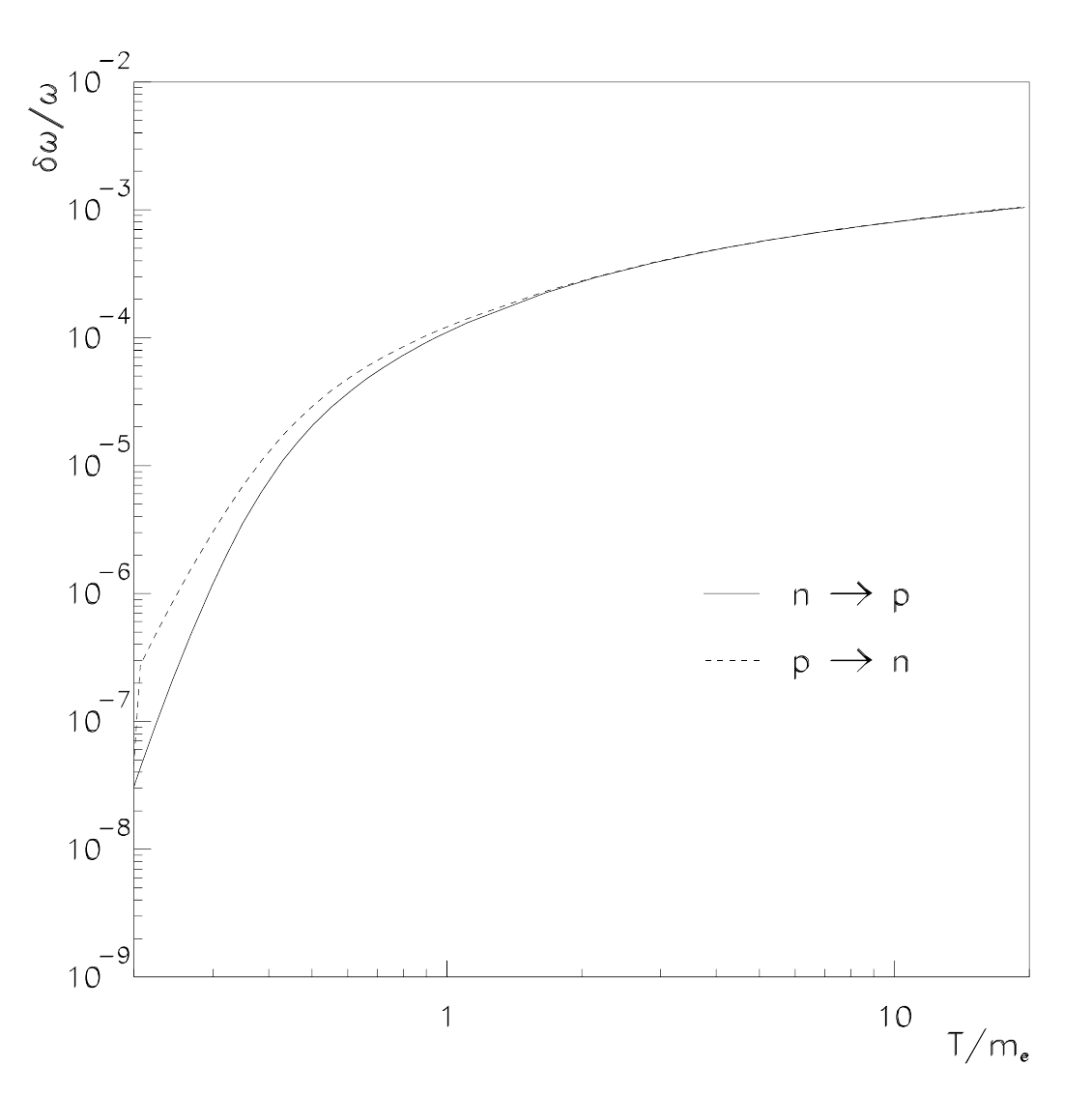}
\end{center}
\caption{The relative correction to the $n \rightarrow p$ (solid
line) and $p \rightarrow n$ (dashed line) total rates due to the
$\gamma + p \rightarrow e^{+}+\nu_e + n$ and inverse processes.}
\label{f:nuovoproc}
\end{figure}

It was correctly pointed out in \cite{pgamma} that when order
$\alpha$ QED corrections are introduced there are new processes
taking place in the plasma which should be added and contribute to
neutron/proton chemical equilibrium \be \gamma +p \rightarrow e^+
+ \nu_e + n ~~~,~~~~~~e^+ + \nu_e  + n \rightarrow \gamma + p \pp
\label{newpgamma} \ee For completeness, since the processes
(\ref{newpgamma}) where not considered in \cite{EMMP1}, we report
the explicit expressions for their rates, evaluated in the
infinite nucleon mass limit
\begin{eqnarray} \fl
\omega(\gamma + p \rightarrow e^{+}+\nu_e + n)= \frac{G_F^2
\left(\cvq+3 \, \caq\right)}{4\pi^3} \frac{\alpha}{\pi}
\int_{0}^{\infty} d|{\bf k'}|\int_{0}^{\infty} d|{\bf k}|
\nonumber\\
{\times} \frac{|{\bf k'}| \, |{\bf k}|^2}{ k_0 \, (\Delta+k_0
+|{\bf k'}|)} \left[\left(\Delta+k_0+|{\bf k'}|\right)^2 \,
\frac{|{\bf k'}|}{|{\bf k}|} \log\left(\frac{k_0+|{\bf
k}|}{k_0-|{\bf k}|}\right) \right.
\nonumber\\
+  \left. 2\, |{\bf k'}| \, \left(\Delta+|{\bf k'}|\right) \right]
f_\gamma(\Delta+k_0+|{\bf
k'}|)\left(1-f_{e^+}(k_0)\right)\left(1-f_{\nu_e}(|{\bf
k'}|)\right) \vv
\end{eqnarray}
where $k_0 \equiv \sqrt{|{\bf k}|^2+m_e^2}$. Inverse process is
simply obtained by changing the statistical factor with the
following one \be \left(1+f_\gamma(\Delta+k_0+|{\bf k'}|)\right)
f_{e^+}(k_0)f_{\nu_e}(|{\bf k'}|) \pp \ee The relative corrections
to the total rates are shown in Figure \ref{f:nuovoproc}.

In the range of temperatures when neutron fraction freezes out,
this rate is not severely suppressed by conservation of
energy-momentum, since photon mean energy is of the order of MeV.
Nevertheless it should be pointed out that the freeze out of
neutron to proton ratio is mainly dictated by two body processes
$(a)-(d)$, which in fact dominate over neutron decay and inverse
process $(e)$ and $(f)$ in the relevant temperature range, so the
inclusion of (\ref{newpgamma}) is very weakly affecting $Y_p$. In
the following we will adopt the results for thermal corrections
obtained in Reference \cite{EMMP1}, to which we refer for all
details, updated with the inclusion of the (\ref{newpgamma}) and
inverse process contributions.
The effects of the QED interaction in the plasma also show up in
modifying the equation of state of electron/positron and photons,
as well as they induce thermal contribution to their mass. These
effects, first considered in \cite{Lopez}, have been already
described in Section \ref{s:neudec}. In particular the thermal
averaged weak rates get an order $\alpha$ corrections when the
electron thermal mass is introduced in the corresponding
distribution function.

\subsubsection{Non instantaneous neutrino decoupling effects on the Born rates.}
\label{s:nut}

As we have already described, neutrino distribution functions get
distorted for the effect of a partial entropy release to neutrinos
by electron/positron pairs. Actually, the $e^+-e^-$ annihilation
phase is not an instantaneous phenomenon, but it partially overlap
in time with $n/p$ ratio freezing, as discussed in
\cite{dodelson}. The overall effect on BBN is due to three
different phenomena. First, the weak rates $(a)-(f)$ are enhanced
by the larger mean energy of electron neutrinos. On the other hand
there is an opposite effect due to the change in the
electron/positron temperature. Finally, since the photon
temperature is reduced with respect to the instantaneous
decoupling value $\bar{z}_{eq}^D= (11/4)^{1/3}$, the onset of BBN,
via $^2$H synthesis, is taking place earlier in time. This means
that fewer neutrons decay from the time of freezing out of weak
interactions and this in turn corresponds to a larger $^4$He yield.

\begin{figure}[!hbtp]
\begin{center}
\includegraphics[scale=0.5]{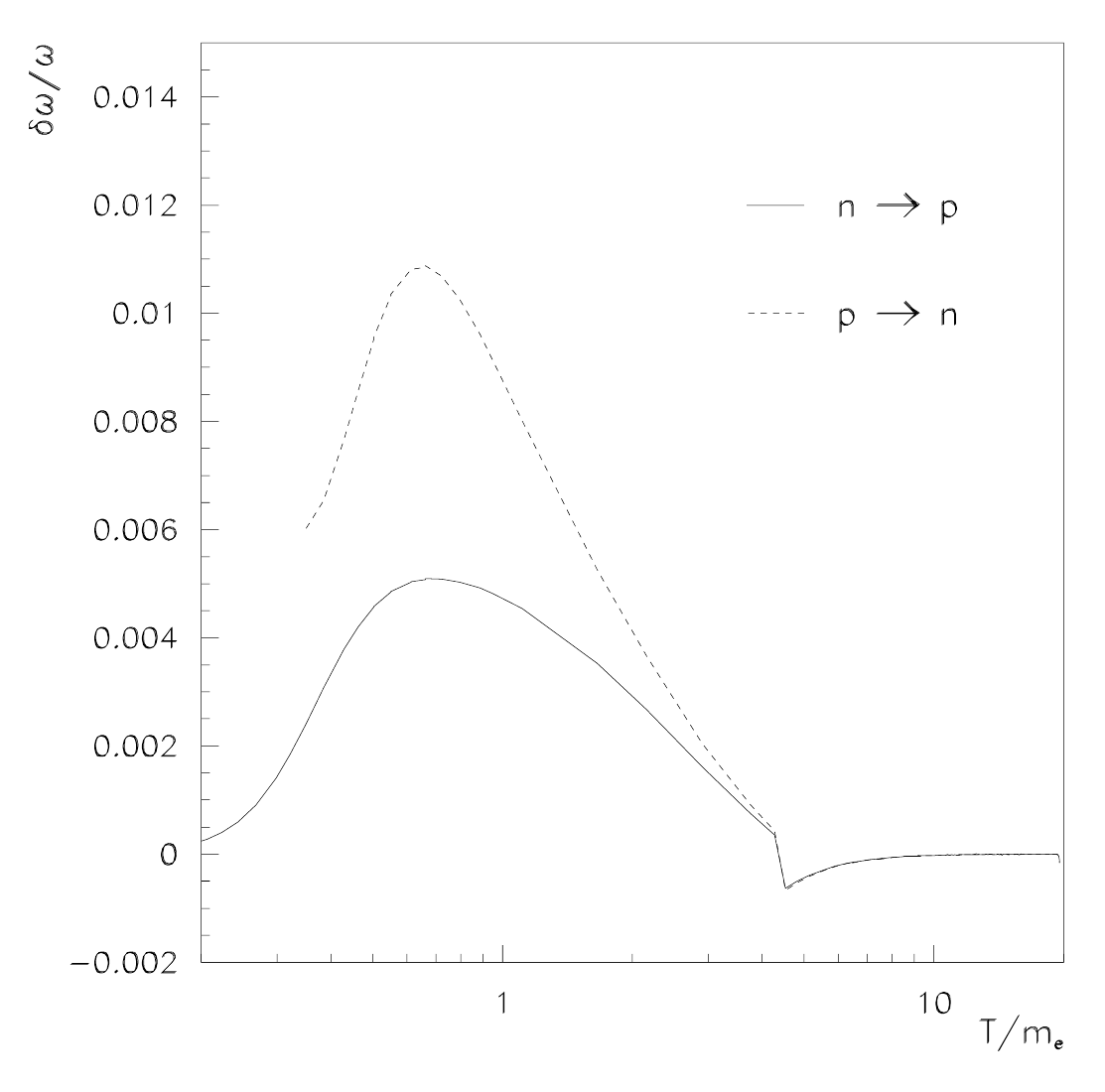}
\end{center}
\caption{The relative correction to the $n \rightarrow p$ (solid
line) and $p \rightarrow n$ (dashed line) total rates, due to the neutrino
distortion. $\delta \omega$ is the difference of the $n \leftrightarrow p$
rates calculated by numerically solving the neutrino decoupling and the
ones for an instantaneous neutrino decoupling at $T_D=2.3$ MeV. For $p
\rightarrow n$ the effect is shown down to $T \sim 0.3$ MeV, since for
lower temperatures the total rate becomes negligibly small.}
\label{f:nucorrrate}
\end{figure}

All these effects have been considered in our analysis. The actual
behavior of $\bar{z}$, shown in Figure \ref{zbarevol}, has been
used in the equation ruling BBN, as described in Section
\ref{s:numsol}. Moreover the Born rates for weak processes are
corrected by using the distortion $\delta f_{\nu_e}(x(z),y)$,
computed numerically. The resulting behaviors of the relative
corrections to the total $n \leftrightarrow p$ rates is shown in
Figure \ref{f:nucorrrate}. In particular $\delta \omega$ is the
difference of the $n \leftrightarrow p$ rates as from present
analysis and the result obtained in the instantaneous neutrino
decoupling limit, with decoupling occurring at $T_D=2.3$ MeV.
Notice the discontinuity in the first derivative at $T_D$. In
fact, $d (T_\nu/T)/dT$ has a discontinuity in the instantaneous
decoupling limit, while it has a smooth behavior when neutrino
decoupling is studied via a full numerical solution of the
corresponding kinetic equations.
We find that the effect of distortion of neutrino distribution
function slightly increases the $Y_p$, by the small amount $\Delta
Y_p \sim 1 {\cdot} 10^{-4}$, in agreement with the result of
\cite{dodelson}.

\subsubsection{The total rates for $n \leftrightarrow p$ reactions.}

Apart from the effect of neutrino distortion and of the $p+\gamma$
process discussed in the previous Sections, the effect of all
remaining corrections listed in Section 3.1 to the weak rates has
been discussed in details in \cite{Lopez,EMMP1}. We here report
the main results, for the sake of completeness. The leading
contribution is given by electromagnetic radiative corrections,
which decrease monotonically with increasing temperature for both
$p \rightarrow n$ and $n \rightarrow p$ processes. Including also
the effect of finite mass corrections due to weak magnetism and
$1/M_N$ corrections to phase space the effect ranges from -2 \% to
8 \% for $n \rightarrow p$ and -3 \% to 7 \% for $p \rightarrow
n$, respectively. In particular they affect the rates for a 2\% at
$T \sim 1 MeV$, {\it i.e.} at the freeze out temperature. Finite
mass corrections $\Delta \omega_K$ (see Section 3.1.2) are less
important and contribute for at most 0.5 \% $\div$ 1.5\% again in
the temperature range where the freezing phenomenon takes place.
Finally the effects due to plasma corrections and thermal
radiative effects are sub-leading, changing the rates at the level
of $(0.3 \div 0.6) \%$ only.

In order to use the $n \leftrightarrow p$ rates in a numerical code, it is
useful to fit their expressions as a function of $z$. The result is
reported in~\ref{tabweak}. The fit has been obtained requiring that the
fitting functions differ by less than $0.01 \%$ from the numerical values.
Notice that it is also a good approximation to consider a vanishing rate
$\omega(p\rightarrow n)$ for $T \leq 0.1 ~{\rm MeV}$, see Equation
(\ref{e:fitpn}), since it is a rapidly decreasing function with $T
\rightarrow 0$.

As a check of the result it is interesting to look whether the $n
\leftrightarrow p$ rates we have obtained, satisfy the detailed
balance condition, namely \be \rho(T) = \frac{\omega(n \rightarrow p)
n_n}{\omega(p \rightarrow n) n_p } =1 \vv \ee where $n_{n (p)}$ are the
neutron (proton) number density. In Figure 5 we plot the ratio $\rho(T)$
versus temperature in unit of $m_e$. We also show the behavior of
$\bar{z}^{-1}$. The result shows indeed that detailed balance condition is
satisfied with very good accuracy for $T \geq m_e$, better than the $1 \%$
level, smaller than the typical order of magnitude of order $\alpha$
radiative corrections in the same temperature range, which set the level of
accuracy of our results.
\begin{figure}[!hbtp]
\begin{center}
\includegraphics[scale=0.5]{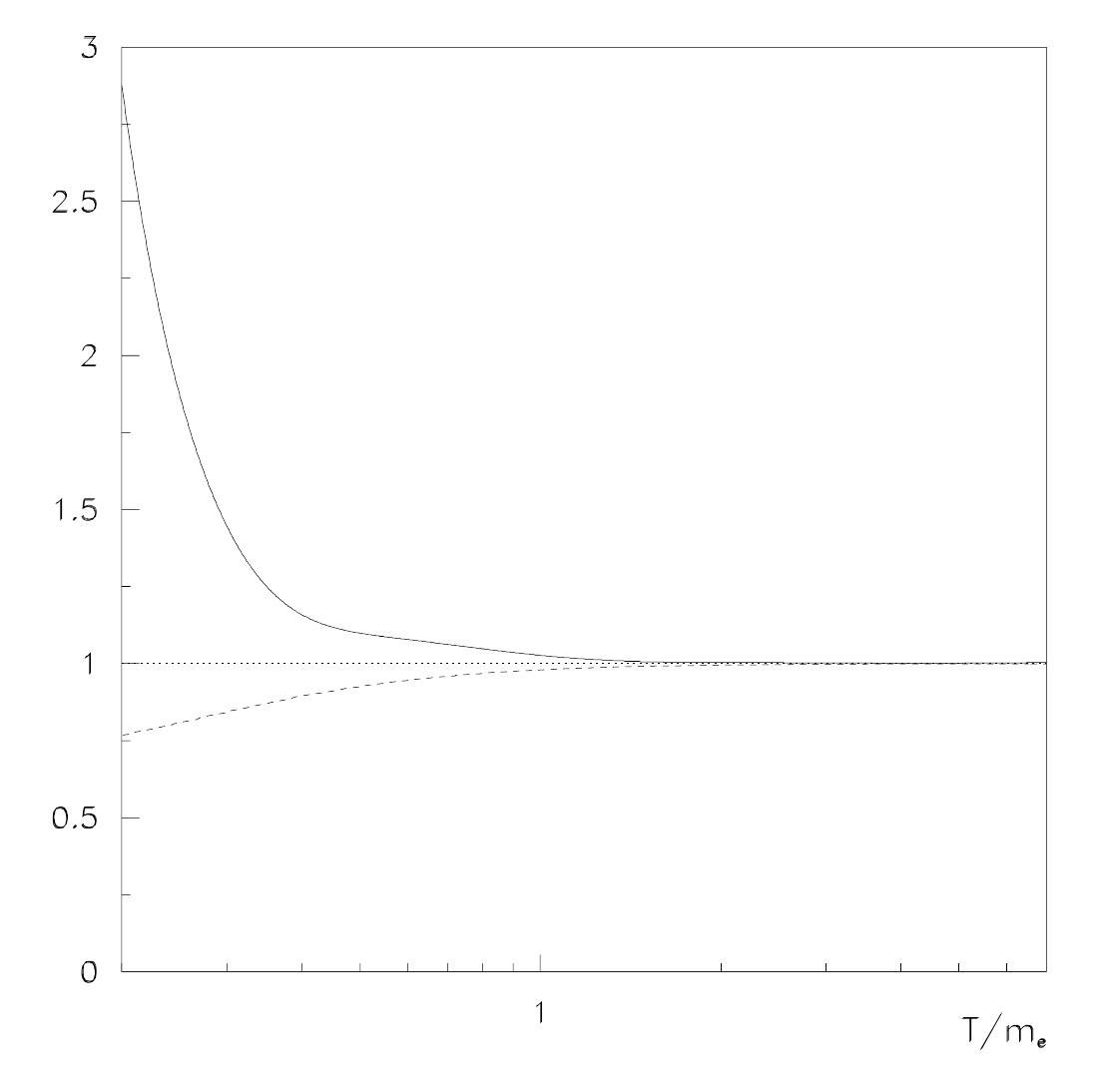}
\caption{The detailed balance
ratio $\rho(T)$ (solid line) versus temperature in unit of $m_e$.
The dashed line is the neutrino to photon temperature ratio
$\bar{z}^{-1}$. Dotted line corresponds to $\rho(T)$ for the case
$T_\nu=T$.}
\end{center}
\label{f:figdb}
\end{figure}
The steep rising of $\rho(T)$ for smaller temperatures, $T \leq
m_e$ is easily understood, since for $T\leq m_e$ pair annihilation
phase starts and neutrino do not share anymore the same
temperature of the electromagnetic plasma, as can be seen by the
behavior of $\bar{z}^{-1}=T_\nu/T$. 
Neutrinos become colder than $e^\pm$ so that all processes with an
ingoing neutrino/antineutrino are suppressed with respect to the reverse
ones. In particular, the detailed balance condition for the neutron
decay, the leading neutron destruction channel for temperatures lower
than 0.3 $m_e$, and the inverse $e^- + \bar{\nu}_e + p \rightarrow n$
reaction does not hold anymore and this fact leads to the increase of
$\rho(T)$. Stated differently, this large
departure at low temperatures of $\rho(T)$ is simply the effect of
the lack of thermodynamical equilibrium due to the freezing of
weak interactions. This temperature region is however irrelevant
for neutron to proton ratio which has been already frozen out well
before the $e^\pm$ annihilation phase.

\subsection{Nuclear reaction rates: analytical calculations.}\label{ratean}

In the primordial plasma during BBN the energy of colliding nuclei
is always much lower than their masses, and their density is very
far from values needed to produce relevant quantum effects. It
follows that the \emph{averaged} reaction rates $\langle \sigma
v\rangle$ for  a typical $a+b$ reaction process can be obtained by
folding the cross section $\sigma(E)$ with a
\emph{Maxwell-Boltzmann} phase space distribution \be
\langle\sigma v\rangle =
\sqrt{\frac{8}{\pi\mu_{ab}}}T^{-3/2}\int_{0}^{\infty}\, d E \,E\,
\sigma(E)\,e^{-E/T} \pp \label{rate2corpi} \ee with $\mu_{ab}$ the
reduced mass of $a$ and $b$ and $E$ denoting the \emph{kinetic
energy} in the Center of Mass (CM) frame. In the following
Sections we will perform this integral for the cases of interest
for the BBN. Though this issue can be found in many excellent
manuals, \cite{blweiss52}-\cite{math94}, we will briefly summarize
the main aspects of it, for the sake of fixing notation and to
summarize the parameterizations adopted for reaction rates, which
in some case have been improved with respect to standard results.

\subsubsection{Non-resonant reactions induced by neutrons.$*$}\label{sigman}

Assuming a s-wave collision, as it is typically the case for low energies
and far from resonances, for a neutron-induced reaction one finds $\sigma
\propto v^{-1}$, from which $\sigma v \simeq$ constant. More generally
one defines the auxiliary function \be R(E)
\equiv (\sigma v) (E) \Rightarrow
\sigma=\frac{R(E)}{v(E)}=\sqrt{\frac{\mu_{ab}}{2E}}R(E) \vv \ee
in terms of which, the rate can be written as \be \langle \sigma v
\rangle=\sqrt{\frac{4}{\pi}}\int_0^\infty dy \sqrt{y} R(y\,T)\,e^{-y}
\vv \label{ratenintR}\ee where $y\equiv E/T$.

For a non-resonant reaction $R$ is a weakly dependent function of energy
$E$, which allows to approximate $R(E)$ by a low order polynomial in
$E^{1/2}$ \be R(E)=\sum_{n=0}^{m}\frac{R^{(n)}(0)}{n !}E^{n/2} \vv \ee
getting
\be \langle \sigma v\rangle(T) =
\sum_{n=0}^{m}b_n T^{n/2}\label{ratenpoly} \vv \ee where \be
b_n=\frac{R^{(n)}(0)}{n!\Gamma(3/2)}\Gamma\bigg(\frac{n+3}{2}\bigg) \pp
\ee

\subsubsection{Non-resonant reactions induced by charged particles.}\label{Rnonrispc}

As well known, the \emph{astrophysical S factor} is defined as \be
S(E)\equiv \sigma(E)\,E\,\exp[\sqrt{E_G/E}] \label{Sgamow} \vv \ee
with $E_G\equiv 2\pi^2\mu_{ab}(Z_aZ_b\alpha)^2$ the \emph{Gamow
energy}. As in the case of neutrons, for a non-resonant cross
section $S(E)$ is a slowly changing function, and thus can be
typically parameterized as a (low-order) polynomial approximation
in $E$. In terms of $S(E)$, one can write \be \fl \langle\sigma
v\rangle=\sqrt{\frac{8}{\pi\mu_{ab}}}T^{-1/2} \int_0^\infty dy
S(y\,T)e^{-\phi(y,y_G)}\equiv \sqrt{\frac{8}{\pi\mu_{ab}}}T^{-1/2}
I \vv \label{ratecpintS} \ee where $y\equiv E/T$, $y_G\equiv
{E_G}/T$ and $\phi(y,y_G)\equiv y+\sqrt{\frac{y_G}{y}}$. Expanding
$\phi(y,y_G)$ up to second order around its minimum at Gamow peak,
$y_0=({y_G}/4)^{1/3}$ and assuming that, in the neighborhood of
$y_0$, $S(E=y\,T)\simeq S(E_0\equiv y_0\,T)\equiv S_0$, one gets
\be \langle \sigma v
\rangle(T)=\sqrt{\frac{32}{4^{1/3}}\frac{3E_G^{1/3}}{\mu_{ab}}}T^{-2/3}S_0\,
\exp\bigg[-\frac{3}{4^{1/3}}\bigg(\frac{E_G}{T}\bigg)^{1/3}\bigg]
\pp\label{ratnocorr} \ee This estimate can be improved in two ways
\begin{enumerate}
\item Gamow Peak Asymmetry\\
Assuming a constant astrophysical factor, let's define \be
\tilde{I}\equiv S_0\,e^{-3y_0}\sqrt{\frac{4\pi}{3}y_0}F(y_0) \vv
\ee where $F(y_0)$ can be evaluated by a term by term integration
of the non-gaussian terms in the series expression of $\phi$; for
$y_0>>1$, which is typically the case for stellar plasmas, this
correction can be written as a polynomial series in $1/y_0$ \be
F(y_0)=1+\frac{f_1}{y_0}+\frac{f_2}{y_0^2}+\ldots=1+\frac{5}{36y_0}+
\frac{35}{2592y_0^2}+\ldots \pp \label{asymmgp}\ee
\item Energy dependence of $S$\\
As $y_0$ depends on $T$, one usually writes $S$ as a (truncated) Taylor
series of starting point $E=0$
\be S(y\,T)=\sum_{n=0}^{m}s_n y^nT^n \vv \label{exps}
\ee where the parameters $s_n\equiv \frac{S^{(n)}(0)}{n!}$
should be fitted from experimental data; it is worthwhile to stress that,
in general, $S(0)$ should not be understood as the zero energy limit of
$S(E)$, but rather as the zero-th order term in the polynomial
approximation of the $S$ factor in the energy range covered by the data,
since the low energy behavior could show very different trends.
\end{enumerate}
Using Equation (\ref{asymmgp}) and (\ref{exps}) one gets \be I
\equiv \int_0^{\infty}dy S(y\,T) e^{-\phi(y)}\simeq \sum_{n=0}^m
I_n\,s_n\,T^n \vv \ee where \be I_n\equiv \int_0^\infty
dy\,e^{-\phi(y)}\,y^n\sim y_0^n I_0 \pp \label{intIn} \ee
Truncating the series at $m=2$ and picking up only the $1/y_0$
correction for $F(y_0)$, one derives \be \langle\sigma
v\rangle=\sqrt{\frac{8}{\pi}}\frac{1}{\sqrt{T
\mu_{ab}}}\sqrt{\frac{4\pi E_0(T)}{3T}}S_{eff}e^{-\frac{3E_0}{T}}
\label{ratecarnonris1} \vv \ee where \be \fl S_{eff}(E_0(T))
=\left[ s_0\bigg(1+\frac{5T}{36E_0}\bigg)+s_1E_0
\bigg(1+\frac{35T}{36E_0}\bigg)
+s_2E_0^2\bigg(1+\frac{89T}{36E_0}\bigg)\right]\label{Seffratecarnonris1}
\pp \ee Finally the rate can be written as \be \langle\sigma
v\rangle(T)=A\,T^{-2/3}\exp(-BT^{-1/3})\bigg(1+\sum_{n=1}^{5}C_nT^{n/3}\bigg)
\vv \ee which is the analytical approximation often used in the
literature.

However, a better, though semi-analytical, approximation
of the rate with respect to the Equations
(\ref{ratecarnonris1}-\ref{Seffratecarnonris1}) can be obtained as
follows. We rewrite the $I_n(T)$ as \be
I_n(T)=\,e^{-3y_0}y_0^n\sqrt{\frac{4\pi }{3}y_0} \alpha_n(T) \vv
\ee where \be \alpha_n(T)\equiv \bigg[\int_0^\infty
dy\,e^{-\phi(y)}\,y^n\bigg]
\bigg[e^{-3y_0}y_0^n\sqrt{\frac{4\pi}{3}y_0} \bigg]^{-1} \pp \ee
Note that $\alpha_0(T)$ coincides with the $F(y_0)$ function
introduced to describe the Gamow peak asymmetry. This expression
can be numerically evaluated in the relevant range and then fitted
assuming the ansatz \be \alpha_n(T)\simeq
a_0^{(n)}+\frac{a_1^{(n)}}{y_0}+\frac{a_2^{(n)}}{y_0^2} \vv
\label{alphanm1} \ee somewhat reminiscent of the series expansion
in the parameter $y_0^{-1}$. The assumption $y_0>>1$ is not always
justified for BBN, nevertheless the previous ansatz reveals to be
a good choice, and with the exclusion of the higher temperatures,
the fitting errors are always less than few percent, typically
less than $0.1 \%$ for $n=0$ and $n=1$, which are the most
relevant terms. The final form of the rate is then \be
\langle\sigma v\rangle=\frac{1}{T}\sqrt{\frac{32
E_0(T)}{3\mu_{ab}}}e^{-\frac{3E_0(T)}{T}} \,
\sum_{n=0}^{m}s_nE_0^n(T)\alpha_n(T) \vv \label{sigmavconalpha}
\ee usually reliable if one cuts at $m\leq 3$. This form, that was
adopted in some cases in our previous work \cite{cuocoetal}, is a good
compromise between accuracy and physical insight. In the present analysis,
however, we numerically integrate the (fitted) $S(E)$ factor, often a
polynomial function, in all the rate evaluations.

Notice that in some cases (e.g. $^4$He + $^3$He $\lrt \gamma$ + $^7$Be)
$S(E)$ is better described by adding an exponential decreasing term. If
\be S_2(E)=S_2(0)\exp[-aE] \vv \ee then \be
I_{(2)}=\frac{S_2(0)}{T}\int_0^\infty dE
\exp\bigg[-E\bigg(a+\frac{1}{T}\bigg)-\sqrt{\frac{E_G}{E}} \bigg]
\vv \ee from which, defining $T_a\equiv T/(1+aT)$, we get \be
I=S(0)\frac{T_a}{T}e^{-3\tilde{y}_0}\sqrt{\frac{4\pi}{3}\tilde{y}_0}
\vv \ee where $\tilde{y}_0\equiv (\tilde{y}_G/4)^{1/3}$ and
$\tilde{y}_G\equiv E_G/T_a$. We finally get in this case for the
rate \be \langle\sigma
v\rangle_2(T)=A_2T^{-3/2}(T_a)^{5/6}\exp[-B_2T_a^{-1/3}] \vv \ee
with $A_2$ and $B_2$ to be (usually) fitted using experimental
data.

\subsubsection{Resonant Reactions.}\label{resreactpar}

The $\sigma(E)$ for an isolated (single-level) resonance is
described by a Breit-Wigner formula \be \sigma(E)=\frac{\pi
\omega}{2\mu_{ab}E}
\frac{\Gamma_1(E)\Gamma_2(E)}{(E-E_R)^2+(\Gamma/2)^2}\label{s} \vv
\ee where $\omega$ is the statistical factor: \be \omega \equiv
\frac{(2J+1)(1+\delta_{ab})}{(2J_a+1)(2J_b+1)} \label{omega} \vv
\ee while $\Gamma_1 $ is the \emph{partial width} (PW) of the
Compound Nucleus (CN) formation from the ingoing channel, and
$\Gamma_2$ the CN decay PW into the outgoing channel. Finally
$\Gamma$ and $E_R$ are respectively the CN total width and
resonance energy.

For $\Gamma<<E_R$ (\emph{narrow resonance}) the integral
(\ref{rate2corpi}) can be rewritten as \be \langle\sigma v\rangle
\simeq \bigg(\frac{2\pi}{\mu_{ab} T}\bigg)^{3/2} (\omega \gamma)_R
e^{-\frac{E_R}{T}} \vv \ee where $(\omega \gamma)_R\equiv \omega
\Gamma_a \Gamma_b /\Gamma$ and $\omega$  is the statistical factor
defined in (\ref{omega}). This expression is a useful guide in fitting
rates with narrow resonances. Note that no simple functional forms exists
for the \emph{broad resonance} case even if formulas like

\be \fl  \langle\sigma v\rangle = \exp\bigg[-\frac{C_0}{T^{1/3}}\bigg]
\frac{1}{[E_0(T)-E_R]+(\Gamma/2)^2}\sum_{i=0}^{m} D_iT^{i/3}
\vv \ee
are sometimes used.

Usually, for both neutron and charged-particle induced reactions,
in presence of narrow resonances we added a lorentzian peak
$[1+(E-E_R)^2/(0.5\Gamma)^2]^{-1}$ to the polynomial component,
while in the case of a single, broad resonance we multiplied the
polynomial term by a similar lorentzian shape. In both cases, the
integration was performed numerically and the fitting form usually
chosen by looking at the past literature.

Other cases are also possible (subthreshold or overlapping
resonances), but they are of minor interest for our purposes and
for further details the interested reader is addressed to the
specialized literature.

\subsubsection{Screening and Thermal Effects.}\label{screenmeas}

When treating nuclear processes in astrophysical/cosmological
environments, one should carefully take into account the different
physical conditions existing there with respect to that explored
in a typical laboratory experiment. For example, the nuclear
reactions are experimentally studied by accelerating partially or
completely ionized species $a$ on atomic or molecular
(electrically neutral) targets $b$. As a consequence, the
projectile feels a \emph{screened} Coulomb potential because of
the presence of the electronic clouds, that can be parameterized
e.g. by a Yukawa-type potential
\begin{equation}
\phi_{screen}(r)=\frac{Z_{b}e}{r}e^{-r/R_{ab}} \vv
\end{equation}
where $R_{ab}$ is an atomic/molecular effective radius depending on the
projectile and (mainly) on the target properties.

The standard phenomenological approach to this problem consists in
introducing the screening energy parameter $U_e\equiv
Z_{a}Z_{b}e^2/R_{ab}$, that represents an effective lowering of the Coulomb
barrier seen by the projectiles, such that
\begin{equation}
\frac{\sigma(E)}{\sigma_{screen}(E)}\simeq \frac{E+U_e}{E}
\exp\left(-\frac{1}{2}\sqrt{\frac{E_G}{E}} \frac{U_e}{E}\right) \pp
\end{equation}
As $U_e$ are of the typical order of magnitude of the atomic energies,
these corrections should be applied to correct the measured cross section
for $E\sim 10 \,keV$ or lower, and thus only marginally affect the physics
of BBN. Nevertheless there are cases, as for the $^2$H + $^2$H $\lrt$ p +
$^3$H or the $^7$Li + p $\lrt$ $^4$He + $^4$He reactions, where this
correction has to be properly taken into account\footnote{We note in fact
that the value deduced for some reactions in deuterated metals seems to be
one or even two orders of magnitude greater (up to $700-800 \,eV$) than
naive upper limits. See e.g. \cite{Raiola:2002} for an analysis of this
phenomenon with several metals for the $^2$H(d,p)$^3$H reaction.}.

During BBN all matter is fully ionized and nuclear reactions take
place in a surrounding plasma. This means that all cross sections
deduced by laboratory experiments should be corrected for the
effect of \emph{target-nuclei thermalization} and
\emph{screening}. In fact, when temperatures are comparable to (or
higher than) the excitation energies of the nuclei, all the
accessible states are soon populated in a statistical way. Since
the experimental determinations refer instead to the ground
states, one has to calculate the needed corrections via
theoretical models. For our purposes, we adopt the model used in
\cite{NACRE}, based on the assumptions of local thermodynamic
equilibrium of the plasma, and the use of the Hauser-Feshbach
model to calculate the ratio between the reaction rate for the
target in the $n$-th excited state and the one for the ground
state nucleus. A further change should be also applied to the
forward/reverse rate relation (see~\ref{tabrat}) where the
statistical factor should be changed as follows \be
g_i\equiv(2J_i^0+1)\longrightarrow\sum_n(2J_i^n+1)
\exp\bigg(-\frac{\epsilon_i^n}{T}\bigg) \vv \ee if $\epsilon_i^n$
and $J_i^n$ are the energy and angular momentum of the $n$-th
excited state of nuclide $i$. Useful fitting formulas can be found
in \cite{NACRE}. Actually, the light nuclides have few (or not at
all) excited states, so that for such corrections to be of some
importance, (say $\sim$1\%) not only high temperatures, but also
high mass numbers $A$ are needed. This explains why, for the BBN
purposes, the inclusion of this effect only marginally (i.e.
$<1\%$) changes the standard predictions of the Lithium isotopes.

We close this Section with some considerations on the screening
effects for the nuclear reactions in the BBN plasma. The treatment
of this problem is simplified since the plasma is non-degenerate
and \emph{weakly coupled}, i.e. the Debye screening length $R_D$
is much greater than the mean separation between particles. When
the thermal $e^{\pm}$ pairs dominate the charged component of the
plasma the enhancement factor $f$ of the thermonuclear reaction 
rates has been calculated in \cite{Itoh:1997}
\begin{equation}
f=\exp\bigg(\frac{Z_1Z_2\alpha}{TR_D}\bigg) =
\exp(1.28\times 10^{-3}Z_1Z_2), \,\,\,\,\,\frac{m_e}{T}<<1
\label{scritoh}
\vv
\end{equation}
so that the effect of screening is of the order of $0.1\%$ at the highest
temperatures involved in the BBN. Even smaller corrections are expected at
lower temperatures ($T\le 0.1$ MeV), when the nuclear chain becomes effective 
and almost all $e^{\pm}$ pairs have
already annihilated. In this case $f$ can be calculated using the standard
Debye-H\"{u}ckel theory, which gives
\be R_D=\sqrt{\frac{T}{4\pi \alpha n_B \zeta}} \vv\ee
with $\zeta \equiv \sum_{j}(Z_j^2+Z_j)X_j\simeq 2X_{p}+6X_{^4{\rm He}}$,
since during the late BBN phase protons and $^4$He nuclei are largely
dominant. Therefore
\begin{equation} \fl
f=\exp\bigg(\frac{Z_1Z_2\alpha}{TR_D}\bigg)\simeq\exp(1.1{\times}10^{-8}
\sqrt{\eta\,10^{10}\zeta}Z_1Z_2)
,\,\,\,\,\,\frac{m_e}{T}>>1 \vv
\end{equation}
that is lower than what is obtained from (\ref{scritoh}), as expected
because of the very low baryon density.

\subsection{Error Estimates for Nuclear rates}\label{ErrEstNucRat}

Unless it is possible to rely on a specific theoretical model, the
$S$ function for various processes\footnote{In the following we
consider the $S$ function, but all results also hold for the $R$
function used for neutron induced reactions.}, both in magnitude
and energy behavior have to be deduced from experimental data.
This would not be a problem at all if were not for the fact that
often, for a single reaction, it is necessary to combine several
data sets affected by different statistical and normalization
errors, and which usually cover different energy ranges with only
a partial overlap. In our discussion we will use the following
notation. With $E_{i_k}$, $S_{i_k}$, $\sigma_{i_k}$, and
$\epsilon_{k}$ we denote, respectively the (center of mass)
energy, $S$ factor, statistical and (relative) normalization
uncertainties of the $i$-th data point of the $k$-th data set.
Whenever only a total error $\sigma_{i_k}^{tot}$ determination is
available for a certain experiment, that error is used instead of
$\sigma_{i_k}$, and an upper limit on the scale error is estimated
as max[$\sigma_{i_k}^{tot}/S_{i_k}$].

In many cases the different data sets do not agree within the
quoted uncertainties, perhaps for the presence of scale/total
errors larger than quoted. There are two possible solution to this
problem. In one case there may be plausible arguments suggesting
that one or more than one data set are seriously affected by
systematics, which are however difficult to quantify. In this case
a conservative approach simply consists in removing that
particular data from the analysis, provided this would not
severely limit our knowledge on the corresponding reaction rate,
i.e. if there is no other information on $S$ in the same relevant
energy range apart from those data. On the other hand, lacking any
hint to decide whether a particular experimental result gives
biased results or not, there are basically no other choices than
keeping all experimental information, but conservatively assume
that data for all experiment may be still affected by an overall,
albeit unknown, normalization factor.

This approach simply means that, differently than in a recent
analysis performed in \cite{Cyburt:2004cq}, we do choose not to
fix the mean normalization for the data sets to unity, but rather
introduce a normalization parameter $\omega_k$, which eventually
have to be included among the fitting parameters, while we still
keep the quoted $\epsilon_k$ as the estimate of the corresponding
relative uncertainties.

This procedure may seems quite arbitrary. Indeed, if we had no
information at all on the possible energy behavior of a definite
$S$ factor, we may potentially introduce a large bias in data
analysis. We do claim however that this is not the case at hand,
since at least qualitatively we $do$ have such a prior knowledge,
namely that the $S$ factor, apart from resonance contribution, is
a smooth function of energy, for which a low order polynomial fit
is a rather good approximation\footnote{We think that a
theoretical $prejudice$ is in any case entering in all possible
data analysis of $S$ factor and, we may say, of any physical
observable. For an intriguing discussion on this controversial
issue see \cite{neural}.}. We are therefore confident that our
method can give quite a reasonable fit of $S$, as well as of the
corresponding uncertainty.

Other approaches have been used in the literature. For example, in
\cite{Nollett:2000fh} a Monte Carlo method for direct
incorporation of nuclear data in the BBN code is adopted, which
has the benefit of simply skipping the intermediate fitting step
and provide a straightforward method for the inclusion of new
data. Unluckily, this method also presents several drawbacks. For
reactions with very little experimental information, the method is
of very limited application. Moreover it may provide unphysical
results in data-poor energy regions, and also underestimate the
errors in intervals where no overlap among data sets exists. See
also \cite{Cyburt:2004cq} for a detailed discussion on these and
related issues.

Very recently another detailed treatment of data for the BBN main
reactions has been considered in \cite{Descouvemont04}, which
makes use of the R-matrix fitting technique. The authors stress
the better parameterization of the cross sections obtained through
this analysis. 
While this may be true in several cases, it is difficult in our opinion 
to claim any statement of generality, since all models rely in fact on some 
assumptions. 
For example, for non-resonant reactions the R-matrix model mimics the smooth
behavior with some high-energy resonance tail, that actually is
not physical, since a direct interaction contribution well described 
in the Born approximation would be more appropriate. 
In this case, the R-matrix approach could require more parameters than a (low-order) 
polynomial fit. 
Moreover, the R-matrix is known to work better in the \emph{extrapolations} to very 
low energies often needed in stellar astrophysics: however this is of poor interest
for BBN, where the reliabilty of the \emph{interpolation} between the data over
a broad energy range is crucial, and that can be equally (if not better) provided
by simpler fits.

Their error analysis, aimed to give a statistical meaning to the fitted quantities, 
makes a somehow arbitrary distinction between the case of underestimated
statistical errors, which are then corrected with the standard
protocol of inflating the $\chi^2$, and that of discrepant
normalization, where instead another procedure is used.

Another issue which should be addressed is how to properly account for the
existing correlation among data obtained by the same experiment (see e.g.
\cite{Nollett:2000fh} and References therein). Again, this point is not as
easy as one may think. One possible approach, as the one used in
\cite{Cyburt:2004cq}, is to directly include the estimated normalization
uncertainties in the covariance matrix which is used to construct the
$\chi^2$ function. However, as discussed at length in
\cite{D'Agostini:1993uj}, this does not seem a fully satisfactory choice,
because of some subtleties related to standard error propagation. In fact,
this procedure produce a negative bias, with the usual definition of bias
in statistics, which systematically underestimates the physical quantities,
in particular in presence of relatively large normalization errors and
large numbers of data. Moreover it typically gives more {\it weight} to
data sets affected by larger systematics, provided they have instead a
smaller statistical error. We address the reader to
\cite{D'Agostini:1993uj} for an extensive discussion on these issues.

Our fitting method can be summarized as follows. We consider the $\chi^2$
function
\begin{equation}
\chi^2(a_l,\omega_k)=\chi_{stat}^2+\chi_{norm}^2 \vv
\end{equation}
where
\begin{eqnarray}
\chi_{stat}^2&=& \sum_{i_k}\frac{\left(S_{th}(E_{i_k},a_l)-
S_{i_k}\omega_k\right)^2}{\omega_k^2\sigma_{i_k}^2}
\vv \label{chistat}\\
\chi_{norm}^2&=&\sum_{i_k}\frac{(\omega_k-1)^2}{\epsilon_{k}^2}\label{chinorm}
\pp
\end{eqnarray}
The $\omega_k$ are as mentioned free parameters of the fit, giving the
renormalization constant for the $k$-th data set, while $S_{th}(E)$ is a
function of energy, typically a polynomial, which is chosen as the $S$
fitting expression. It depends on some parameters $a_l$ to be determined by
standard minimization procedures, along with the $\omega_k$. For the choice
of the fitting function, usually done by looking at the pre-existing
(theoretical and experimental) literature, see also
section~\ref{resreactpar} and \ref{S_Rfactors}. The sensitivity to the
fitting form was checked by varying the number of free parameters (e.g.,
the polynomial degree) until one finds a ``plateau'' of the minima of the
reduced $\chi^2$ in the model space around some configuration of
parameters. The model lying in this region with the lowest number of $a_l$
was finally chosen, in order to minimize unwanted fluctuations between the
data. Notice that correlations among the data of the same experiment are
taken into account in two ways. First of all, all data of some data set
share the same value of $\omega_k$. Secondly, we add a $penalty$ $factor$
to the $\chi^2$, given by $\chi_{norm}^2$, that disfavors a choice for
$\omega_k-1$ greater than the estimated normalization or total error
$\epsilon_k$. The fitting parameters coming from the minimization procedure
are expected not to suffer of the bias problem discussed above, as it will
be clear from its explicit application to several reaction rates in the
next Sections. Note that our method is the most natural generalization of
the {\it unbiased} one presented in \cite{D'Agostini:1993uj} (see Equation
(3)), and the one actually suggested to be used (\cite{D'Agostini:1993uj},
Section 4). Notice that there are as many terms in $\chi_{norm}^2$ norm as
there are in $\chi_{stat}^2$. This takes into account that both the
normalization and the shape uncertainty (in principle, on an equal footing)
do contribute to the overall indetermination.

Let us now discuss our estimate of rate uncertainties. From
results of the previous Section, we recall that, through $S(E)$ or
$R(E)$, the rate is a function of the parameters $a_l$ whose best
estimate $\hat{a}_l$ and covariance matrix $cov(a_i,a_j)$ are
obtained via $\chi^2$ minimization. The rate $f$ is then obtained
by numerical integration of $S(E)$ (or $R(E)$) convoluted with the
appropriate Boltzmann/Gamow Kernel (see Equations
(\ref{ratecpintS}) and (\ref{ratenintR}))
\begin{equation}
f(T)=\int_{0}^{\infty}dE K(E,T)S(E,\hat{a})\label{rate_f(S)} \vv
\end{equation}
while its (squared) error $\delta f$ trough the standard error
propagation as
\begin{equation} \fl
\delta f^2 =\int_{0}^{\infty} dE^{'} \,K(E^{'},T)
\int_{0}^{\infty} dE \,K(E,T) \sum_{i,j}\frac{\partial
S(E^{'},a)}{\partial a_i}\bigg{|}_{\hat{a}} \frac{\partial
S(E,a)}{\partial a_j}\bigg{|}_{\hat{a}} cov(a_i,a_j) \vv
\end{equation}
thus fully including the correlations among the fitted parameters.
As we said, a dominant role could be played by the systematic
error. The systematic discrepancy between the data is shown by
values of the reduced chi squared $\chi_\nu^2$ significantly
greater than 1. Note that its value has been also partially
increased because of our choice to add $\chi_{norm}^2$, while on
the other hand further parameters $\omega_k$ have to be estimated
from the data thus reducing the number of degrees of freedom.
Since in our approach the bulk of the systematic uncertainty
related to the different normalization of the data has been taken
into account, any residual discrepancy can be considered as due to
some unidentified/underestimated source of error in one or several
experiments. In this case we simply use the standard prescription
of inflating the estimated error by the factor $\sqrt{\chi_\nu^2}$
(see for example \cite{PDG}). For illustrative purposes, for each
reaction we will also quote the quantity $\varepsilon$ defined as
\begin{equation}
\varepsilon^2\equiv
\frac{\sum_{k}^{K}w_k(\omega_k-1)^2}{\sum_{k}^{K}w_k} \vv
\label{varepsilon}
\end{equation}
the sum being on the $K$ different data sets, and the weights
$w_k$ chosen as $w_k=(\chi_k^2/N_k)^{-1}$, where $\chi_k^2$ is the
contribution of the $n$-th data set (with $N_k$ data) to the
$\chi^2$. With this choice we assign a greater weight to the
renormalization factors $\omega_k$ of the data sets closer to the
fitted function, which indeed is characterized by a lower value of
$\chi_k^2/N_k$. The value of $\varepsilon$ represents the typical
renormalization of data, and is a qualitative estimator of the
scale disagreement among several data set, but it should not be
confused with a discrepancy error (that is already taken into
account in our approach and already discussed) or a scale error,
that we are now going to define.

The \emph{overall scale error} used in the analysis, was chosen to
be equal to the {\it lowest experimentally determined}
$\epsilon_k$ for that reaction. As it is by definition independent
on $E$, and given the equation~(\ref{rate_f(S)}), it applies both
to the S-factor and to the rate. It was added in quadrature to the
statistical error in the fits presented in~\ref{tabrat} and in our
following numerical analysis. The rationale under our choice for
it is that combining several data set one should always expect a
\emph{better} estimate of the $S$ factor, with a smaller error.
Indeed this is not the case for data sets showing some serious
discrepancy but, as we said already, this source of systematic
error is separately taken into account. A different approach has
been adopted in \cite{Cyburt:2004cq}. Apart from suffering in some
cases of a negative bias on the best value determination, as we
already emphasized, which typically leads to overestimate the {\it
discrepancy error} as defined in Equation (22) of
\cite{Cyburt:2004cq}, in this analysis an {\it intrinsic
normalization error} is introduced (see Equation (23) again in
\cite{Cyburt:2004cq}). The latter represents an average of the
normalization error, which is always greater than the best
determination of the scale error. This typically results in larger
estimates of the errors. 

A comparison with previous compilations
generally shows a decrease in the error estimates (when
available), and a shift in the best value of the rates for some
reactions, typically of few percent. For the pre-NACRE era, this
is likely due to significant differences in the datasets included,
or to the use of analytical method instead of numerical one in the
integration of the rates, and will not be discussed further. On
the contrary, when a comparison is possible, our results
essentially agree with the all-purpose NACRE compilation, though
the inclusion of new data, the focusing on the BBN energy regime
and the more  ``statistically motivated'' method we used instead
of the ``minimum-maximum'' rate given in \cite{NACRE} naturally
explain the lower uncertainties found in our work. Qualitatively
similar conclusions were found in~\cite{Descouvemont04}, and the
residual quantitative differences with respect to our compilation
can be fairly attributed to the slightly different databases and
the different regression protocol used.

\subsection{Leading processes}

In this Section and in the following one we discuss in details all
leading reactions in the BBN network, as well as a selection of
those reactions which, though presently playing a sub-leading
role, are affected by uncertainties large enough so that they
still may contribute to some extent to the eventual nuclide
abundances, or that present some historical or peculiar importance
that make them worth to be discussed. The leading or sub-leading
role of a reaction was firstly established graphically, by looking
at the temperature behavior of their contribution to the right
hand side of the corresponding Boltzmann equation for $X_i$ (see~\ref{nucvst} 
and~\cite{SKM93} for a similar approach),
and then checked numerically. In particular for these two classes
of reactions we report our results on adopted best estimates and
uncertainties. A complete list of the full BBN network used in our
study is presented in~\ref{nucnetwork}. For the sake of
brevity, we do not report in this work all the statistical
quantities entering each of the reaction analysis (see e.g.
equations (\ref{chistat}), (\ref{chinorm}), (\ref{varepsilon})).
The interested reader can obviously require further information by
addressing a mail to one of the authors.

\subsubsection{Reaction \emph{pn$\gamma$}: p + n $\lrt$ $\gamma$ + $^2$H}.

It is the main channel for $^2$H production and, indirectly, of
all other nuclides. There is a lack of experimental information
about this non-resonant process, at least in the energetic range
relevant for BBN. Except for the measurements of the thermal
neutron capture cross section $\sigma_{th}$ (see in particular
\cite{cokmelk77}), the only low-energy determinations are the ones
in \cite{Su95} and \cite{Na97}. One can also rely on some deuton
photo-disintegration data at energy near the threshold, as the
historical ones of \cite{bi50,sn50} (see also the review
\cite{as91}), that can be related to the direct process through
the detailed balance principle. Recently, new photo-destruction
measurements have been performed \cite{hara03}, that allow for an
evaluation of this crucial reaction rate with an uncertainty
smaller than $6\%$ in the relevant range for the BBN, thus
comparable with the estimate of \cite{SKM93}, but only based on
experimental data.

It is worth pointing out that this process has been theoretically
studied well enough to allow for even better results. In fact,
whenever a comparison has been possible, experiments have shown a
substantial agreement with theoretical calculations
\cite{Tornow:2003ze}.

An estimate of the rate based on the calculations of \cite{BeMo56}
and \cite{Ev55}, normalized to $\sigma_{th}= 0.332{\pm}0.002\,b$
(as measured by \cite{HuSchw58}) was already presented in
\cite{FCZI}.\\ In \cite{SKM93} and in the previous BBN analysis
the calculation of \cite{Hale91} has been used, which took into
account the thermal capture evaluation of \cite{Mugh81}
($\sigma_{th}=0.3326 {\pm} 0.0007\,b$), on the high energy data of
\cite{Bos79} and on the ones of deuton photo-disintegration, with
a $7\%$ total error on the rate obtained by adding in quadrature
the evaluation errors ($\le 5\%$), the fitting ones ($5\%$) and
the numerical integration approximations ($2\%$). Note that in
\cite{SKM93} it is erroneously quoted $ 0.006\,b$ instead of
$0.0007\, b$ for the uncertainty on this cross section evaluation.

In the present work, we evaluate the rate on the basis of the
(few) available direct and inverse low-energy data and, in
particular, the theoretical calculation of Rupak \cite{rupak00}.
As in the previous work of Chen and Savage \cite{CS99}, a
\emph{pionless effective field theory} is used, but the
calculation is pushed to the next order, thus lowering the
relative uncertainty to less than $1\%$. In this approach, since
the relevant energy scale is much lower than the pion mass $(\sim
140\,{\rm MeV})$, it is meaningful to describe the strong
interaction among nucleons without explicitly introducing the pion
degrees of freedom, and using effective four-nucleons local
operators, while the electromagnetic coupling is obtained via the
gauge principle. It can be shown that, at the energies relevant
for BBN, the transition amplitude for the \emph{pn$\gamma$}
process is dominated by the ${M1}_V$ (iso-vectorial magnetic
dipole transition) and ${E1}_V$ contributions (iso-vectorial
electric dipole transition), respectively at lower and higher
energies; the ${M1}_V$ amplitude was calculated up to the
next-to-next-to-leading order ($N^2LO$) and normalized to the
thermal neutron capture cross section $\sigma_{th}$ taken from
\cite{cwc95}; the ${E1}_V$ amplitude was instead computed up to
the $N^4LO$ order and normalized to the nucleon-nucleon scattering
data and using the deuton photo-destruction measurements (in the
$2.6\div 7.3$ MeV energy range of the $\gamma$). When sub-leading
effects are neglected, one obtains the \emph{effective range
theory} standard results.

\begin{figure}[!htbp]
\begin{center}
\includegraphics[scale=0.5]{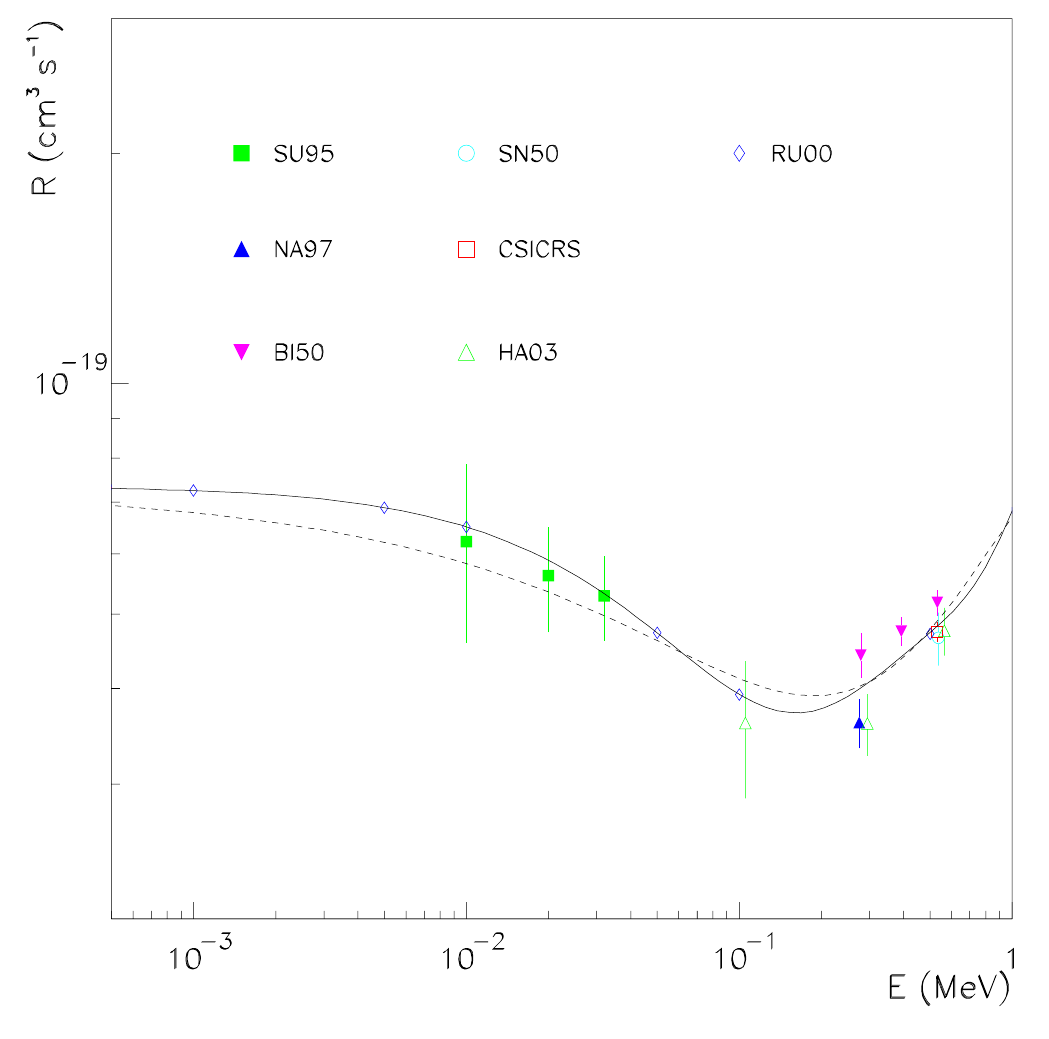}
\end{center}
\caption{The low energy data and the fit of the auxiliary function
$R(E)$ for the p + n $\lrt$ $\gamma$ + $^2$H reaction. The solid
line is the overall best fit, the dashed curve is based only on
experimental data. The theoretical points taken from
\cite{rupak00} have a negligible error bar on this scale. The
error bars for the experimental points also include the
\emph{propagated} uncertainty on the energies.}\label{tabpngd}
\end{figure}

The fit of the $R(E)$ function is almost completely determined by the Rupak
results. The reaction rate was calculated analytically according to
Equation (\ref{ratenpoly}). Linear error propagation leads to an estimate
of uncertainty $\le$ 0.4\% for $T_9\le 1.5$.

Since the database was limited to $E\le 1\,{\rm MeV}$, we expect
that our rate will disagree with the previous estimates at high
temperature; actually, starting from $T_9\sim 2$, with $T_9 =
T/10^9 \,K$, the difference with respect to the rate published in
\cite{SKM93} has almost a linear growth; the inclusion of higher
energy data in an auxiliary fit ($\sim 3\div 6 \,{\rm MeV}$)
allows to check that $0.1\%$ variations appear from $T_9\sim 2$.
For this reason we choose to use our rate in the range $T_9 \le
1.5$, where it is certainly more accurate, and matching it with
the rate evaluated in \cite{SKM93}, that is still used at higher
temperatures $T_9 \ge 1.5$. The overall uncertainty has been
conservatively estimated as $1.2\%$ for $T_9 \le 1.5$, and
includes the theoretical ($\le 1\%$), the fitting ($\le 0.18\%$),
statistical ($\le 0.4\%$) and normalization errors ($\sim
0.48\%$). The latter is due to the disagreement in the thermal
capture cross sections \cite{cokmelk77}. The uncertainty grows to
$\sim 8\%$ for $T_9 > 1.5$, being based on the error budget of
\cite{SKM93} and the matching and normalization errors. For
comparison, the analysis performed in \cite{cfo01} gives a
\emph{sample variance} error of $4.45\%$, while the uncertainty
quoted in \cite{Cyburt:2004cq} is $2.5\%$. All the available rates
agree at about 1 $\sigma$ or better within the quoted errors.

\subsubsection{Reaction \emph{dp$\gamma$}: $^2$H + p $\lrt $ $\gamma$ + $^3$He}.

It is one of the main responsible of the $^3$He synthesis both at
high and low temperatures. When the deuterium formation channel is
opening, this role is played instead by the strong reaction
\emph{ddn}, $^2$H + $^2$ H $\lrt$ n + $^3$He (see Figure
\ref{synth_he3}). The \emph{dp$\gamma$} is a direct capture
reaction, for which it exists quite a recent theoretical model
\cite{SVK96}. The data sets we consider are \cite{Casella02},
\cite{Griffiths62}-\cite{Ma97}. Some discrepancies exist for the
lower energies data, but in \cite{Schmid96} the authors stress on
the presence of a systematic error in their previous data (see
e.g. \cite{Schmid95}), with an upward correction that reduces the
compatibility problem with the older data reported in \cite{gr63}.
Moreover, the latter data set by Griffiths is likely to be
affected by a 10\%-15\% normalization error due to the wrong
stopping powers used for the heavy-ice targets
\cite{Schmid97,Ma97}, so that the disagreement is now reduced with
respect to the first claims. Finally, the recent data from the
LUNA collaboration \cite{Casella02} allow to firmly establish the
low energy behavior of the astrophysical factor: for example, the
older analysis performed in \cite{cfo01}, not including them and
with a less accurate treatment of the other experiments
systematics gave a \emph{sample variance} error of $13.2\%$, while
our present study suggests an overall uncertainty for the rate
less than 3\%. In the recent compilation \cite{Cyburt:2004cq} the
estimated error is close to 7\%, but this number is dominated by
the normalization error and so it is likely to be overestimated
(see our discussion in Section 3.3). We get a data normalization
spread of $\varepsilon \sim 1 \%$.
\begin{figure}
\begin{center}
\includegraphics[scale=0.5]{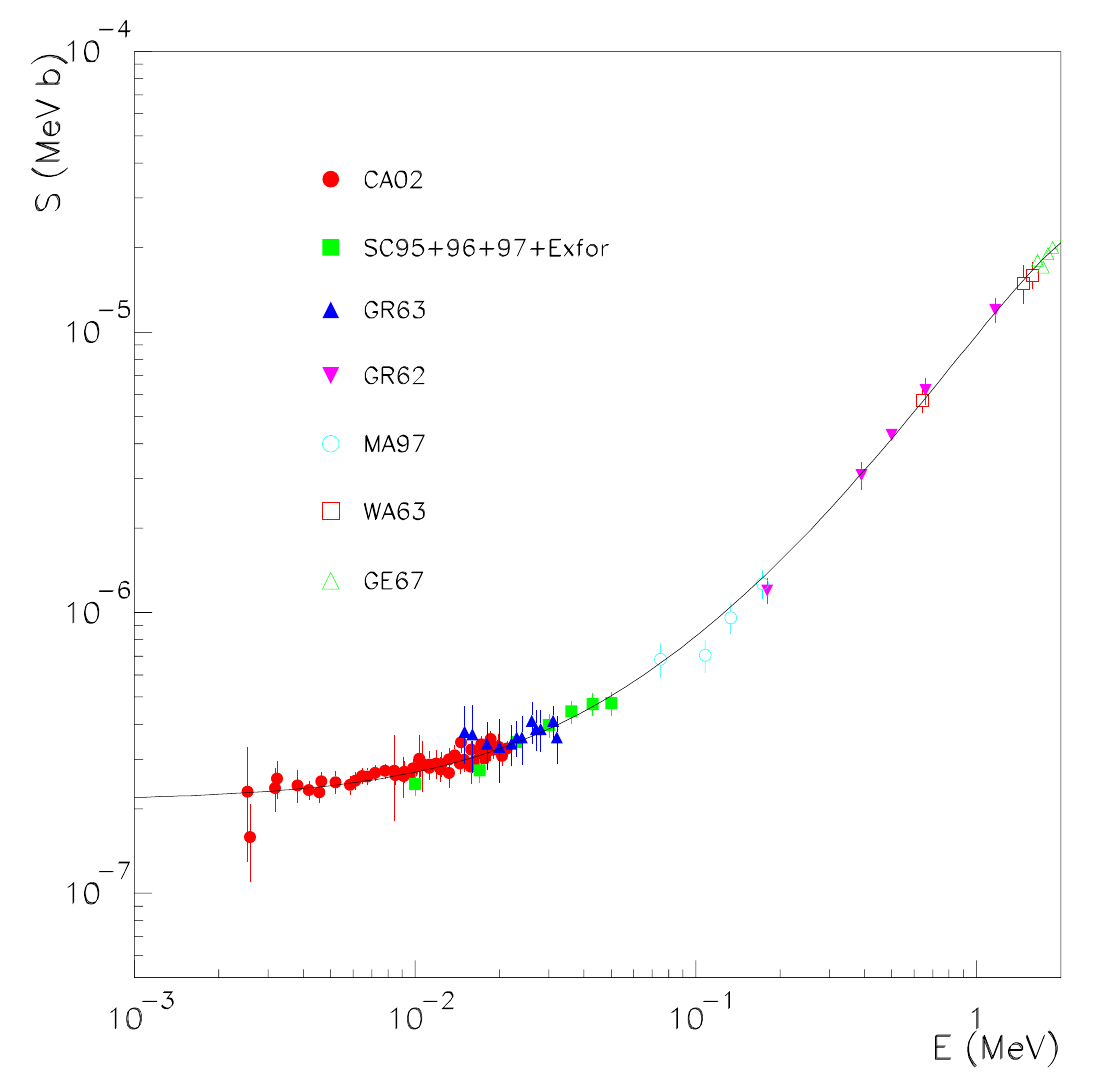}
\end{center}
\caption{The data and fit for the astrophysical factor of ${^2H} +
p \rt \gamma + {^3He}$. The fit of $S(E)$ is with a cubic
polynomial.} \label{plotdpg}
\end{figure}
As in our previous work \cite{cuocoetal}, the fit of all the data
available has been performed with a cubic polynomial, and both the
rate and uncertainty were obtained by numerical integrations, in
fair agreement with the semi-analytical formula introduced in
Section 3.2.2.

\subsubsection{Reaction \emph{ddn}: $^2$H + $^2$H $\lrt$ n + $^3$He}.

At $T_9\sim 1$, almost all $^3$He is produced through this typical
Direct Interaction channel (see Figure \ref{synth_he3}). The
energy range of main interest for the BBN is $0.01 \div 2.5$ MeV,
with a particular relevance of the values $E\le 0.4\,{\rm MeV}$.
Apart from some windows, the experimental situation for both the
\emph{ddn} and \emph{ddp} (see later) reactions is quite good, if
compared with other processes: they are strong interactions, their
Coulomb barrier is low, and their relevance as thermonuclear
fusion energy routes made their study quite appealing.
Nevertheless their importance for BBN requires further efforts. We
used the data \cite{mcneill51}-\cite{Greife95}; note that the data
of \cite{Jarmie85}, used in \cite{Cyburt:2004cq}, are only a
preliminary report of the measurements quoted in \cite{Brown90}, 
so they are not an independent data set and should
be excluded\footnote{Incidentally we note that there is a misprint
in \cite{Cyburt:2004cq}, as the data of \cite{Jarmie84} referring
to the \emph{tdn} reaction are erroneously quoted.}. The data of
\cite{Schulte72} quite accurately fix the high energy behavior of
this astrophysical factor, and the data in \cite{Brown90} allow to
quote a normalization of the $S$ factor (and then of the rate) of
1.3\%.
\begin{figure}
\begin{center}
\includegraphics[scale=0.5]{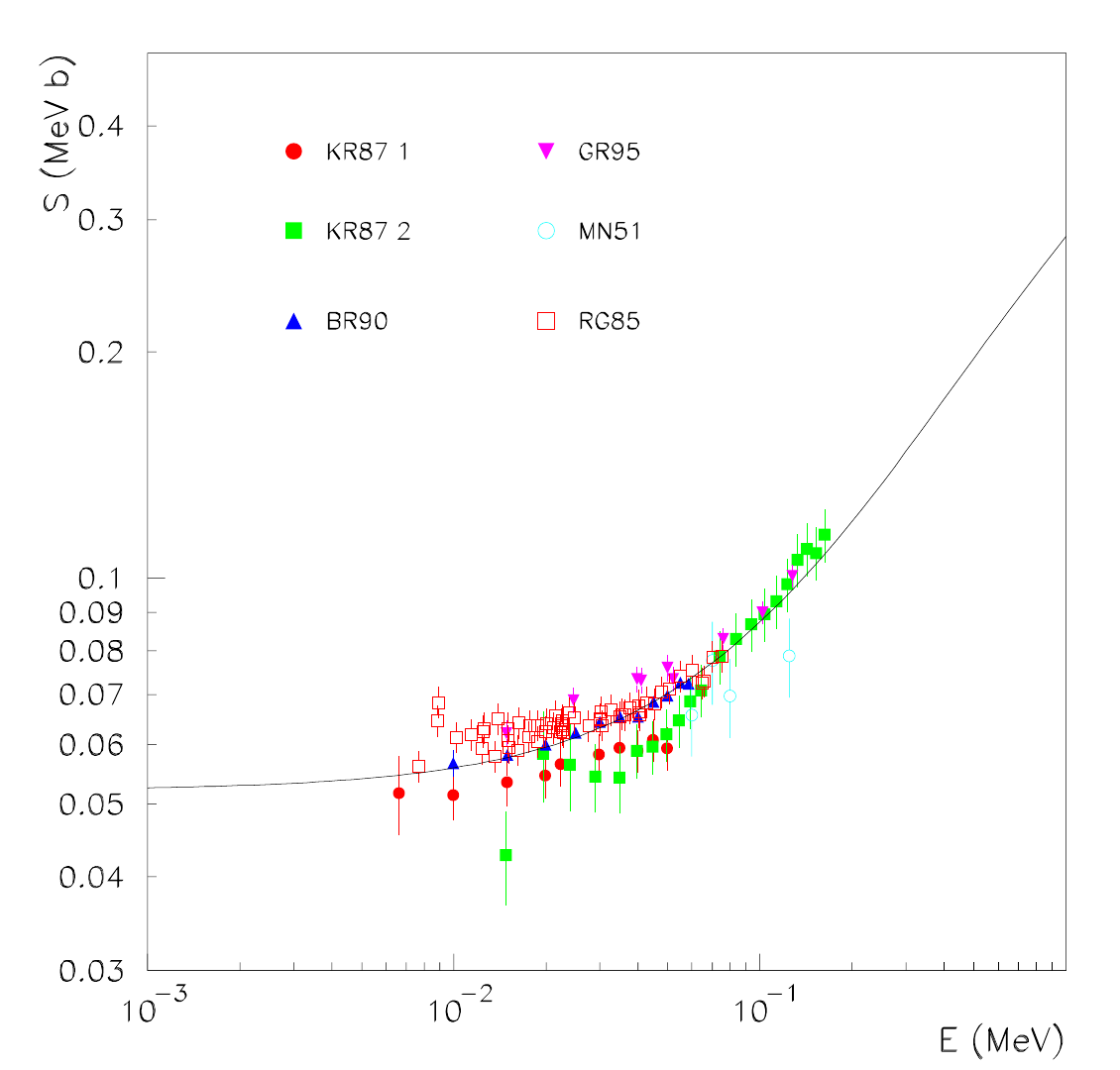}
\end{center}
\caption{The data and fit for the $S$ factor of the
$^2$H + $^2$H $\lrt$ n + $^3$He reaction.}\label{tabddn}
\end{figure}
As many data are available, and they are generally affected by a
small normalization or overall uncertainties (at the level of few
percent), even relatively low discrepancies on the scale and/or
the behavior of the $S(E)$ will show up in the value of the
reduced $\chi^2$, which is of the order of $10$ or greater at its
minimum. Our method gives an overall uncertainty at the level of
$1.3\%$, while for comparison the analysis performed in
\cite{cfo01} gives a \emph{sample variance} error of $3.1\%$. It
should be pointed out that, while the uncertainty expected on the
basis of the best determinations should be even lower, when
combining several data sets the experimental situation suggests
the presence of systematics. Indeed some experiments are likely to
have underestimated the errors or could be affected by some
unknown bias. In this situation, every regression method makes
poor sense, and we continue to follow our approach just for
consistency. As a confirmation of how difficult the analysis is in
this case, we observe that the error budget estimated in a
completely different approach in \cite{Cyburt:2004cq} is greater
for the \emph{ddp} ($\simeq 6.9\%$) than for the
\emph{ddn}($\simeq 5.5\%$) reaction, despite of the fact that the
quality and the source of data are comparable, when not better and
even more abundant for the first reaction.

As this reaction and the conjugate \emph{ddp} strongly influence the $^2$H
error budget, so that it affects many of the nuclides predictions, we
strongly recommend a new experimental campaign to determine accurately (say
at the 1\% level) both the magnitude and shape of the $S(E)$ in the wide
range useful for BBN studies (E$\le$ 2.5 {\rm MeV}), and in particular for
$0.2\leq E \leq 1$ MeV, where actually no data presently exist.

\subsubsection{Reaction \emph{ddp}: $^2$H + $^2$H $\lrt$ p + $^3$H}.

This process is the leading source of direct tritium synthesis
(see Figure \ref{synth_h3}). The discussion of its experimental
situation closely follows our analysis of the previous reaction
\emph{ddn}, apart from the fact that very low energy data are
available \cite{Greife95}, which have to be corrected for
screening effects in order not to bias the result. The data we
used for this reaction are contained in the same References quoted
for the \emph{ddn}, as these processes are usually studied
together. Since these two reactions are non-resonant, in both
cases a polynomial fit for $S(E)$ is used, while the rates and
errors were obtained by numerical integration. The uncertainty in
the relevant temperature range is less than $1 \%$.
\begin{figure}
\begin{center}
\includegraphics[scale=0.5]{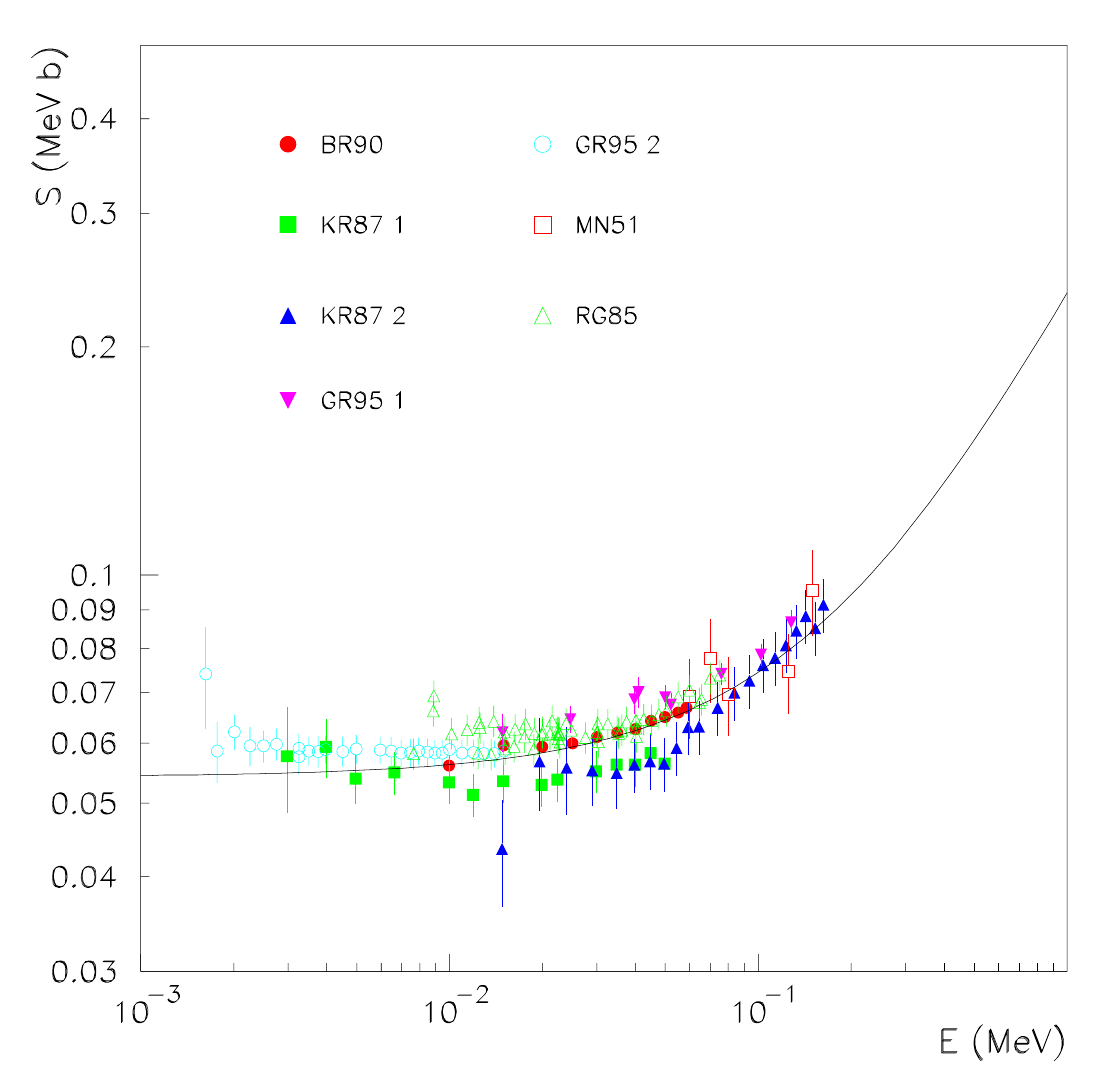}
\end{center}
\caption{The data and fit for the $S$ factor of the
$^2$H + $^2$H $\lrt$ p + $^3$H reaction.}\label{plotddn}
\end{figure}

\subsubsection{Reaction \emph{tdn}: $^3$H + $^2$H $\lrt$ n + $^4$He}.

The \emph{tdn} is the fundamental channel for $^4$He synthesis
during BBN. Many data are available for this process, also because
it is the most promising process for the thermonuclear fusion
research (low Coulomb barrier and high Q-value). Actually it is
the energy production mechanisms considered for the next
generation controlled fusion reactor, ITER \cite{ITER}. A broad
resonance in $^5$He, having $E_R = 0.064$ MeV and $\Gamma = 0.076
{\pm} 0.012$ MeV is superimposed to the non-resonant background.
This allows to use a Breit-Wigner shape to extrapolate with some
accuracy the $S$ factor also to energies below the range covered
by the data. The data sets we used are
\cite{Argo52}-\cite{Brown87}, with the Conner data \cite{Conner52}
excluded at energies greater than $240$ keV, where the isotropy
assumption for $\sigma(E)$ fails. The rate and error estimate were
both calculated numerically.
\begin{figure}
\begin{center}
\includegraphics[scale=0.5]{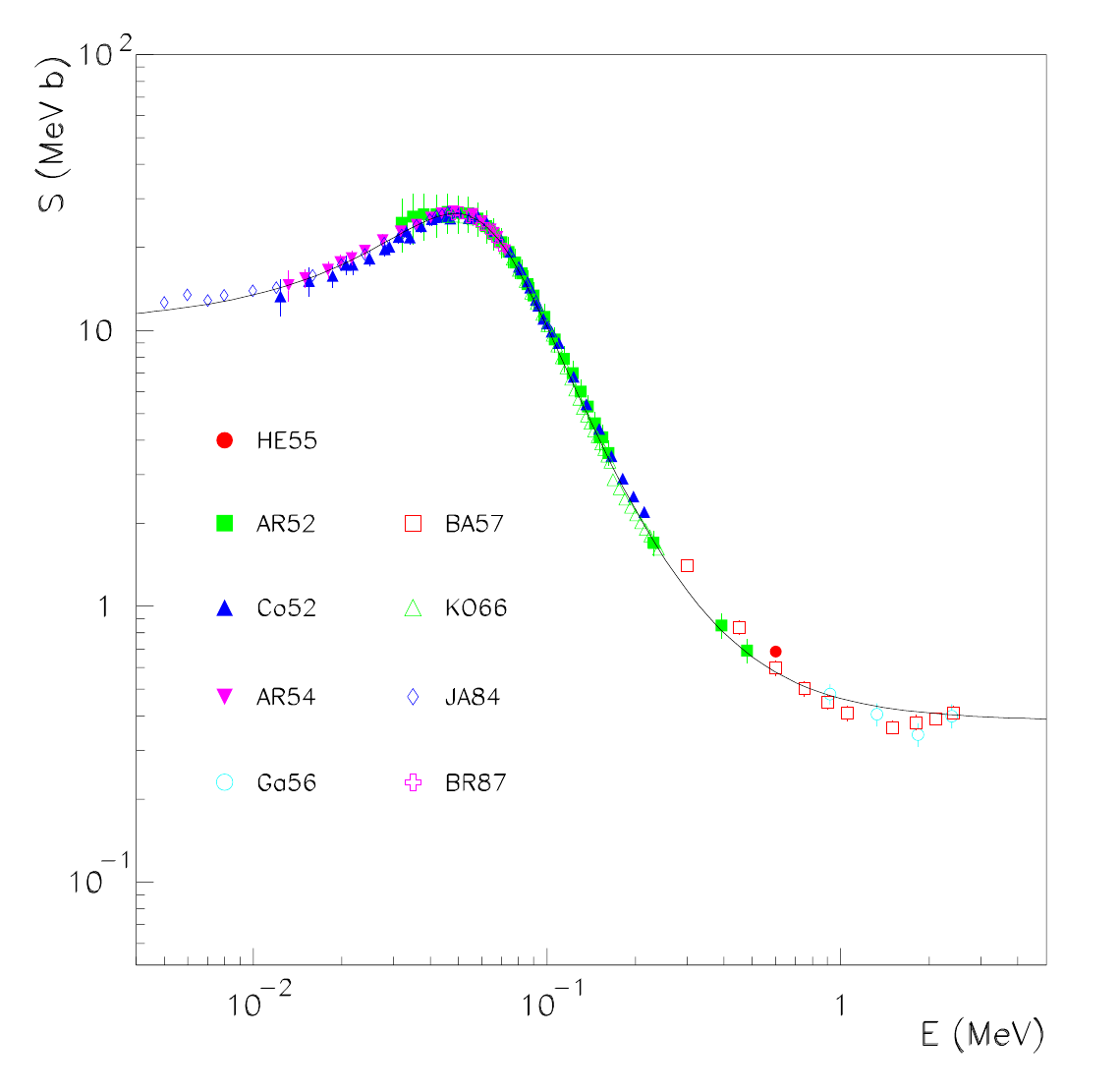}
\end{center}
\caption{The $S(E)$ data and fit for the $^3$H + $^2$H $\lrt$ n + $^4$He channel}\label{plottdn}
\end{figure}
Our error estimate is $\simeq 0.6 \%$, and $\varepsilon \sim 0.02$
(for a $\chi_\nu^2\simeq 1.4$), while for comparison the analysis
performed in \cite{cfo01} gives a \emph{sample variance} error of
$\sim 4\%$, while the discrepancy error estimated in
\cite{Cyburt:2004cq} is about of 2.2\%.

\subsubsection{Reaction \emph{he3dp}: $^3$He + $^2$H $\lrt$ p + $^4$He}.

Like the conjugate reaction \emph{tdn}, it is dominated by a broad
$^5$He resonance, in this case at $\simeq 0.2$ MeV. There is quite
a satisfactory experimental knowledge of this process, even if the
dispersion of the data grows for $E \le 0.2\,{\rm MeV}$,
introducing some uncertainty on the resonance parameters. It is
the second route to $^4$He production after the \emph{tdn}
reaction, but its indetermination mainly plays a role in the
$^3$He and $^7$Li yields, as it essentially controls the burning
of $^3$He. The data considered in our analysis are the ones
reported in \cite{Bonner52}-\cite{Aliotta01}; the two recent data
sets \cite{Costantini00,Aliotta01}, firstly included in this
compilation, once corrected for the screening effect allow for a
better coverage of the low-energy region. Our new rate agrees
within the errors with the rate published in \cite{SKM93}, where
the total error is estimated to be 8\% (1 $\sigma$). For
comparison, the analysis performed in \cite{cfo01} gives a
\emph{sample variance} error of 9.15\%, while in
\cite{Cyburt:2004cq} a 7.3\% result is quoted. We estimate a rate
uncertainty of the order of $0.6 \%$, while $\varepsilon \sim
0.03$.
\begin{figure}
\begin{center}
\includegraphics[scale=0.5]{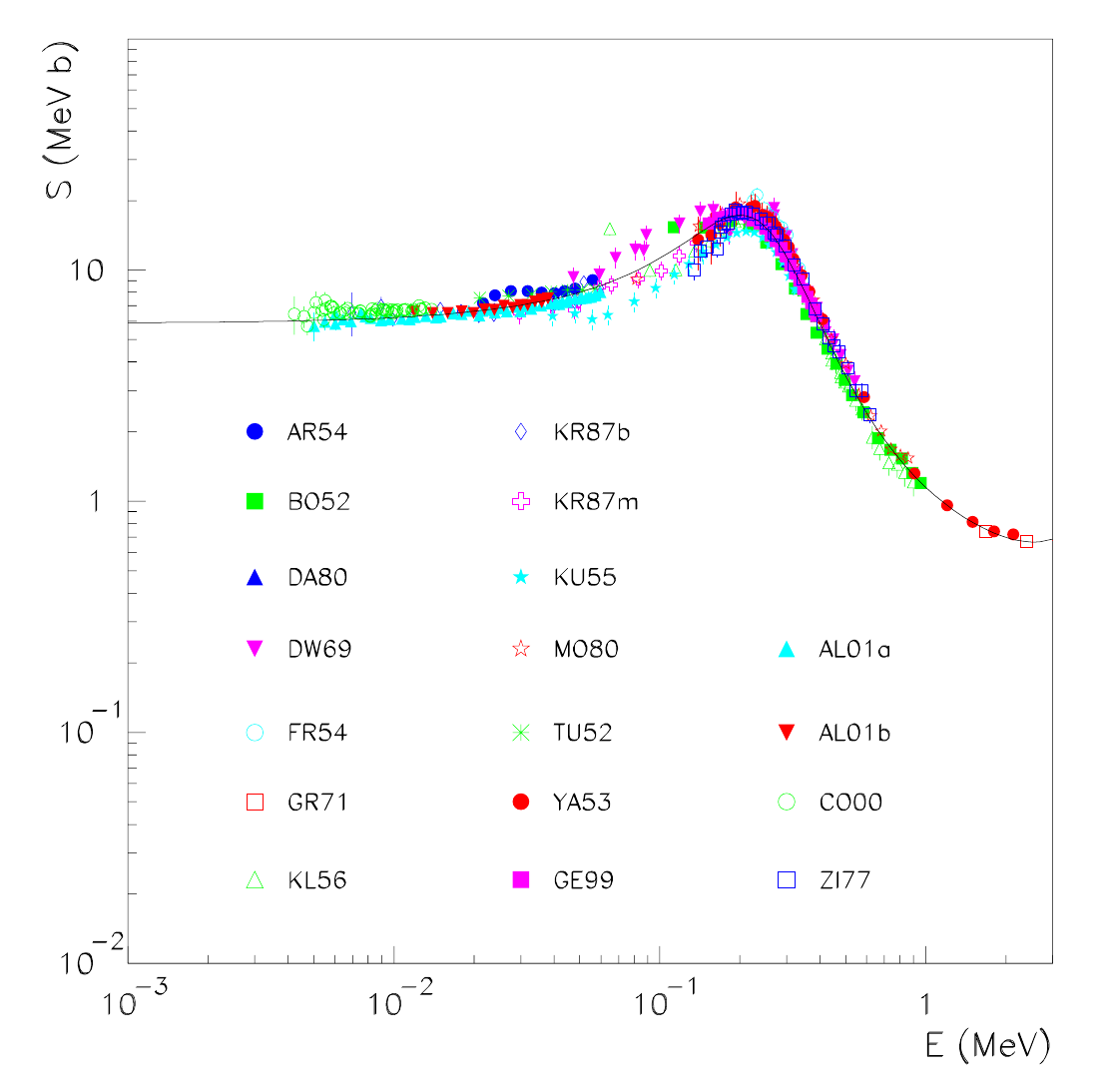}
\end{center}
\caption{The $S(E)$ data and fit for the $^3$He + $^2$H $\lrt$ p + $^4$He reaction}\label{plothe3dp}
\end{figure}

\subsubsection{Reaction \emph{he3np}: $^3$He + n $\lrt$ p + $^3$H}.

This process keeps $^3$He and $^3$H in equilibrium at (relatively)
high temperatures. During BBN it plays an important role in
determining the final abundances of $^3$He and $^7$Li. In our
regression we used the data reported in
\cite{Costello70}-\cite{breal99}, by limiting the analysis to
$E\le 1 {\rm MeV}$ as in \cite{Cyburt:2004cq}: this allows to
cover the most interesting range for the BBN by significantly
reducing the numbers of parameters needed to have a good fit.
Thanks to the new measurements of the reverse reaction cross
section near the threshold reported in \cite{breal99} and to the
accurate knowledge of the thermal capture rate \cite{ANi64}, both
the normalization and the shape of the $R(E)$ factor are well
known, so that its contribution to the final error budget is low
enough now not to require further measurements. In particular we
find a statistical uncertainty of at most $0.2 \%$, and a
normalization spread of $\varepsilon \sim 0.002$.
\begin{figure}
\begin{center}
\includegraphics[scale=0.5]{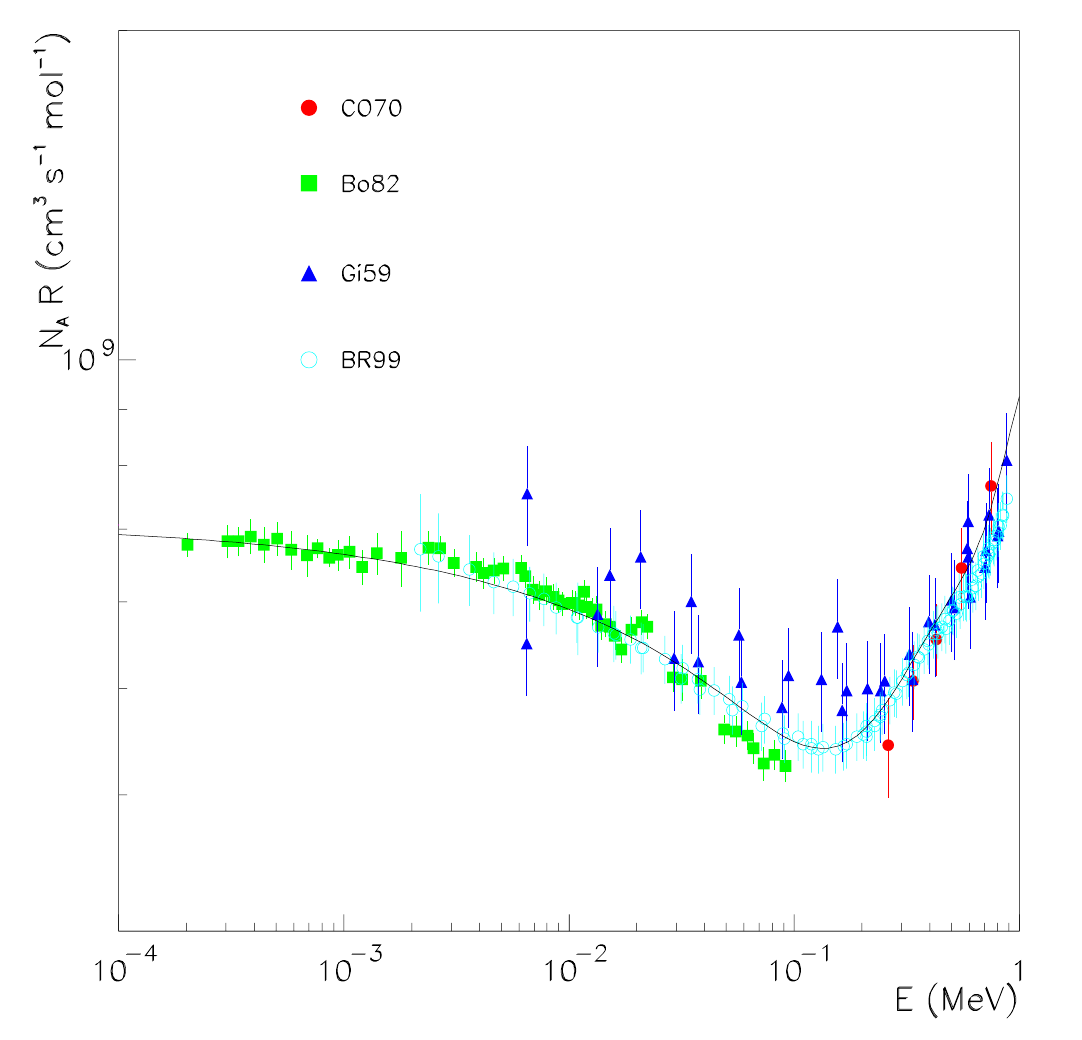}
\end{center}
\caption{The $S$ factor data and fit of the $^3$He +
n $\lrt$ p + $^3$H.}\label{plothe3np}
\end{figure}

\subsubsection{Reaction \emph{be7np}: $^7$Be + n $\lrt$ p + $^7$Li}.

It is the main responsible for $^7$Be/$^7$Li conversion during the
relevant phases of the BBN (in the final stage, the $^7$Be
electron capture is actually more important). Only for very low
energies, from $25\;{\rm meV}$ to $13.5$ keV, data are available
on the \emph{direct} process \cite{Ko88}, while for higher
energies one has to rely on the data for the reverse reaction
\cite{Ta48,Se76}. To avoid the introduction of significant errors
due to the use of the detailed balance conversion near the
threshold, we considered indirect data with energy $\ge 40$ keV
only. As for the previous \emph{he3np} reaction, we also
restricted the analysis at energies $E\le 2$ MeV, that are
essentially the only ones relevant for BBN. However, differently
than the non-resonant \emph{he3np} reaction, whose rate was
obtained analytically, this process has a resonance at $E\simeq
0.32$ MeV that suggests to use a fully numerical approach, both
for the rate and the uncertainties. Note that despite of the few
data set available, the Koeler data \cite{Ko88} fix the overall
scale error to the $\simeq 2.1\%$ level, thus making this process
quite accurately known for the purpose of BBN studies. The
statistical error is of the order of $0.7 \%$.
\begin{figure}
\begin{center}
\includegraphics[scale=0.5]{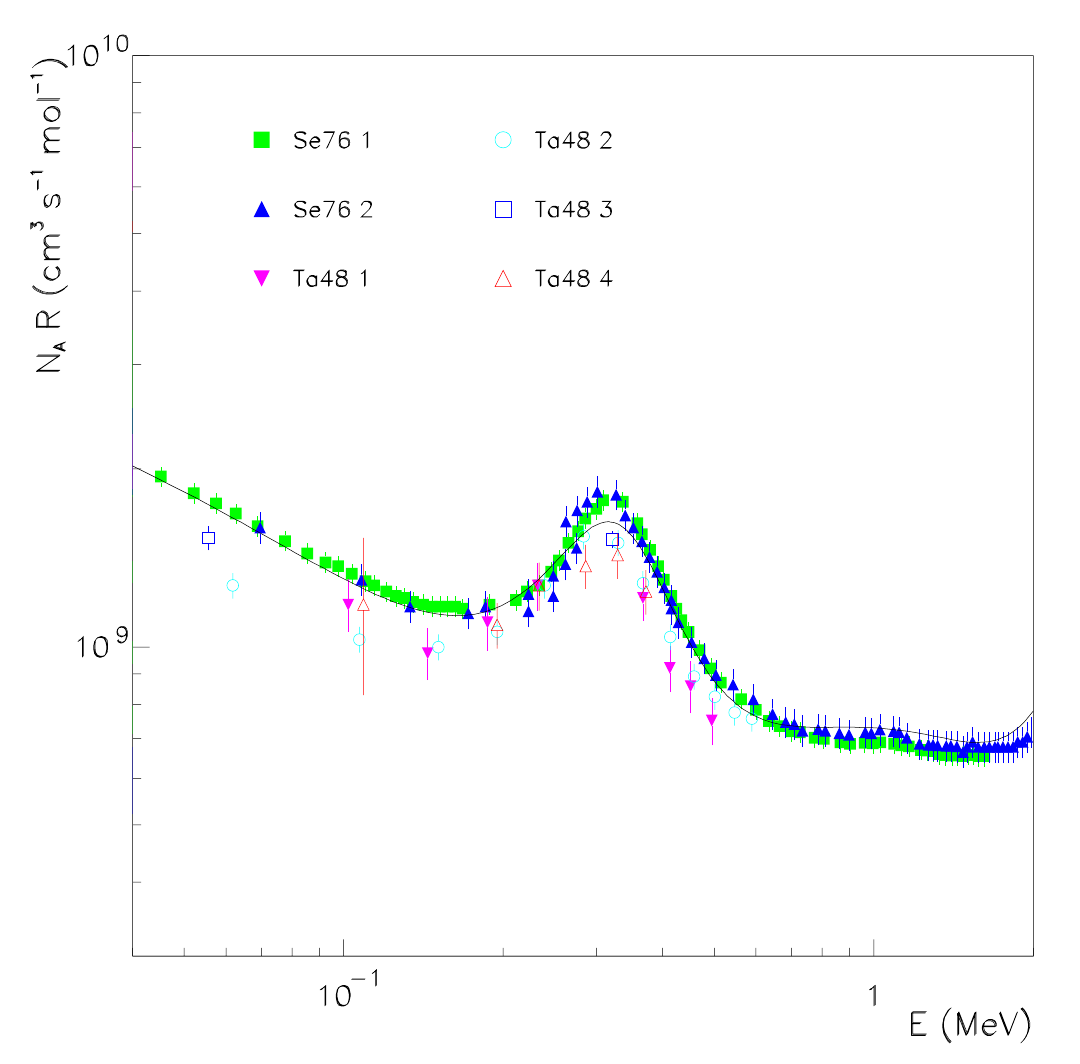}
\end{center}
\caption{The $S$ factor data and fit of the $^7$Be
+ n $\lrt$ p + $^7$Li reaction.}\label{plotbetnp}
\end{figure}

\subsubsection{Reaction \emph{$\alpha$he3$\gamma$}: $^4$He + $^3$He $\lrt \gamma$ +
$^7$Be}.

It is the dominant channel for direct ${^7Be}$ production and, in
the (relatively) high-$\eta$ Universe suggested by WMAP data,
practically all $^7$Li synthesis is controlled by this reaction.
Its importance for the prediction of the solar neutrino spectra
has motivated several theoretical and experimental efforts in the
past years to obtain a better estimate of the cross section. A
constant plus a decreasing exponential times a polynomial was used
to fit its non-resonant $S$ factor, and the data used are
\cite{Holmgren59}-\cite{Hilgemeier88}, where the data in
\cite{Krawinkel82} were renormalized by a factor 1.4 to correct
for the Helium gas density. Our regression method estimates an
error of less than 3\%, mainly dominated by the scale uncertainty.
Both the value of the reduced $\chi^2$, $\chi_\nu^2=2.1$, and the
high average renormalization we find from the fit, of the order of
$20 \%$, in agreement with the majority of the quoted errors,
strongly suggest to undergo a new measurement campaign, to finally
establish both the shape and the scale of this process at a few
percent accuracy level. Of course the solar neutrino predictions
would benefit of this new data, too.
\begin{figure}
\begin{center}
\includegraphics[scale=0.5]{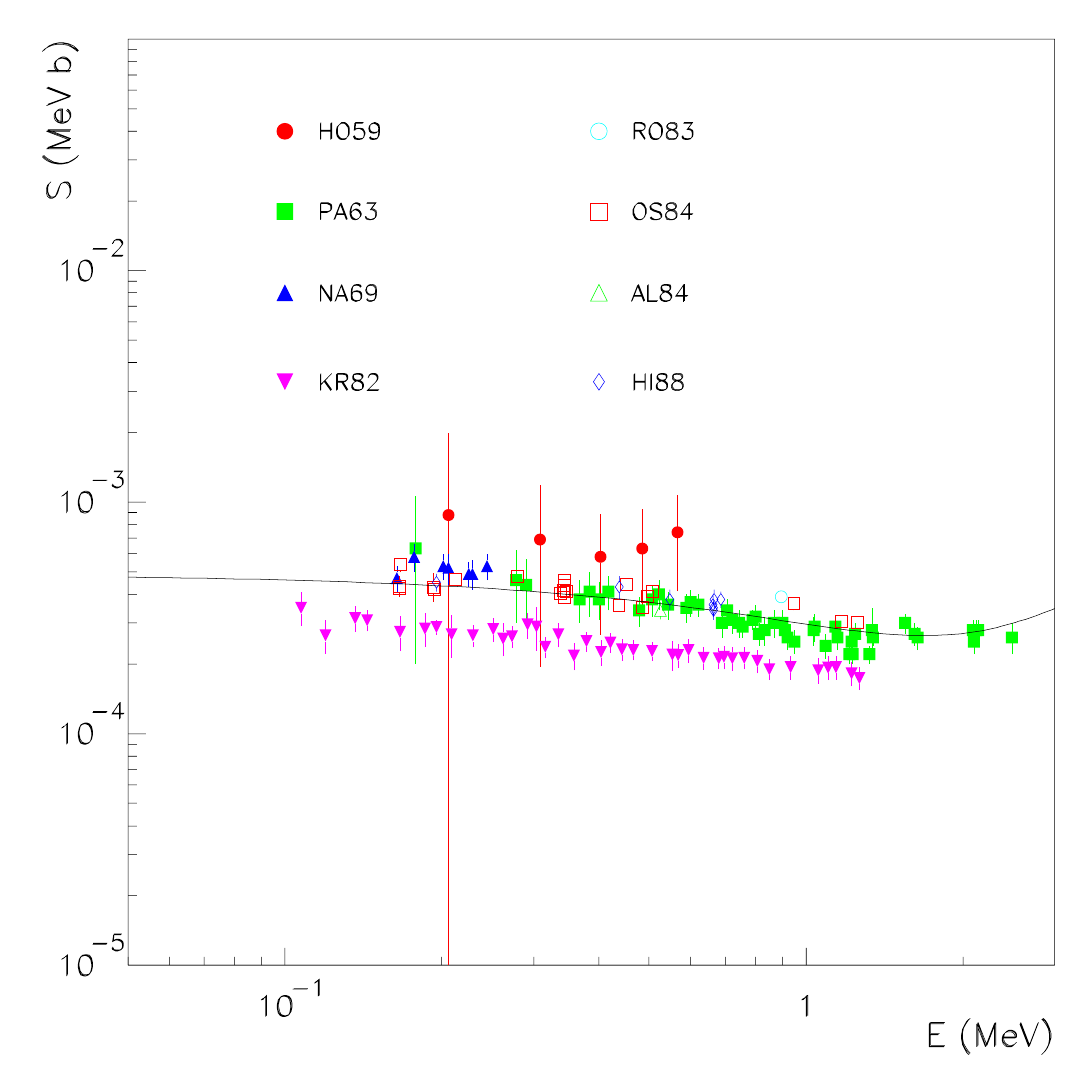}
\end{center}
\caption{The $S$ factor fit and data for the $^4$He + $^3$He $\lrt$ $\gamma$ + $^7$Be reaction.}
\label{plothe4he3g}
\end{figure}

\subsubsection{Reaction \emph{li7p$\alpha$}: $^7$Li + p  $\lrt$ $^4$He + $^4$He}.

This is the main process destroying $^7$Li during the BBN. The
data used in our analysis are the ones quoted in
\cite{Fiedler67}-\cite{Engstler92b}. A self-consistent correction
for the screening potential was also performed, whose effect is
particularly significant for the low-energy data of Engstler \etal
\cite{Engstler92a,Engstler92b}. Our estimate for the overall error
is of the order of 2\% in the relevant temperature range, while
$\varepsilon \sim 0.01$.
\begin{figure}
\begin{center}
\includegraphics[scale=0.5]{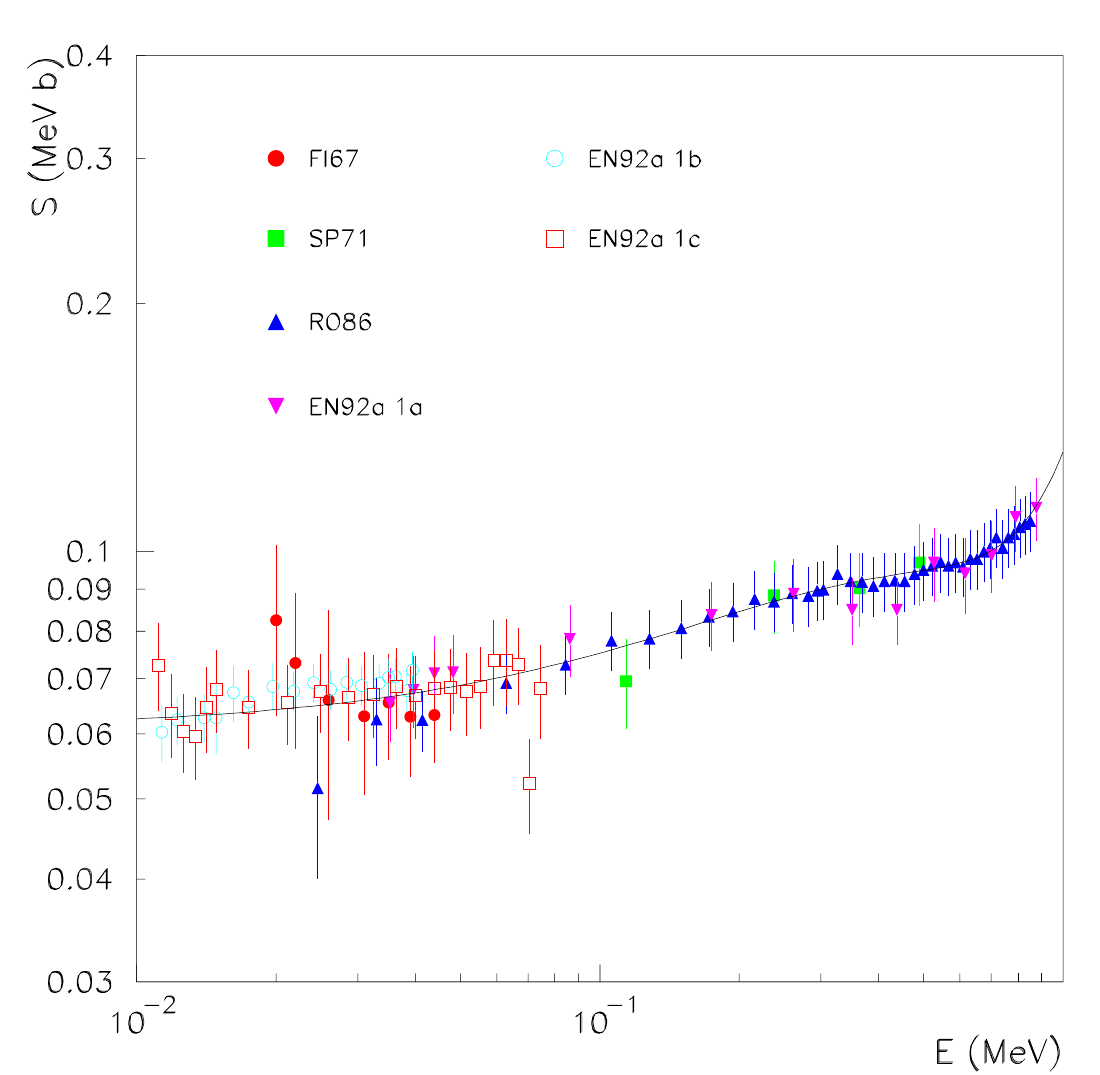}
\end{center}
\caption{The data and fit for the $S$ factor of the $^7$Li + p $\lrt$ $^4$He + $^4$He reaction.}
\label{plotli7pa}
\end{figure}

\subsubsection{Reaction \emph{$\alpha$d$\gamma$}: $^4$He + $^2$H $\lrt \gamma$ + $^6$Li}.

Even if it is a weak electric quadrupole transition, this reaction
is important as represents the main $^6$Li production reaction. It
is experimentally studied only at E $>$ 1 MeV and around the 0.711
MeV resonance. The low energy range is only weakly constrained by
an upper limit, while the theoretical estimates for the
non-resonant rate compiled in \cite{NACRE} show differences of
$1\div 2$ orders of magnitude and were used to establish upper and
lower limits. These features suggest the introduction of a
temperature-dependent asymmetric uncertainty. It would be useful
to get new data at E $\le 0.6$ MeV in order to establish a
reliable estimate of the standard BBN production of $^6$Li; in
fact one cannot presently rule out the possibility that a relevant
fraction of the observed $^6$Li in PopII stars has primordial
origin (see \cite{Nollett:1996ef} and our discussion in Section
4.1.5). In view of the serious lack of data, we simply updated the
code with the inclusion of the NACRE evaluation for rate and
errors.

\subsubsection{Reaction \emph{li6phe3}: $^6$Li + p $\lrt$ $^3$He + $^4$He}.

It is the main $^6$Li destruction channel. Available data for the
cross section are quite abundant and accurate

We used the data \cite{Marion56}-\cite{Kwon89} as well as those in
References \cite{Spinka71,Engstler92a}, already cited for the
\emph{li7p$\alpha$} process, with the last one self-consistently
corrected for the screening effect. The estimated error is lower
than $2\%$ and, quite negligibly contributing to the eventual
accuracy of $^6$Li estimate, with respect to the one coming from
the \emph{$\alpha$d$\gamma$} reaction. We estimate $\varepsilon
\sim 0.09$.
\begin{figure}
\begin{center}
\includegraphics[scale=0.5]{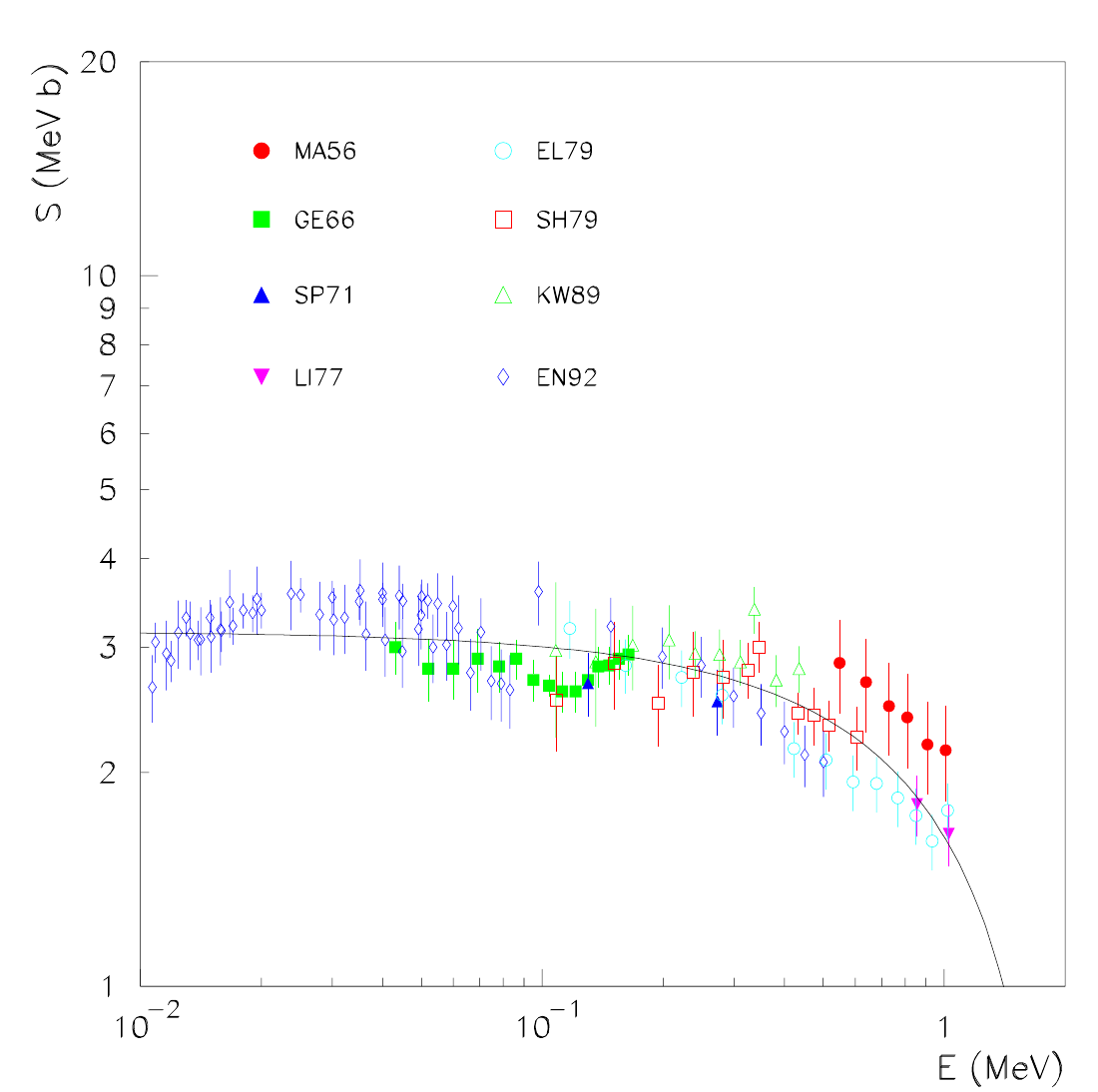}
\end{center}
\caption{The data and fit for the $S$ factor of the $^6$Li + p $\lrt$ $^3$He + $^4$He reaction.}
\label{plotli6phe3}
\end{figure}

\subsection{Main sub-leading processes}

\subsubsection{Reaction \emph{$\alpha$t$\gamma$}: $^4$He + $^3$H $\lrt$ $\gamma$ +
$^7$Li}.

This is the main channel for the \emph{direct} synthesis of
$^7$Li, and has been known since a long time as a crucial process
for the $^7$Li predictions of the BBN (see \cite{SKM93}). However,
especially thanks to the recent data of \cite{Brune94}, its rate
is relatively well known and, within the uncertainties, it only
contributes at the percent level to the $^7$Li yield. This is one
of the main consequences of the present suggested range for
$\omega_b$, that indeed points out the leading role of the
\emph{$\alpha$he3$\gamma$} reaction as the main route to A=7
elements production.
\begin{figure}
\begin{center}
\includegraphics[scale=0.5]{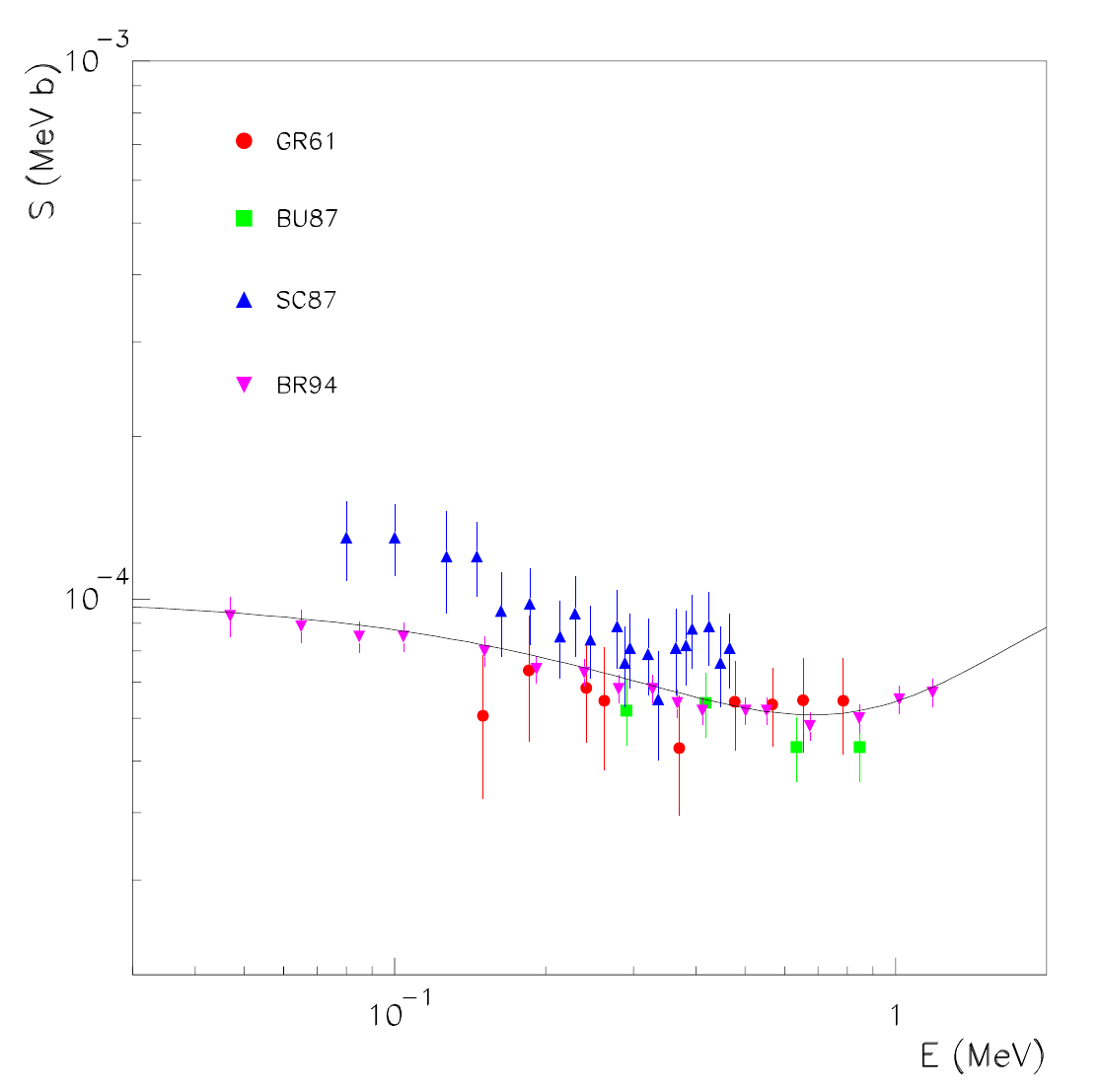}
\end{center}
\caption{The data and fit for the $S$ factor of the $^4$He + $^3$H
$\lrt$ $\gamma$ + $^7$Li reaction.} \label{plothe4tg}
\end{figure}
We parameterize the $S$ factor as in the case of the
\emph{$\alpha$he3$\gamma$} reaction. The data used are
\cite{Brune94}-\cite{Schroder87}, and our regression protocol
gives $\chi_\nu^2\simeq 0.71$, with an overall uncertainty
estimated at the level of 15\% and $\varepsilon \sim 0.02$.

\subsubsection{Reaction \emph{tp$\gamma$}: $^3$H + p $\lrt$ $\gamma$
+ $^4$He}.

This non-resonant process is the third one in order of importance
involved in the $^4$He synthesis, but its influence on the final
error budget of all nuclides is marginal. The $old$ rate for this
reaction is that published in the compilation \cite{ca88}.
Meanwhile, new data were taken \cite{hahnetal95,canonetal02}, and
a new fitting of the astrophysical $S$ factor is now available
\cite{canonetal02}
\bea S(E)=S_0+S_1E+S_2E^2 \vv\\ S_0=2.0{\pm} 0.2 \:keV\,mb \vv\\
S_1=(1.6{\pm} 0.4){\times} 10^{-2}\:mb \vv\\ S_2=(1.1{\pm} 0.3){\times} 10^{-4}\:mb/keV \pp \eea
We used these parameters to estimate the best rate, while its uncertainty
is assumed to be $\Delta S(0)/S(0) \simeq 10\%$. Notice that correlations
between the parameters haven't been published, so our error propagation
method on this rate cannot be used without a full re-analysis of the data.

\subsubsection{Reaction \emph{li7p$\gamma$}: $^7$Li + p$\stackrel{^8{\rm Be}}{\longrightarrow}
\gamma$ + $^4$He + $^4$He}.

This reaction mainly proceeds trough the resonant term at
$E_{8{\rm Be}}=19.86$ MeV, but also has a non-negligible
non-resonant contribution. Both were quite recently re-determined
in \cite{ZA95}. The relative importance of this reaction as an
additional burning channel for the $^7$Li was pointed in
\cite{SKM93}, where its contribution was estimated to change the
destruction rate up to $\sim$ 8\% at $T_9=1$, but this seems to be
neglected in all recent studies. Notice that the new data
collected in \cite{ZA95} move upwards the estimate of the
non-resonant contribution by more than a factor 10, thus further
increasing its role. Tough sub-leading, it is worthwhile to note
that within the actual uncertainties and assuming the WMAP range
for $\omega_b$, this process acquires a role comparable to that of
$^7$Li +p $\rt$ $^4$He + $^4$He and $^4$He+$^3$H$\rt\gamma$+$^7$Li
reactions in determining the final prediction of $^7$Li. The data
sets used in our analysis are \cite{ZA95}-\cite{Ja57}. We estimate
an overall error of $3 \%$ and a large value of the normalization
spread parameter, $\varepsilon \sim 0.14$.
\begin{figure}
\begin{center}
\includegraphics[scale=0.5]{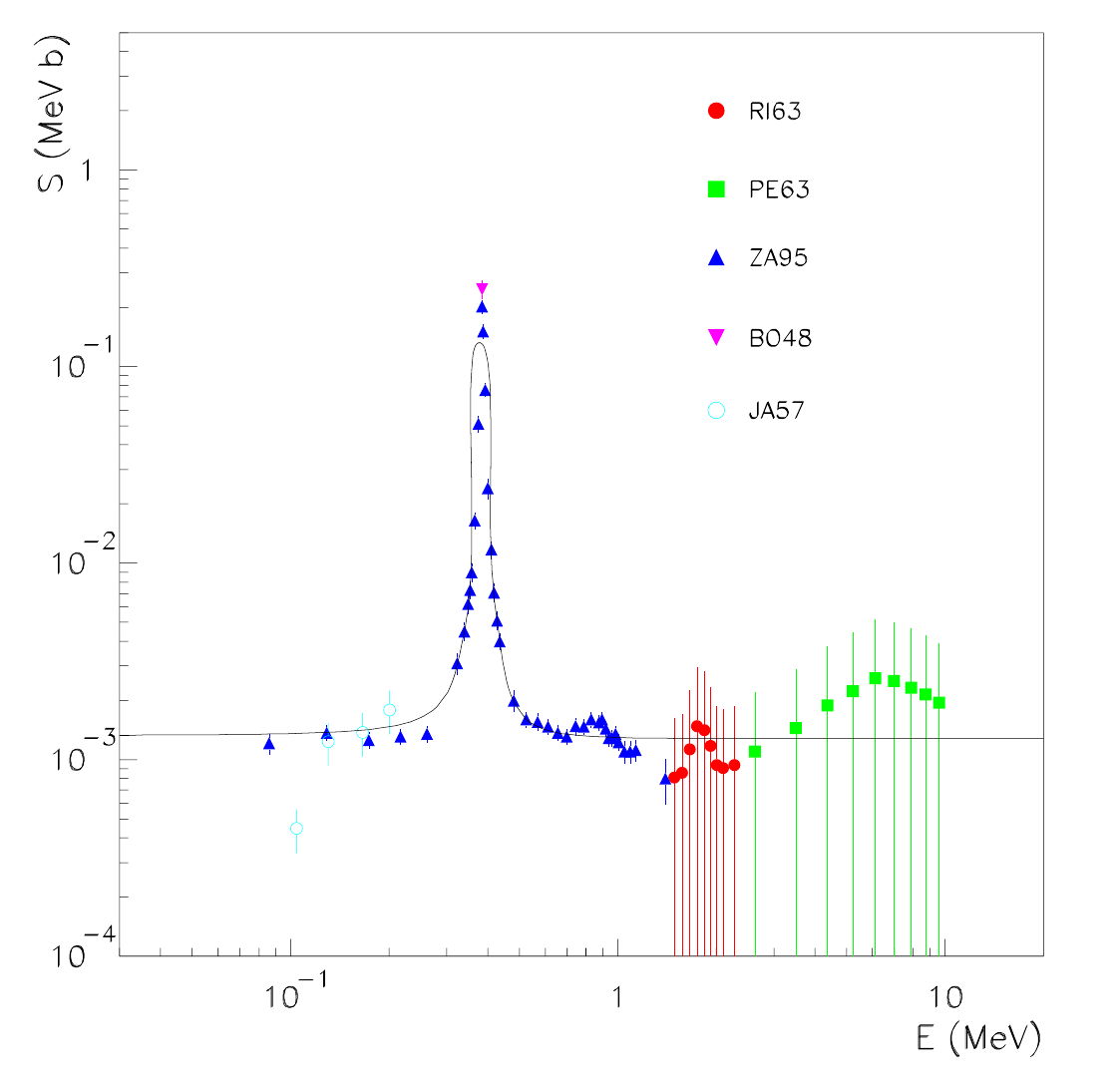}
\end{center}
\caption{The data and fit for the $S$ factor for the $^7$Li + p $\lrt$ $\gamma$ + $^8$Be
reaction.} \label{plotli7pg}
\end{figure}

\subsubsection{Reaction \emph{be7n$\alpha$}: $^7$Be + n $\lrt$
$^4$He + $^4$He}.

To our knowledge, evaluations for the rate of this reaction have
only been published in \cite{wag69} and \cite{Wagoner}, without
information on the sources of the data and error estimate. We did
not find further analysis in the following compilations by Fowler
\etal. \cite{FCZI}. The two data sets of the reverse process
published in \cite{king77,mercer96} refer to center of mass
energies of the direct one greater than 0.6 MeV, thus leaving a
great uncertainty in the BBN window. They seem to be roughly
consistent with the old estimate of the rate, and a new one in
view of so scarce data would make little sense. For this reason we
adopted Wagoner's rate, assuming a factor 10 uncertainty, as he
suggested as a typical conservative value. Within this allowed
range, this reaction could play a non-negligible role in
\emph{direct} $^7$Be destruction, so it would be fruitful to have
a new experimental determination. Apart from the role of unknown
or little known $^8$Be resonances, it is however unlucky that the
used extrapolation may underestimate the rate by more than one
order of magnitude, as this process mainly proceeds through a
p-wave.

\subsubsection{Reaction \emph{li7dn$\alpha$}: $^7$Li + $^2$H $\lrt$ n + $^4$He +
$^4$He}.

The deuterium induced destruction processes of $^7$Li are
dominated by this reaction, and a relatively recent analysis on
this issue has been published in Reference \cite{boydeal93}.
However, within a conservatively assigned 50\% uncertainty, the
new evaluation produces almost no change in the Lithium amount
with respect to the compilation \cite{ca88}, so it seems difficult
that this process could play a role in reducing the $\sim 3\sigma$
discrepancy between predicted and observed $^7$Li yields.

\subsubsection{Reaction \emph{be7dp$\alpha$}: $^7$Be + $^2$H $\lrt$ p + $^4$He +
$^4$He}.

The rate for this reaction is taken from \cite{ca88}, and is
probably derived from the Parker estimate \cite{parker72} based on
the data published in \cite{kav60}. It has been recently pointed
out \cite{Coc:2003ce} that an enhancement of this process at low
energies ($E_{CM}\le 0.5 \:{\rm MeV}$) of a factor $100$ or larger
could reduce the discrepancy between the predicted and observed
values of $^7$Li. Even if this possibility cannot be ruled out on
the basis of the available data, we stress that there are no
compelling reasons, both theoretical and experimental, suggesting
such an anomalous enhancement at low energies. New measurements
will greatly help in clarifying this issue.

\section{Light nuclei abundances}
\setcounter{equation}0

In this Section we first review the present data on primordial
abundances of light nuclides. The aim is to provide a sufficiently
self-contained description of the issue, mainly collecting the
most recent results more than entering into the complexity of the
measurement procedures and of all possible sources of systematics.
For this reason it should not be considered as a complete
discussion on such an involved topic. More details on the
experimental techniques can be found in the original papers we do
quote in the following. The second part of this Section is then
devoted to compare the theoretical expectations, obtained by
numerically solving the BBN system of equations, with experiments.
Particular stress is given to quantify uncertainties, in
particular those due to propagation of nuclear rate measurement
errors.

\subsection{Experimental results}

\subsubsection{$^2$H}.

The deuterium is known to be the nuclide that better allows for a precise
determination of the value of $\eta$. The most recent experimental results
are discussed in \cite{Kirkman:2003uv}, where the authors deduce the
estimate
\begin{equation}
X_{^2{\rm H}}/X_p= (2.78^{+0.44}_{-0.38}){\cdot} 10^{-5} \pp
\label{qsodeut}
\end{equation}
This value is the average of five measurements of DI/HI column
ratio in different QSO absorption systems at high red-shift,
obtained via deuterium isotopic shift from hydrogen. In particular
this analysis takes advantage of the new detection towards
Q1243+3047, based on an improved modeling of the continuum level,
the Ly$\alpha$ forest and the velocity structure of the
absorption, as well as on a better relative flux calibration.

In spite of the fact that  the reduced $\chi^2$ is pretty large
$\chi^2_\nu=4.1$, when using the set of QSO measurements
considered in \cite{Kirkman:2003uv}, so that unidentified
systematics may be still affecting some of the data (correlations
with the column density, local peculiar depletion, etc.), it is
nevertheless interesting to note that the preferred range for
$X_{^2{\rm H}}/X_p$ is in quite a good agreement with a standard
BBN scenario with three neutrinos and WMAP result for $\omega_b$.
We will come back to this point later.

\subsubsection{$^3$He}.

The $^3$He could provide other constraints on the BBN scenario, but here
the experimental study is complicated by the fact that a relevant
contribution is expected from stellar destruction and production, making
difficult to extrapolate the primordial abundance of this nuclide. Apart
from Solar System measurements, $^3$He is usually detected through the
8.665 GHz spin flip transition of $^3$He$^{+}$, but data are limited to
(chemically evolved) galactic HII regions and planetary nebulae (for this
issue, see \cite{bania02} and References therein.) Assuming the simplest
scenario of a typically limited stellar enrichment of $^3$He, the best
observed system (S209 region) suggests $X_{^3{\rm He}}/X_p \le (1.1 {\pm} 0.2){\cdot}
10^{-5}$, while a more conservative limit based on a larger sample gives
$X_{^3{\rm He}}/X_p \le (1.9 {\pm} 0.6){\cdot} 10^{-5}$. As discussed in
\cite{Vangioni-Flam:2002sa}, at the moment one cannot really consider the
$^3$He as a strong ``baryometer'' because of the uncertainty on the role of
stellar processing, the limited range in metallicity explored, and also the
weaker dependence of $X_{^3{\rm He}}$ on $\eta$ with respect to $X_{^2{\rm
H}}$. On the other hand, if combined with the increasing precision of the
BBN predictions, and using the value of $\eta$ as determined by present and
forthcoming CMB data, these observations could clarify several issues
related to the Galactic Chemical evolution.

Finally, we note that the evidence collected till now for a sort
of `plateau' with respect to metallicity \cite{bania02} and the
order of magnitude of the observed abundance at least
qualitatively support the idea of a common, primordial origin for
the bulk of this nuclide abundance.

\subsubsection{$^4$He}.

The value of $Y_p$ is usually derived by extrapolating to zero metallicity
measurements done in dwarf irregular and blue compact galaxies (BCGs), that
are among the least chemically evolved galaxies. The values obtained have a
typically low statistical uncertainty (at the level of $0.002$, i.e. less
than $1\%$) because of the large number of data, but are affected by
several systematical uncertainties, whose origin has been considered in
many recent works (see e.g. References \cite{Olive:2000qm,Thuan:2001zc}).
The estimate given in \cite{Fields:1998gv}, based on a wide database and on
a careful study of the chemical evolution of BCGs, is
\begin{equation}
Y_p=0.238{\pm}(0.002)_{stat}{\pm}(0.005)_{sys} \vv
\end{equation}
while in the recent work \cite{Izotov:2003xn} the authors quote
\begin{equation}
Y_p=0.2421{\pm}(0.0021)_{stat} \vv
\end{equation}
when the linear regression is performed vs. O/H on a restricted,
detailed-studied sample of galaxies. These observations are obtained with
high signal-to-noise ratio, and a self-consistent approach on several He
spectral lines was used in order to fit simultaneously as many parameters
as possible. They also found that the typical size of the systematic errors
is of the order of few percent, but the best estimate of the combined
effect is at the $0.5\%$ level. Slightly higher values (from 0.243 to
0.247) are found neglecting some corrections, using a wider data sample,
performing  regression versus N/H, or finally using different Equivalent
Widths (EWs).

Very recently, in~\cite{Olive:2004kq} a new, detailed analysis of the
budget of systematics in $Y_p$ was performed. With a different method and
sample selection, this analysis of the same data of Izotov and Thuan gives
$Y_p$ = $0.249 {\pm} 0.009$, quite different indeed, even if consistent within
1 $\sigma$. If we adopt the conservative range quoted in
\cite{Olive:2004kq}, (0.232 $\div$ 0.258), and viewing these bounds as a
reasonable 2 $\sigma$ interval for $Y_p$, we have (mimicking a ``1 $\sigma$'' error)
\begin{equation}
Y_p=0.245 {\pm} 0.007 \vv \label{conservhe}
\end{equation}
which indeed is the value we will use to compare our theoretical
results. Even if a knowledge of $Y_p$ at the level of 3\% may seem
quite satisfactory, in order to use $^4$He to bound possible BBN
scenarios, it would be quite important to improve the $Y_p$
measurement accuracy at the level of the quoted {\it statistical}
uncertainty, which requires a clear understanding of main sources
of systematics.

We would briefly point out that other methods have been proposed
to estimate $Y_p$, that hopefully could help in understanding the
role of unknown systematics in the previous results. For example,
in \cite{Salaris:2004xd} and References therein, $Y_p$ is
determined from indirect studies of Galactic Globular Clusters.
The value they found
$Y_p=0.250{\pm}(0.006)_{stat}{\pm}(0.019)_{sys}$ is still far from
the required accuracy, but at the moment it is a promising
contribution to clarify the debated issue of the primordial $^4$He
determination. Finally in \cite{Trotta:2003xg} a first attempt was
made to determine $Y_p$ from CMB anisotropies data, but the 1
$\sigma$ confidence level $0.160<Y_p<0.501$ is still too broad.

\subsubsection{$^7$Li}.

The role of possible depletion mechanisms of the \emph{Spite
plateau} found in PopII dwarfs halo stars in the seminal paper
\cite{Spite*2:82} by Spite \& Spite, is still under debate. For
example, the authors of \cite{ryaneal99} found evidence for a
positive dependence of $X_{^7{\rm Li}}$ on metallicity, estimating
\be X_{^7{\rm Li}}/X_p = 1.23 {\pm} 0.06 ^{+0.68}_{-0.32} \vv \ee
at 95 $\%$ C.L., with uncertainty dominated by systematics effect.
In other cases \cite{pinseal01,salweis01,thevaucl01} a negative
trend and/or a dispersion of the measurements around the plateau
greater than the experimental error was suggested. A huge number
of mechanisms has been invoked to explain such a $^7$Li depletion,
among which rotational mixing, diffusion, stellar winds, etc (see
e.g.\cite{Talon:2004jq} and References therein for an introduction
to the current status of PopII stellar models). However, according
to \cite{Boneal02AA}, these effects are second order, and a
general consensus on the primordial origin of (at least) the bulk
of the $^7$Li plateau has been achieved. The value quoted in
\cite{Boneal02AA} is the following \be X_{^7{\rm Li}}/X_p
=2.19^{+0.46}_{-0.38} \pp \ee An important step is the recent
check that other objects like the Globular Cluster NGC 6397
turn-off stars share essentially the same $X_{^7{\rm Li}}$
\cite{Boneal02AA,bon97MNRAS}, and even the troublesome
observations in the globular cluster M92 reported in
\cite{boesgal98} seem to agree much better with other
determinations in the reanalysis performed in \cite{bon02AA} with
a new temperature calibration.

At the moment, how to compare the extracted value of  $X_{^7{\rm Li}}$ with
the one obtained from the standard BBN scenario is still quite a
problematic issue. As the only traditionally observed line is the 6708 {\AA}
doublet of the neutral Lithium (Li I) and the fraction of ionized Lithium
(Li II) is estimated to be larger than $ 99.7\%$, a critical parameter for
the extraction of A(Li)$\equiv \log_{10}(n_{{\rm Li}}/n_{{\rm H}})+12$ is
the effective temperature of stellar atmospheres $T_{\rm eff}$. Other
parameters as surface gravity, metallicity or microturbulence are much less
important \cite{bon97MNRAS}. As $T_{\rm eff}$ is only estimated indirectly,
recently it was also pointed out the possible bias in extracting the
plateau abundance from one spectral line only. Actually when the
subordinate line at 6104 {\AA} is tentatively used, some disagreement is found,
and values obtained for A(Li) could be high enough, A(Li)$\sim 2.5$, to
reduce the discrepancy with BBN predictions (further details can be found
in \cite{Ford02} and References therein).

As in previous studies, we do not use $X_{^7{\rm Li}}$ in our BBN
likelihood analysis, waiting for a clearer understanding of
possible systematics in stellar modeling. However, further
efforts should be also devoted in trying to reduce the large
theoretical uncertainty which is still affecting $X_{^7{\rm Li}}$,
mainly by lowering the uncertainty on the leading reactions,
$^7$Be + n $\lrt$ $^4$He + $^4$He, $^4$He + $^3$He $\lrt$ $\gamma
+{^7}$Be, and $^7{\rm Be}$ + $^2{\rm H}$ $\lrt$ p + $^4{\rm He}$ +
$^4{\rm He}$, which are responsible for a large fraction of the
overall error on $X_{^7{\rm Li}}$. We will come back to this point
in Section 4.2.3.

\subsubsection{$^6$Li}.

Among the remaining light nuclides, the most abundant (for the
best range for $\eta$) is $^6$Li, with $X_{^6{\rm Li}}/X_p \sim
10^{-14}$. One has to conclude that, in the more viable scenarios,
there is no hope at the present for its detection. Actually, some
positive measurements in halo stars at the level of
$^6$Li/$^7$Li$\simeq 0.05$  were obtained in the last decade
\cite{li6obs}, but they reflect the post-primordial production of
this nuclide in Cosmic Ray spallation nucleosynthesis. This does
not mean that this issue is unrelated to BBN studies; on the
contrary, as discussed e.g. in \cite{Olive:1999kq}, the study of
the chemical evolution of the fragile isotopes of Li, Be and B
could constraint the $^7$Li \emph{primordial} yield, hopefully
further clarifying the observational situation in the Spite
Plateau. For example, the survival of such a (relatively) high
fraction of  $^6$Li in halo PopII stars severely limits the
possible depletion factor for the more tightly bound $^7$Li.

\subsection{Theoretical predictions and uncertainties}

\subsubsection{General considerations}.

The goal of a theoretical analysis of BBN is to obtain a reliable
estimate of the nuclide abundances $X_i$ and of their
uncertainties, once the best values and errors on the standard
physics inputs are known. These inputs are
\begin{itemize}
\item $\tau_n$, the neutron lifetime; \item $G_N$, the Newton
gravitational constant; \item $\eta$, the baryon to photon number
density ratio; \item the nuclear rates.
\end{itemize}
The first two parameters are now known with a satisfactory accuracy
\cite{PDG}
\begin{eqnarray}
\tau_n&=&885.7{\pm} 0.8 \,s \vv \\
G_{\rm N}&=&6.7087{\pm} 0.0010 {\cdot} \, 10^{-39}\, GeV^{-2} \pp
\end{eqnarray}
Until recently, BBN has been the only probe of $\eta$, and the best fitting
with the observed abundances was used to guess its value plus some insights
on neutrino or non-standard physics. Thanks to the precise determinations
of $\omega_b$ obtained by the WMAP Collaboration results on CMB
anisotropies, even $\eta$ is now independently known with an increasing
precision, fixing the predictions of the standard scenario of BBN within a
relatively narrow range. The value of the baryon parameter $\omega_b=
\Omega_b h^2$ from CMB measurements is presently the following
\cite{wmap03} \be \omega_b = 0.023 {\pm} 0.001 \pp \ee As these
first three sources of error are independent and uncorrelated, one
can easily work out the propagated uncertainty on each $X_i$, by
adding in quadrature the three contributions.

The error matrix due to nuclear rate uncertainties are calculated
as in our previous work \cite{cuocoetal}
\begin{equation}
\sigma_{ij}^2=\frac{1}{4}\sum_{k}[(X_i(\Gamma_k^{+})-
X_i(\Gamma_k^{-})][(X_j(\Gamma_k^{+})-X_j(\Gamma_k^{-})] \vv
\end{equation}
where the sum is over all reactions and with all other parameters set at
their best values. The value $\Gamma_k^{\pm}$ is the average $\Gamma_k$ rate
plus (minus) the estimated $1$ $\sigma$ uncertainty. Thus the contribution
to the uncertainty on $X_i$ is given by
\begin{equation}
\sigma_i=\sqrt{\sigma_{ii}^2} \vv \label{sisi}
\end{equation}
and the correlation between $X_i$ and $X_j$
\begin{equation}
\rho_{ij}(\eta)=\frac{\sigma_{ij}^2(\eta)}{\sigma_{i}(\eta)\,\sigma_{j}(\eta)}
\pp \label{rhoij}
\end{equation}

Being a generalization of the method described in \cite{flsv98}, this
approach avoids the use of (time consuming) Monte Carlo simulations, while
it does not require the existence of the functionals
$\lambda_{ij}(\eta)=\partial ln\,X_i/\partial ln\Gamma_k$. This means that
it can be applied even to the case of asymmetric or temperature dependent
errors, which indeed is the case for several process rates. The reliability
of the method is supported by the Central Limit Theorem (at least when
many reactions contribute to the error almost on the same level)
and the fact that
there is typically a linear relation between the $X_i$ and the parameters
they depend on. Furthermore a direct comparison with Monte Carlo
simulations results done in \cite{flsv98} shows that the method gives
accurate results.
We expect that this agreement is even improved for the typically smaller
errors we found, and the allowance for the temperature-dependence of these
quantities.

Apart from the inputs we have discussed so far, a BBN scenario does depend
on several parameters which set a specific theoretical framework. As a main
example, the eventual abundances of (mainly) $^4$He is influenced by the
fact that extra particle species contribute to the overall energy density
at the MeV temperature scale. This increases the value of the Hubble
parameter, so that neutron to proton ratio freezes out at larger
temperatures, where the neutron fraction is higher. It is customary to
parameterize their contribution in terms of the effective number of extra
degrees of freedom, $\Delta N$, defined as follows \be \rho =
\rho_{e,\gamma,\nu} + \rho_B+ \frac{7}{8} \,
\left(\frac{T_\nu}{T}\right)^4\, \Delta N \, \rho_\gamma \vv
\label{deltan} \ee where $\rho$ is the total energy density.
Actually $\Delta N$ is a simple constant term for species which
are fully relativistic at the BBN epoch, while it is a scale factor
depending function otherwise: for example, for particles with mass much 
larger than MeV, $\Delta N$ increases proportionally to the scale factor.
For the case of extra relativistic degrees of freedom
decoupled from the electromagnetic plasma, the maximal allowed range 
$0.232\leq Y_p \leq 0.258$ quoted in~\cite{Olive:2004kq} would 
constrain $-1.14\leq \Delta N \leq +0.73$ for $\omega_b=0.023$.
Since $\Delta Y_p\simeq 0.013\Delta N$, a shift of 0.001 in $\omega_b$
would produce only a change of the order of 0.04 in 
the previous limits (see the Tables~\ref{cmbpriorxi} and~\ref{addendum} and~\ref{fitrate}). 

In addition to $\Delta N$, light nuclei yields are in general
sensitive to several parameters characterizing particle
interactions or rather features of their distribution function.
Most studied examples are the possible value of neutrino chemical
potential \cite{ks92} (for most recent bound on this parameter see
for example \cite{bargeretal,cuocoetal}), or neutrino lifetime
(for a review on this issue see \cite{dolgovrep}).

In what follows we will consider only the {\it minimal standard
scenario}. We therefore assume no extra parameter at all, and so
all theoretical predictions for $^2$H, $^4$He and $^7$Li
abundances are determined by the value of $\eta$. A careful study
of this case seems to us of particular relevance since, as we
mentioned already, the baryon to photon density is independently
fixed by CMB anisotropy measurements, so there are no free
parameters left. Comparing experimental result with theoretical
expectations is therefore a clean test of consistency of the
simplest BBN dynamics, based on our present understanding of
fundamental interactions and of the cosmological model. It may
also give useful hints to point out possible systematics in the
experimental results reviewed in the previous Section.

To compare the present sensitivity to $\omega_b$ of BBN with the one of
WMAP, we will also study how this parameter is constrained by observed
nuclide abundances, i.e. leaving its value as the only free parameter to be
fitted, rather than using the WMAP value prior. To do this, or more
generally to bound the values of any other set of parameters of the kind
discussed so far, the standard procedure is to construct a likelihood
function \be {\cal L}(\eta)\propto \exp\left( -\chi^2(\eta)/2\right)\ee
with
\begin{equation}
\chi^2(\eta)=\sum_{ij}[X_i(\eta)-X_i^{obs}]W_{ij}(\eta)[X_j(\eta)-X_j^{obs}]
\vv
\end{equation}
where the proportionality constant can be obtained by requiring
normalization to unity, and with $W_{ij}(\eta)$ denoting the inverse
covariance matrix \be
W_{ij}(\eta)=[\sigma_{ij}^2+\sigma_{i,exp}^2\delta_{ij}+\sigma_{ij,other}^2]^{-1}
\vv \ee where $\sigma_{i,exp}$ is the uncertainty in experimental
determination of nuclide abundance $X_i$ and by $\sigma_{ij,other}^2$ we
denote the propagated squared error matrix due to all other input parameter
uncertainties ($\tau_n$, $G_{\rm N}$, etc.).

\subsubsection{Some remarks on $\eta$}.

There are two different way to parameterize the baryon content of
the Universe today, which can be used interchangeably, namely
$\eta$ and $\omega_b$. It is interesting to review the simple
relationship between these two quantities, in order to asses how
uncertainties propagate from one parameter to the other. It is
straightforward to get
\begin{equation}
\omega_b=\rho_b\frac{8 \pi G_N}{3 (H_0/h)^2}=\frac{8 \pi G_N
m_u}{3(H_0/h)^2}\, \eta \, \frac{2\zeta(3)}{\pi^2}\, T_{\gamma
0}^3 \, f \vv
\end{equation}
where
\begin{equation}
f=\frac{\rho_b}{m_u n_b}\simeq \left[
\frac{m_H}{m_u}-Y_p\left(\frac{m_H}{m_u}-\frac{m_{4He}}{4m_u}\right)\right]
\pp
\end{equation}
This expression leads to
\begin{equation} \fl
\eta {\cdot} 10^{10} = 273.49 \,\omega_b \frac{1}{1- 0.007 \,Y_p}
\bigg(\frac{6.707{\times} 10^{-45}
{\rm MeV}^{-2}}{G_N}\bigg)\bigg(\frac{2.725 K}{T_{\gamma 0}}\bigg)^3 \pp
\end{equation}
The uncertainties on the parameters ($Y_p,G_N,T_{\gamma 0}$)
generate a $\Delta \eta/\eta$ at the level of $0.1\%$ (errors
added in quadrature), which in view of the present determination
of $\omega_b$, $\Delta \omega_b /\omega_b \sim 4 \%$, is
sub-leading and can be neglected in what follows.

As a further remark we recall that the value of $\eta$ lowers from early
times to the end of nucleosynthesis, because of the entropy transfer to
photons during $e^{\pm}$ annihilation phase. Typically \cite{KawCode92} the
final value $\eta_f$ is chosen as the input parameter, while the initial
value, at $T\simeq 10\,{\rm MeV}$, is then deduced as
\begin{equation}
\eta_{i}= \eta_{f}\left(\frac{\bar{z}_f}{\bar{z}_i}\right)^3 \vv
\end{equation}
where we recall  $\bar{z}=a T$. A careful estimate of this
relation should take into account the mass of the electron even at
initial condition, so that entropy is not simply proportional to
$\bar{z}^3$. Furthermore the value of photon temperature $T$ get
modified by the QED thermodynamical corrections to energy
densities and pressures of $e^{\pm}$ and $\gamma$, as well as from
the the small effects of the residual neutrino coupling to the
electromagnetic plasma. Including all these one gets the estimate
$\eta_i\simeq 2.73 \eta_f$, slightly different than the simplest
result one would get neglecting these effects, and which is for
example used in \cite{KawCode92}, $\eta_i = 2.75 \eta_f$. Actually
the corresponding corrections on the nuclide abundances are quite
small. For $\omega_b=0.023$ we estimate
\begin{equation} \fl
10^5\Delta X_{^2{\rm H}}/X_p =0.03,\,\,\Delta Y_p=-0.6{\times}
10^{-4},\,\,,10^{10}\Delta X_{^7{\rm Li}}/X_p=-0.07 \vv
\end{equation}
In what follows $\eta$ always stands for $\eta_f$.

\subsubsection{Results with the WMAP prior on $\omega_b$}.

\begin{table*}
\begin{center}
\begin{tabular}{|c|c|c|c|}
\hline nuclide $i$ & central value & $\sigma_{ii}$ & $\sigma_{\omega_b}$ \\
\hline &&& \\
 $X_{^2{\rm H}}/X_{{\rm H}}$ \, ($10^{-5}$) & 2.44 & ${\pm} 0.04$ & $^{+0.19}_{-0.16}$ \\
 &&& \\
\hline &&& \\ $X_{^3{\rm He}}/X_{{\rm H}}$ \,($10^{-5}$) & 1.01 &
${\pm} 0.03$ & $^{+0.02}_{-0.03}$ \\ &&& \\ \hline &&& \\ $Y_p$ & 0.2481 &
$^{+0.0002}_{-0.0001}$ & $^{+0.0004}_{-0.0004}$
\\ &&& \\ \hline &&&
\\
$X_{^6{\rm Li}}/X_{{\rm H}}$ \, ($10^{-14}$)& 1.1 & ${\pm} 1.7$ & $ {\pm} 0.07$ \\
&&&
\\ \hline  &&& \\ $X_{^7{\rm Li}}/X_{{\rm H}}$ \, ($10^{-10}$) & 4.9 & ${\pm} 0.4$
& ${\pm} 0.4$
\\ &&& \\ \hline
\end{tabular}
\end{center}
\caption{The light nuclide abundances for $\omega_b=0.023$ (second
column). The (1 $\sigma$) uncertainties due to nuclear rates
errors ($\sigma_{ii}$) and $\omega_b$ ($\sigma_{\omega_b}$) are
also shown in the third and fourth columns, respectively.}
\label{cmbpriorxi}
\end{table*}

As we mentioned, in the standard BBN scenario with no extra effective
degrees of freedom or neutrino chemical potential, the light nuclide
abundances are determined by the baryon density only. In Table
\ref{cmbpriorxi} we report the values for the most relevant nucleosynthesis
products, {\it assuming} the WMAP result, $\omega_b=0.023 {\pm} 0.001$. We also
show the corresponding uncertainties due to the propagated errors
$\sigma_{ii}$ on nuclear rates, as well as $\sigma_{\omega_b}$, the one
induced by $\omega_b$ 1 $\sigma$ error. As a general comment on these
results we can say that our analysis of experimental rates, and in
particular the adopted method for error estimates, led to a sensible
reduction of $\sigma_{ii}$ compared to our previous results in
\cite{cuocoetal}. The reason for this is easily understood. In
\cite{cuocoetal} we used a much more conservative approach, since we
accounted for the fact that, for several rates, experimental results were
often in disagreement by introducing a large maximal error defined
similarly to the parameter $\varepsilon$ of Equation (\ref{varepsilon}).

As we already mentioned in the Introduction the value of
$\omega_b$ adopted in our analysis is the result given by the WMAP
Collaboration combining their findings with those on CMB
anisotropies of CBI and ACBAR experiments as well as with 2dFGRS
data on power spectrum assuming a $\Lambda$CDM model. Slightly
different values are obtained by including Lyman $\alpha$ forest
data or introducing a running spectral index (see Tables 7 and 8
in \cite{wmap03}). In particular the best value for $\omega_b$
varies in the small range $0.022 \div 0.024$ when combining
different data sets, while the $1\sigma$ uncertainty is always of
the order of $0.001$, or slightly smaller. The impact of these
different determinations of $\omega_b$ on the nuclei abundance is
shown in Table \ref{addendum} where we report the best values for
$\omega_b=0.022$, which is obtained combining WMAP+CBI+ACBAR in
the running spectral index scenario \cite{wmap03}, and for
$\omega_b=0.024$, which is instead the preferred value of WMAP
data only with a power law $\Lambda$CDM model. Notice that
Deuterium and $^7$Li show appreciable changes, at the level of 10
\%. We also note that the uncertainty due to nuclear rates, which
in principle depends on the value of $\omega_b$, have very tiny
variations and are still given by the values shown in the third
column of Table 3. See also~\ref{fitrate} where we report a fit of
both nuclei abundances and the covariance matrix for $^2$H, $^4$He
and $^7$Li.

\begin{table*}
\begin{center}
\begin{tabular}{|c|c|c|}
\hline nuclide $i$ & $\omega_b=0.022$ & $\omega_b=0.024$  \\
\hline && \\
 $X_{^2{\rm H}}/X_{{\rm H}}$ \, ($10^{-5}$) & 2.63 & 2.28  \\
 && \\
\hline && \\ $X_{^3{\rm He}}/X_{{\rm H}}$ \,($10^{-5}$) & 1.03 &
0.98  \\ && \\ \hline && \\ $Y_p$ & 0.2477 & 0.2485
\\ && \\ \hline &&
\\
$X_{^6{\rm Li}}/X_{{\rm H}}$ \, ($10^{-14}$)& 1.1 & 1.0 \\
&&
\\ \hline  && \\ $X_{^7{\rm Li}}/X_{{\rm H}}$ \, ($10^{-10}$) & 4.5 &
5.3
\\ && \\ \hline
\end{tabular}
\end{center}
\caption{The light nuclide abundances for $\omega_b=0.022$
(WMAP+ACBAR+CBI with a running spectral index) and
$\omega_b=0.024$ (WMAP data only with a power law $\Lambda$CDM
model).} \label{addendum}
\end{table*}

Comparison with the analysis of \cite{Cyburt:2004cq} shows a
similar trend, though the uncertainties quoted in \cite{Cyburt:2004cq} are  
typically larger for the reasons discussed in Section~\ref{ErrEstNucRat}. 
As a main exception we mention the fact that we
perfectly agree on the uncertainty quoted for $Y_p$, since in this
case (almost) all error is due to neutron lifetime statistical
uncertainty.

We now consider in more details our results for each nuclide.
\begin{itemize}

\item[--]$^2$H --

The Deuterium abundance is in quite a good agreement with the
experimental result quoted in Section 4.1.1. This is perhaps the
main indication that indeed the standard picture of BBN is fully
satisfactory. We know in fact that $^2$H is extremely sensitive to
the value of baryon density. Thus it is rather remarkable that the
value found for $\omega_b$ from a completely independent
observable, such as the CMB anisotropies, leads to a $^2$H yield
very close (1 $\sigma$) to measurements.

In our findings the uncertainty due to nuclear rates is now
reduced to $1.6 \%$, mainly because of our analysis of the two
leading reactions \emph{ddn} and \emph{ddp}. It is interesting in
fact to study the contribution of each rate to the total error
$\sigma_{^2{\rm H}^2{\rm H}}$. The results are reported in Table
\ref{errord}.
\begin{table*}
\begin{center}
\begin{tabular}{|c|c|}
\hline rate & $\delta \sigma_{^2{\rm H}^2{\rm H}}^2/ \sigma_{^2{\rm H}^2{\rm H}}^2 (\%)$ \\
\hline \emph{dp$\gamma$}& 49 \\
\emph{ddn}& 37 \\
\emph{ddp}& 14 \\
\hline
\end{tabular}
\end{center}
\caption{The contribution of reaction rate errors to the total
(squared) uncertainty on $X_{^2{\rm H}}/X_p$} \label{errord}
\end{table*}
The role of \emph{ddn} and \emph{ddp} is now of the order of $50
\%$, while in our previous study it was largely dominant
\cite{cuocoetal}. The reduced estimated uncertainties on these two
processes, which are very efficient $^2$H burning channels, is
mainly responsible for that. Notice that $^2$H uncertainty is now
dominated by $\sigma_{\omega_b}$. Even more than in the past, the
role of $^2$H as a baryometer is clearly established.

\item[--]$^3$He --

The estimated value for $^3$He almost saturates the bound reported
in Section 4.1.2, $X_{^3{\rm He}}/X_p \le (1.1 {\pm} 0.2){\cdot}
10^{-5}$, while is a factor two smaller than the conservative
limit, $X_{^3{\rm He}}/X_p \le (1.9 {\pm} 0.6){\cdot} 10^{-5}$. It
would be of great impact for standard BBN, as well as for theories
on galactic chemical evolution, to have further experimental
information on the abundance of this nuclide, trying to reduce the
present bound. The error budget study shows that the theoretical
uncertainty is equally shared by the nuclear rate errors and the
uncertainty on $\omega_b$. Actually it is worth stressing that the
value of the error due to nuclear rates is very close to what is
found for $^2$H. It receives the main contributions from the
processes \emph{he3dp} and \emph{dp$\gamma$}, as can be read from
Table \ref{errorhe3}. On the other hand the fact that $^3$He is
less critically depending on $\omega_b$ with respect to $^2$H
shows up in a relatively smaller value of $\sigma_{\omega_b}$.
\begin{table*}
\begin{center}
\begin{tabular}{|c|c|}
\hline rate & $ \delta \sigma_{^3{\rm He}^3{\rm He}}^2/ \sigma_{^3{\rm He}^3{\rm He}}^2 (\%)$ \\
\hline \emph{he3dp}& 80.7 \\
\emph{dp$\gamma$}& 16.8 \\
\emph{ddp}& 1.3 \\
\emph{ddn} & 1.2 \\ \hline
\end{tabular}
\end{center}
\caption{The contribution of reaction rate errors to the total
(squared) uncertainty on $X_{^3{\rm He}}/X_p$} \label{errorhe3}
\end{table*}

\item[--]$^4$He --

The value of the Helium mass fraction parameter is in very good
agreement with previous theoretical determinations (see for
example \cite{Cyburt:2004cq}, \cite{cuocoetal}, \cite{cyburtpl}).
Comparison with the experimental values only shows a satisfactory
agreement if we adopt the conservative estimate discussed in
Section 4.1.3. The value given by \cite{Fields:1998gv} (see
Equation 4.2) is instead compatible at the 3-$\sigma$ level, even
considering their estimate of systematic error in the measurement.
Similarly the result of \cite{Izotov:2003xn} reported in Equation
4.3 is significantly lower with respect to the theoretical value. If we
assume that Deuterium results points towards the consistency of
both standard BBN and the WMAP result on $\omega_b$, we are led to
the conclusion that the uncertainty of present measurements of
$Y_p$ is largely dominated by systematic effects, which lead to an
underestimation of $^4$He density. Of course a better agreement
between data and theoretical estimates is obtained in less
conservative theoretical frameworks with exotic physics like the
presence of extra particles or large chemical potentials. These
scenarios have been recently studied in \cite{bargeretal},
\cite{cyburtpl} and \cite{cuocoetal}.
\begin{table*}
\begin{center}
\begin{tabular}{|c|c|}
\hline rate & $\delta \sigma_{^4{\rm He}^4{\rm He}}^2/ \sigma_{^4{\rm He}^4{\rm He}}^2 (\%)$ \\
\hline \emph{wnp}& 98.5 \\
\emph{ddn}& 1 \\
\emph{ddp}& 0.25 \\
\emph{dn$\gamma$} & 0.25 \\ \hline
\end{tabular}
\end{center}
\caption{The contribution of reaction rate errors to the total
(squared) uncertainty on $Y_p$} \label{errorhe4}
\end{table*}
We show in Table \ref{errorhe4} the main contribution of rate
uncertainties to the squared error $\sigma_{^4{\rm He}^4{\rm
He}}$. We see in this case again how the new estimate for
\emph{ddn} and \emph{ddp} uncertainties has lowered their role in
the theoretical uncertainties on light nuclides. For $^4$He this
uncertainty is now almost entirely due to neutron lifetime error.
This is also the finding of \cite{Cyburt:2004cq}.

\item[--]$^6$Li --

The best estimate of $^6$Li, $X_{^6{\rm Li}}/X_p = 1.1 {\cdot} 10^{-14}$ gives
for the ratio $^6{\rm Li}/^7{\rm Li}$ the result $0.22 {\cdot} 10^{-4}$, well
below the result obtained for this ratio in halo stars, which, as discussed
in Section 4.1.5, is three order of magnitude larger. Theoretical
uncertainty on this nuclide is still very large, at the level of $100 \%$,
and is entirely due to the effect of \emph{$\alpha$d$\gamma$}, whose rate
is still quite poorly known (see Section 3.4).

\item[--]$^7$Li --

There are several reactions contributing to the total uncertainty
for this nuclide, as shown in Table \ref{errorli7}. Leading
contributions to the $8\%$ error on the theoretical estimate for
$\omega_b=0.023$ comes from \emph{be7n$\alpha$},
\emph{he4he3$\gamma$} and \emph{be7dp$\alpha$}. $^7$Li still has
the largest theoretical error among the observed nuclides, of the
same order of the one induced by $\omega_b$ uncertainty. In view
of this it is worth stressing that new experimental results in
particular on the \emph{be7n$\alpha$} and  \emph{he4he3$\gamma$}
would be extremely important. However, unless unexpected new data
will change the present picture of the main production and
destruction channels for this nuclide, it seems quite hard to
reconcile the theoretical result with the experimental data, which
are smaller by a factor $2 \div 3$. This difference is at the
level of $4 \,\sigma$, more severe than in our previous analysis
in \cite{cuocoetal} because of the different treatment of nuclear
rate errors, and is possibly an evidence for POPII star
atmospheric Lithium depletion.
\begin{table*}
\begin{center}
\begin{tabular}{|c|c|}
\hline rate & $\delta \sigma_{^7{\rm Li}^7{\rm Li}}^2/ \sigma_{^7{\rm Li}^7{\rm Li}}^2 (\%)$ \\
\hline \emph{be7n$\alpha$}& 40.9 \\
\hline \emph{he4he3$\gamma$} & 25.1 \\
\hline \emph{be7dp$\alpha$}& 16.2 \\
\hline \emph{he3dp}& 8.6 \\
\hline \emph{dp$\gamma$}& 4.0 \\
\hline \emph{be7np}& 2.7 \\
\hline \emph{ddn}& 1.8 \\
\hline \emph{others}& 0.7\\
 \hline
\end{tabular}
\end{center}
\caption{The contribution of reaction rate errors to the total
(squared) uncertainty on $X_{^7{\rm Li}}/X_p$} \label{errorli7}
\end{table*}

\end{itemize}

\subsubsection{Results for $\omega_b$ from $^2$H and $^4$He}.

We now study how the value of $\omega_b$ is constrained by
standard BBN alone, without using the WMAP results for this
parameter, but rather determining its best value and uncertainty
from light nuclide abundances only. The aim of this analysis is to
compare the present capability of BBN to fix the baryon content of
the observable Universe with respect to that of CMB anisotropies.
We consider two different analysis. We first use the $^2$H
abundance only, to check the role of deuterium as a good
baryometer. To this aim we construct the likelihood function
 \be {\cal L} \propto \exp \left[-
\left(X_{^2{\rm H}}(\omega_b)-X^{obs}_{^2{\rm H}}\right)^2
W_{^2{\rm H}^2{\rm H}}(\omega_b) /2 \right] \vv \ee where notation
has been described in Section 4.2.1. Using the QSO averaged result
reported in (\ref{qsodeut}) the best fit value and 68\% C.L.
uncertainty are found to be $\omega_b= 0.021 {\pm} 0.002$. Notice
that $W_{^2{\rm H}^2{\rm H}}(\omega_b)$ is dominated by the
experimental uncertainty in QSO measurements. Improved
experimental measurements of $^2$H abundance would correspond to a
more refined determination of $\omega_b$, since the role of
theoretical uncertainty due to nuclear rate errors is presently
quite small.
\begin{figure}[!hbtp]
\begin{center}
\includegraphics[scale=0.5]{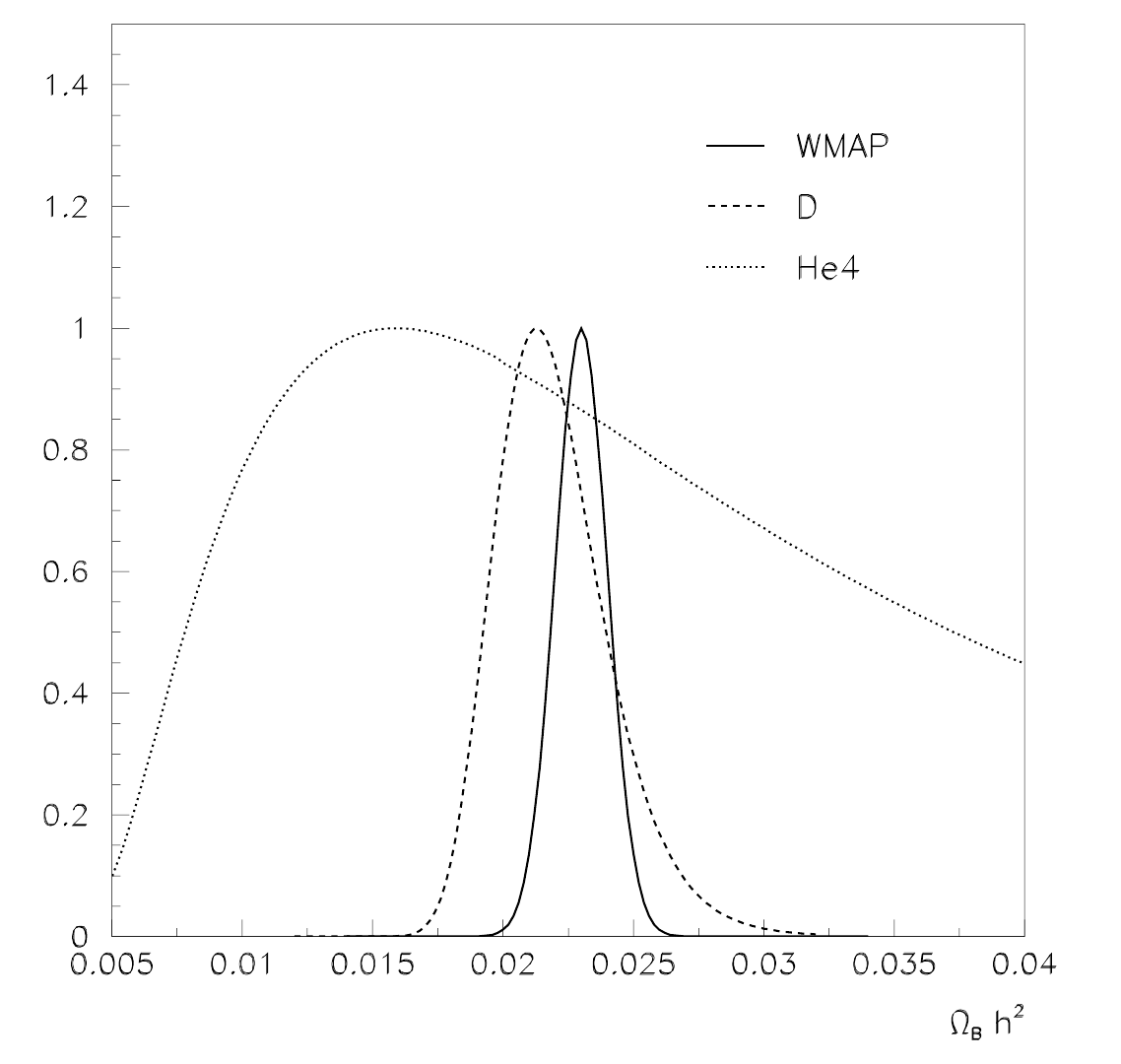}
\end{center} \caption{The $\omega_b$ determination using
different cosmological observables. We report the likelihood
function for WMAP result on CMB anisotropies (full line), $^2$H
measurement (dashed line) and $^4$He analysis (dotted line).}
\label{likelyomegab}
\end{figure}

A similar analysis can be performed using $^4$He. In this case we
get $\omega_b = 0.016^{+0.026}_{-0.009}$, at 68 \% C.L., which is
not very constraining in view of the generous uncertainty we have
adopted for the experimental error on $Y_p$.

A comparison of all results is shown in Figure
(\ref{likelyomegab}), where we report the likelihood profiles for
the two determination of $\omega_b$ from $^2$H and $^4$He, as well
as the one by WMAP. We read from this figure that indeed CMB
anisotropies presently measure $\omega_b$ with the best accuracy,
though $^2$H is also able to fix its values with a comparable
precision. We are confident that the big effort performed by many
groups in refining the theoretical modeling of BBN, along with
new data and a better understanding of systematics may lead to a
more accurate estimate of this important cosmological parameter by
BBN alone. In this case comparison with present and very accurate
determination from CMB experiments such as PLANCK will represent a
{\it precision} test of the cosmological model describing the
evolution of the Universe.

\section{Conclusions}

In this paper we have studied in details the results of standard
primordial nucleosynthesis in view of the fact that baryon density
is now independently measured using the Cosmic Microwave
Background anisotropies. The WMAP Collaboration result for
$\omega_b$, with its 4\% accuracy, represents indeed an important
step toward a detailed understanding of nuclide formation in the
early Universe. Fixing the value of $\omega_b$ in fact implies
that standard BBN scenario becomes a parameter free model, and
comparison of theoretical expectations with data is therefore
particularly significant.

Much effort have been done in the recent years by several groups
in order to increase the level of accuracy of theoretical
prediction on nuclide abundances, in particular by improving the
estimate of the neutron to proton weak conversion rates, and the
nuclear rate network. We think that the present paper represents a
further contribution along this line, and we here summarize the
main results of our study.

We have considered in great details the $n-p$ weak rates, which
are responsible for neutron/proton chemical equilibrium, including
the effect of distortion of neutrino distribution and QED effects.
These rates are now known with a great accuracy, at the level of
$0.1\%$. This result is now quite well established, and compatible
results have been obtained by several groups, as we mentioned in
Section 3. The main benefit of all these studies is that we
presently have a remarkably accurate estimate of $^4$He mass
fraction, which is mainly depending on the value of the neutron
density at freeze out of weak reactions.

We have critically reviewed the whole set of nuclear reaction
rates, and we have reported a detailed analysis of the main rates
contributing to nuclide formation, as well as some sub-leading
channels which, however, in view of their large uncertainty, may
still play a role in BBN. Similar studies have appeared in the
recent literature \cite{Cyburt:2004cq,Descouvemont04} which as the
present paper largely benefit of the NACRE Collaboration database.
Our main goal has been to get a consistent data regression method,
which may be also applied to all cases where data shows a clear
evidence for systematic errors, and which can also account for
correlations among data belonging to the same data set without
introducing the bias discussed in \cite{D'Agostini:1993uj}.

Our study of the main nuclear reaction rates and their
corresponding uncertainties has been used to quantify their
contribution to the total error on the estimate of light nuclides,
in particular $^2$H, $^3$He, $^4$He, $^6$Li and $^7$Li. With our
present regression method typically the role of nuclear rate
uncertainty is now smaller than in previous analysis
\cite{SKM93,Cyburt:2004cq,cuocoetal} and is comparable or smaller
than the corresponding uncertainty due to the present $\omega_b$
error. It is remarkable that indeed for $^2$H this improvement is
particularly relevant. Refined experimental measurements of the
ratio $X_{^2{\rm H}}/X_p$ as well as of the baryon fraction from
CMB anisotropies would be a clean test of consistency of standard
BBN. Primordial $^6$Li, whose theoretical estimate is still
affected by a very large error of the order of $100\%$, is not
presently measured, since it is likely that the observed value in
POPII halo stars is due to cosmic ray spallation nucleosynthesis.

The $^4$He mass fraction uncertainty is estimated at the level of
$0.1 \%$, and is almost completely due to the neutron lifetime
error. The nuclear rate uncertainties of the leading \emph{ddn},
\emph{ddp} and \emph{dn$\gamma$} processes contribute in fact for
only $1.5\%$ of the total theoretical uncertainty for
$\omega_b=0.023$.

The $^7$Li estimate is still affected by an order $10\%$
uncertainty mainly due to the role of the reactions
\emph{be7n$\alpha$}, \emph{he4he3$\gamma$} and
\emph{be7dp$\alpha$}. Further experimental results on these rates
would be therefore highly desirable. In particular a better
measurement of the \emph{he4he3$\gamma$} channel would also have
important implications on the estimate of solar neutrino flux.

Apart from these processes, the BBN uncertainty budget is
dominated by the effect of \emph{ddn}, \emph{ddp} and
\emph{dp$\gamma$}, which largely contribute to both $^2$H and
$^3$He error, \emph{he3dp} which represents the main source of
uncertainty for $^3$He, and finally the \emph{$\alpha$d$\gamma$}
process, which is entirely responsible for the large $^6$Li
theoretical error.

The theoretical estimates for standard BBN with experimental
determinations of $^2$H, $^4$He and $^7$Li show several
interesting features. First of all it is remarkable that the value
found for $^2$H for the WMAP best estimate of $\omega_b$ is in
quite a good agreement with the results of QSO measurements, since
they are compatible at 1 $\sigma$ level. The value found in the
same scenario for $Y_p$ is instead slightly larger than what
expected from present data, though agreement is found at 1
$\sigma$ level if we adopt a conservative estimate of the
uncertainty, as in \cite{Olive:2004kq}, largely dominated by
systematics. The fact that the theoretical uncertainty for $Y_p$
is very small has important consequences in this respect. New
experimental campaigns aimed to detect $Y_p$ and in particular a
further effort to identify sources of systematics will clearly
tell us if the standard BBN scenario is a satisfactory description
or rather we have to consider more exotic possibilities.

Further constraints on this may come from future experimental
measurements of $^3$He, which may be an independent way of fixing
the baryon density. Finally, despite of the fact that $^7$Li
theoretical predictions is still affected by a $10\%$ error,
nevertheless there is a clear indication of a possible depletion
mechanism which lowers the primordial abundance down to the value
measured with Spite plateau. A different picture might emerge if
new experimental results on some key reaction, as for example the
\emph{be7dp$\alpha$} suggested in \cite{Coc:2003ce}, will change
our present understanding of the hierarchy of the nuclear
processes contributing to light nuclide production in the early
Universe.

\ack

We are pleased to thank Richard H. Cyburt for stimulating
discussions and useful comments, and G. Farrar for spotting some typos. 
P.D. Serpico was supported in part by the Deut\-sche For\-schungs\-ge\-mein\-schaft under grant
SFB 375.

\appendix

\section{}
\label{nucnetwork}

{\bf BBN reaction network}\\

We report in this Appendix the full list of reactions adopted in
our numerical study of BBN responsible for the synthesis or the destruction of the
$A\leq 7$ nuclides.\\

Weak Processes.\\

\begin{tabular}{|c|c|}
\hline \boldmath{Name} & \boldmath{Reaction} \\
\hline
\hline  \emph{wnp} & n $\lrt$ p \\
\hline  \emph{h3}  & $^3$H $\rightarrow$ $\bar{\nu}_e$ + $e^{-}$ + $^3$He\\
\hline  \emph{be7} & $^7$Be + $e^{-}$ $\rightarrow$ $\nu_e$ + $^7$Li\\
\hline
\end{tabular}\\
\\
(n,$\gamma$) Reactions\\

\begin{tabular}{|c|c|}
\hline \boldmath{Name} & \boldmath{Reaction}\\
\hline
\hline  \emph{pn$\gamma$}  & p + n $\lrt$ $\gamma$ + $^2$H\\
\hline  \emph{dn$\gamma$}  & $^2$H + n $\lrt$ $ \gamma$ +$^3$H\\
\hline  \emph{he3n$\gamma$}  & $^3$He + n $\lrt$ $ \gamma$ + $^4$He\\
\hline  \emph{li6n$\gamma$}  & $^6$Li + n $\lrt $ $\gamma$ + $^7$Li \\
\hline  \emph{li7n$\gamma$} & $^7$Li + n $\lrt $ $\gamma$ + $^8$Li \\
\hline
\end{tabular}\\
\\
(p,$\gamma$) Reactions\\

\begin{tabular}{|c|c|}
\hline \boldmath{Name} & \boldmath{Reaction}\\
\hline
\hline  \emph{dp$\gamma$} & $^2$H + p $\lrt$ $ \gamma$ + $^3$He\\
\hline  \emph{tp$\gamma$} & $^3$H + p $\lrt$ $ \gamma$ + $^4$He\\
\hline  \emph{li6p$\gamma$}  & $^6$Li + p $\lrt$ $ \gamma$ + $^7$Be \\
\hline  \emph{li7p$\gamma$}  & $^7$Li + p $\lrt$ $ \gamma$ + $^8$Be \\
\hline  \emph{be7p$\gamma$} & $^7$Be + p $\lrt $ $\gamma$ + $^8$B\\
\hline
\end{tabular}\\
\\
(d,$\gamma$) Reactions\\

\begin{tabular}{|c|c|}
\hline \boldmath{Name} & \boldmath{Reaction}\\
\hline
\hline  \emph{dd$\gamma$} & $^2$H + $^2$H $\lrt$ $ \gamma$ + $^4$He \\
\hline  \emph{$\alpha$d$\gamma$}  & $^4$He + $^2$H
$\lrt$ $ \gamma$ + $^6$Li \\
\hline  \emph{li7d$\gamma$} & $^7$Li + $^2$H $\lrt$ $ \gamma$ + $^9$Be\\
\hline
\end{tabular}\\
\\
(t,$\gamma$) and ($^3$He,$\gamma$) Reactions\\

\begin{tabular}{|c|c|}
\hline \boldmath{Name} & \boldmath{Reaction}\\
\hline
\hline  \emph{he3t$\gamma$} & $^3$He + $^3$H $\lrt$ $ \gamma$ + $^6$Li\\
\hline  \emph{$\alpha$t$\gamma$}  & $^4$He + $^3$H
$\lrt$ $ \gamma$ + $^7$Li \\
\hline  \emph{$\alpha$he3$\gamma$}  & $^4$He + $^3$He
$\lrt$ $ \gamma$ + $^7$Be \\
\hline  \emph{li6t$\gamma$} & $^6$Li + $^3$H $\lrt$ $ \gamma$ + $^9$Be \\
\hline  \emph{be7t$\gamma$} & $^7$Be + $^3$H $\lrt$ $ \gamma$ + $^{10}$B \\
\hline  \emph{li7he3$\gamma$} & $^7$Li + $^3$He $\lrt$ $ \gamma$ + $^{10}$B\\
\hline
\end{tabular}\\
\\
($\alpha, \gamma$) Reactions\\

\begin{tabular}{|c|c|}
\hline \boldmath{Name} & \boldmath{Reaction}\\
\hline \hline  \emph{li6$\alpha\gamma$} & $^6$Li + $^4$He $
\lrt$ $ \gamma$
+ $^{10}$B\\
\hline  \emph{li7$\alpha\gamma$} & $^7$Li + $^4$He
$\lrt$ $ \gamma$ + $^{11}$B\\
\hline  \emph{be7$\alpha\gamma$} & $^7$Be + $^4$He
$\lrt$ $ \gamma$ + $^{11}$C\\
\hline
\end{tabular}\\
\\
Charge exchange reactions\\

\begin{tabular}{|c|c|}
\hline \boldmath{Name} & \boldmath{Reaction}\\
\hline
\hline  \emph{he3np} & $^3$He + n $\lrt$  p + $^3$H\\
\hline \emph{be7np}  & $^7$Be + n $\lrt$  p + $^7$Li\\
\hline
\end{tabular}\\
\\
$^2$H Stripping/Pickup (S/P) Reactions\\

\begin{tabular}{|c|c|}
\hline \boldmath{Name} & \boldmath{Reaction}\\
\hline
\hline \emph{ddn} & $^2$H + $^2$H $\lrt$ n + $^3$He\\
\hline \emph{ddp}  & $^2$H + $^2$H $\lrt$ p + $^3$H\\
\hline \emph{tdp}  & $^3$H + $^2$H $\lrt$ n + $^4$He\\
\hline \emph{he3dp} & $^3$He + $^2$H $\lrt$ p + $^4$He\\
\hline \emph{li6dn} & $^6$Li + $^2$H $\lrt$ n + $^7$Be\\
\hline \emph{li6dp}  & $^6$Li + $^2$H $\lrt$ p + $^7$Li\\
\hline  \emph{li8pd} & $^8$Li + p $\lrt$ $^2$H + $^7$Li\\
\hline  \emph{b8nd} & $^8$B + n $\lrt$ $^2$H + $^7$Be \\
\hline
\end{tabular}\\
\\
$^3$H and $^3$He S/P Reactions\\

\begin{tabular}{|c|c|}
\hline \boldmath{Name} & \boldmath{Reaction}\\
\hline
\hline  \emph{li6nt}  & $^6$Li + n $\lrt$ $^3$H + $^4$He\\
\hline  \emph{li6phe3}  & $^6$Li + p $\lrt$ $^3$He + $^4$He\\
\hline
\hline  \emph{he3td}  & $^3$He + $^3$H $\lrt$ $^2$H + $^4$He \\
\hline  \emph{li6td}  & $^6$Li + $^3$H $\lrt$ $^2$H + $^7$Li  \\
\hline  \emph{li6he3d}  & $^6$Li + $^3$He $\lrt$ $^2$H + $^7$Be \\
\hline
\hline  \emph{li6tp}  & $^6$Li + $^3$H $\lrt$ p + $^8$Li \\
\hline  \emph{b8nhe3}  & $^8$B + n $\lrt$ $^3$He + $^6$Li \\
\hline  \emph{li7tn}  & $^7$Li + $^3$H $\lrt$ n + $^9$Be \\
\hline  \emph{li7he3p}  & $^7$Li + $^3$He $\lrt$ p + $^9$Be \\
\hline  \emph{be7tp}  & $^7$Be + $^3$H $\lrt$ p + $^9$Be\\
\hline  \emph{li8dt}  & $^8$Li + $^2$H $\lrt$ $^3$H + $^7$Li\\
\hline  \emph{b8dhe3}  & $^8$B + $^2$H $\lrt$ $^3$He + $^7$Be\\
\hline
\end{tabular}\\
\\
$^4$He Pickup Reactions\\

\begin{tabular}{|c|c|}
\hline \boldmath{Name} & \boldmath{Reaction}\\
\hline
\hline \emph{be7n$\alpha$}  & $^7$Be + n $\lrt$ $^4$He + $^4$He\\
\hline \emph{li7p$\alpha$}  & $^7$Li + p $\lrt$ $^4$He + $^4$He\\
\hline  \emph{be9p$\alpha$}  & $^9$Be + p $\lrt$ $^4$He + $^6$Li\\
\hline  \emph{b10n$\alpha$}  & $^{10}$B + n $\lrt$ $^4$He + $^7$Li\\
\hline  \emph{b10p$\alpha$}  & $^{10}$B + p $\lrt$ $^4$He + $^7$Be\\
\hline
\hline  \emph{li6d$\alpha$}   & $^6$Li + $^2$H $\lrt$ $^4$He + $^4$He\\
\hline  \emph{be9d$\alpha$}  & $^9$Be + $^2$H $\lrt$ $^4$He + $^7$Li \\
\hline
\hline  \emph{be7t$\alpha$}   & $^7$Be + $^3$H $\lrt$ $^4$He + $^6$Li\\
\hline  \emph{li7t$\alpha$} & $^7$Li + $^3$He $\lrt$ $^4$He + $^6$Li \\
\hline  \emph{b8t$\alpha$}  & $^8$B + $^3$H $\lrt$ $^4$He + $^7$Be\\
\hline  \emph{li8he3$\alpha$}  & $^8$Li + $^3$He
$\lrt$ $^4$He + $^7$Li\\
\hline \emph{be9d$\alpha$} & $^9$Be + $^2$H $\lrt$ $^4$He + $^7$Li\\
\hline
\end{tabular}\\
\\
Ingoing 2-body Reactions.\\

\begin{tabular}{|c|c|}
\hline \boldmath{Name} & \boldmath{Reaction}\\
\hline
\hline  \emph{ttnn} & $^3$H + $^3$H $\lrt$ n + n + $^4$He \\
\hline \emph{he3tnp} & $^3$He + $^3$H $\lrt$ p + n + $^4$He\\
\hline \emph{he3he3pp} & $^3$He + $^3$He $\lrt$ p + p + $^4$He \\
\hline \emph{li7dn$\alpha$} & $^7$Li + $^2$H $\lrt$ n + $^4$He + $^4$He\\
\hline  \emph{be7dp$\alpha$} & $^7$Be + $^2$H
$\lrt$ p + $^4$He + $^4$He \\
\hline \emph{li7he3d$\alpha$} & $^7$Li + $^3$He
$\lrt$ $^2$H + $^4$He + $^4$He \\
\hline \emph{be7td$\alpha$} & $^7$Be + $^3$H
$\lrt$ $^2$H + $^4$He + $^4$He \\
\hline \emph{li7tnn$\alpha$} & $^7$Li + $^3$H
$\lrt$ n + n + $^4$He + $^4$He \\
\hline \emph{be7tpn$\alpha$} & $^7$Be + $^3$H
$\lrt$ p + n + $^4$He + $^4$He \\
\hline \emph{li7he3np$\alpha$} & $^7$Li + $^3$He
$\lrt$ n + p + $^4$He + $^4$He \\
\hline \emph{be7he3pp$\alpha$} & $^7$Be + $^3$He
$\lrt$ p + p + $^4$He + $^4$He\\
\hline
\end{tabular}\\
\\
Ingoing 3-body Reactions.\\

\begin{tabular}{|c|c|}
\hline \boldmath{Name} & \boldmath{Reaction}\\
\hline
\hline  \emph{pnnd}  & p + n + n $\lrt$ n + $^2$H \\
\hline  \emph{ppnd}  & p + p + n $\lrt$ p + $^2$H \\
\hline
\hline \emph{pnn$\gamma$}  & p + n + n $\lrt$ $ \gamma$ + $^3$H \\
\hline \emph{ppn$\gamma$}  & p + p + n $\lrt$ $  \gamma$ + $^3$He\\
\hline \emph{pn$\alpha\gamma$}  & p + n + $^4$He $\lrt$ $ \gamma$ + $^6$Li \\
\hline
\end{tabular}\\

\section{}
\label{nucvst}

{\bf Graphical analysis reaction rates}\\

A qualitative understanding of the role of each nuclear reaction
to the creation/destruction of a definite light nuclide $i$ can be
grasped by looking at the temperature behavior of their
contribution to the right hand side of the corresponding Boltzmann
equation for $X_i$. To this end we collect in this Appendix the
log-log plots of the combinations \be \Gamma_{kl \rt ij}\,
\frac{X_l^{N_l}\, X_k^{N_k}}{N_l!\, N_k !} \vv \ee versus
$T_9=T/10^9\,K$, see (\ref{e:dXdt}). Baryon density is chosen as
$\omega_b=0.023$. We show results only for those processes which
contribute to change $X_i$ at leading level or for some
interesting sub-leading reaction.\\

\begin{figure}[!hbtp]
\begin{center}
\includegraphics[scale=0.5]{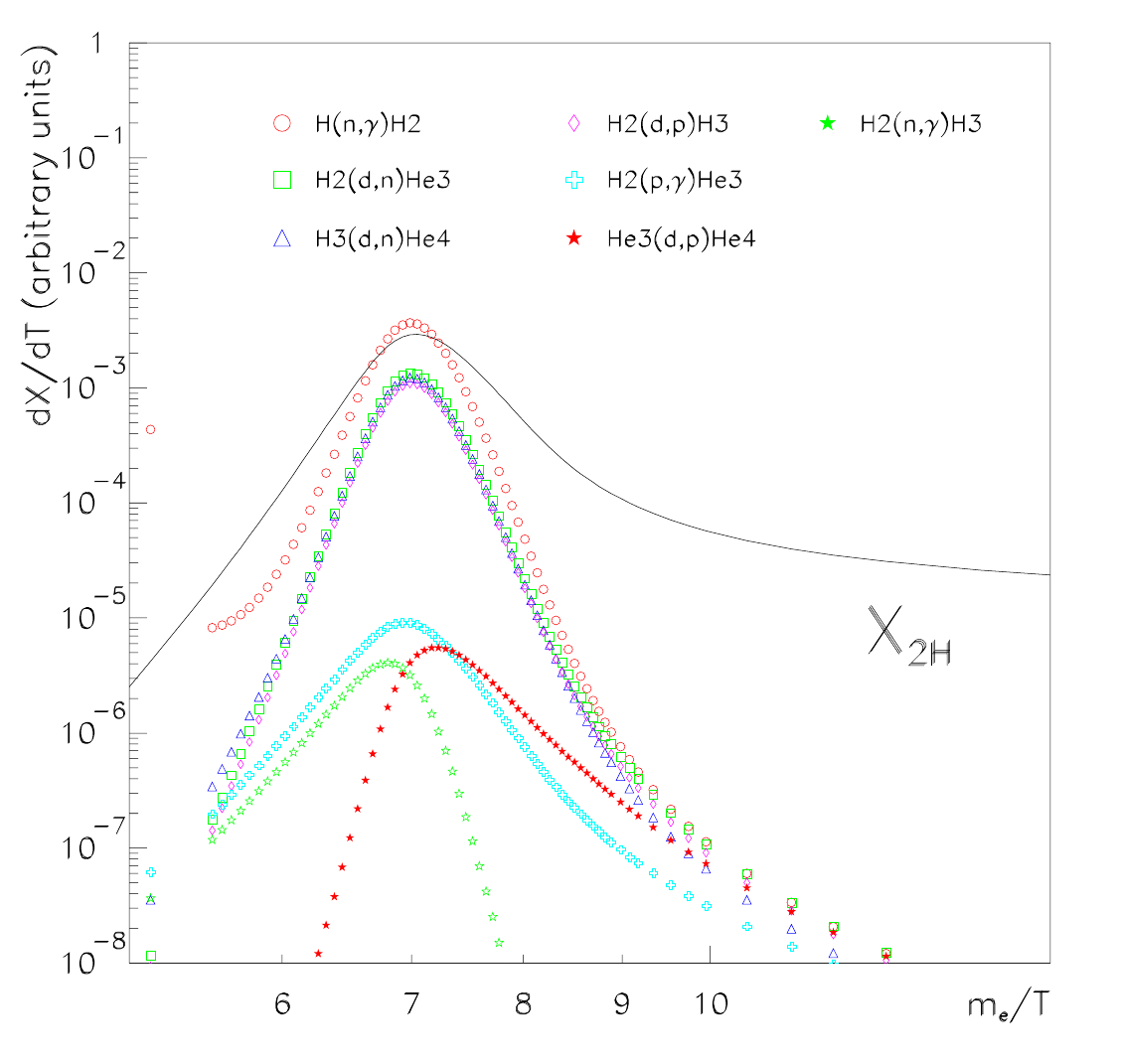}
\end{center}
\caption{Leading processes for production and destruction of
$^2$H.} \label{synth_h2}
\end{figure}

\begin{figure}[!hbtp]
\begin{center}
\includegraphics[scale=0.5]{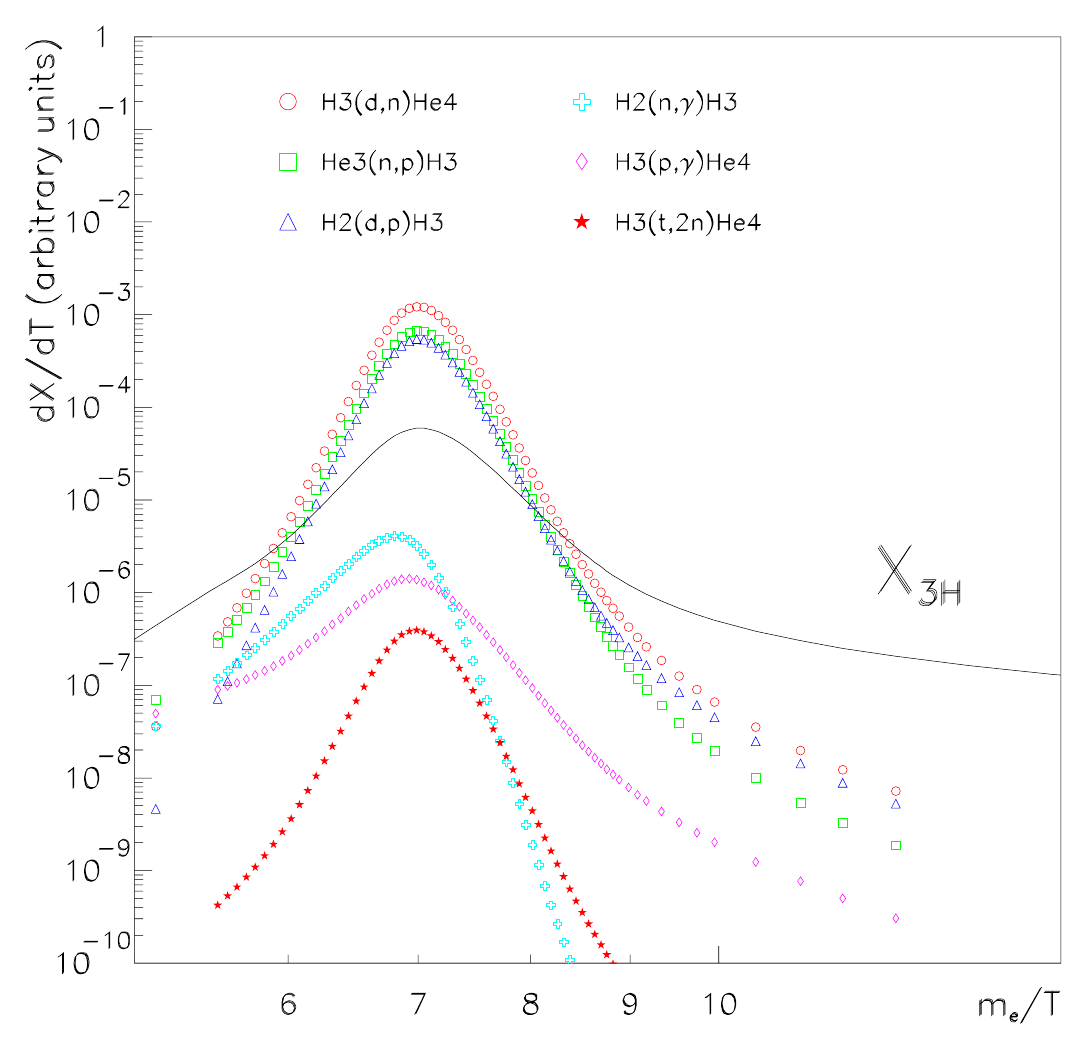}
\end{center}
\caption{Leading processes for production and destruction of
$^3$H.} \label{synth_h3}
\end{figure}

\begin{figure}[!hbtp]
\begin{center}
\includegraphics[scale=0.5]{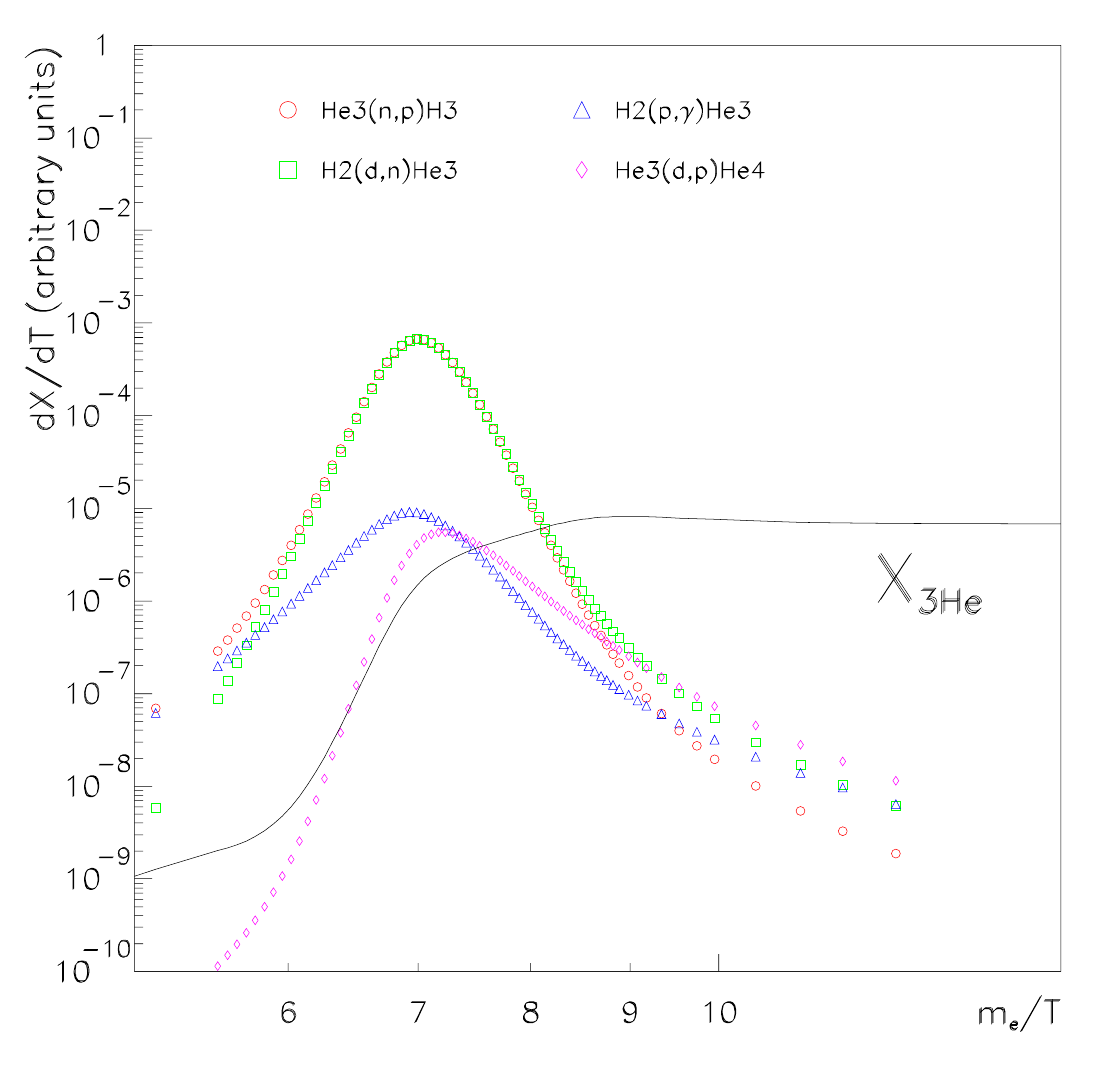}
\end{center}
\caption{Leading processes for production and destruction of
$^3$He.} \label{synth_he3}
\end{figure}

\begin{figure}[!hbtp]
\begin{center}
\includegraphics[scale=0.5]{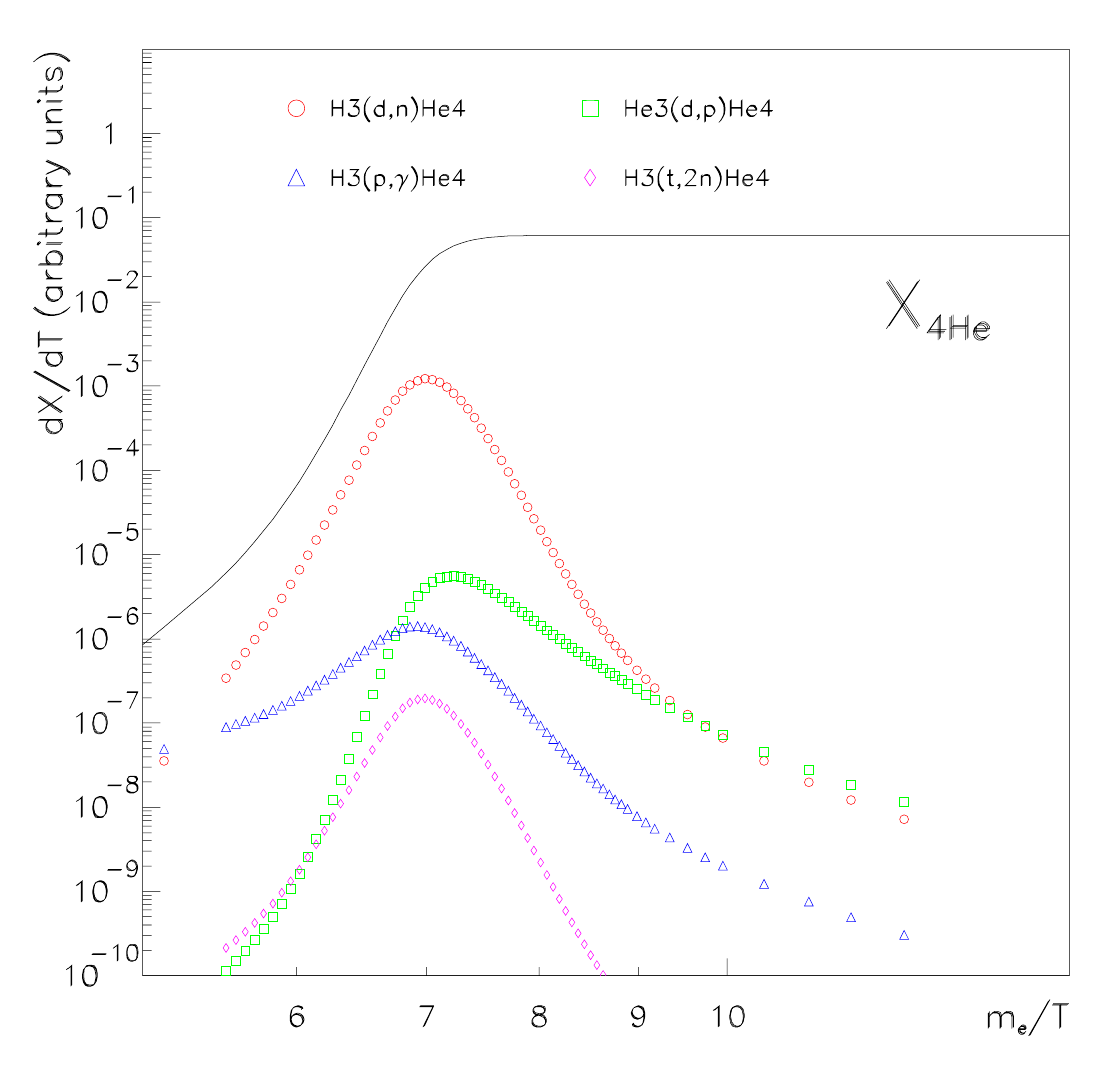}
\end{center}
\caption{Leading processes for production and destruction of
$^4$He.} \label{synth_he4}
\end{figure}

\begin{figure}[!hbtp]
\begin{center}
\includegraphics[scale=0.5]{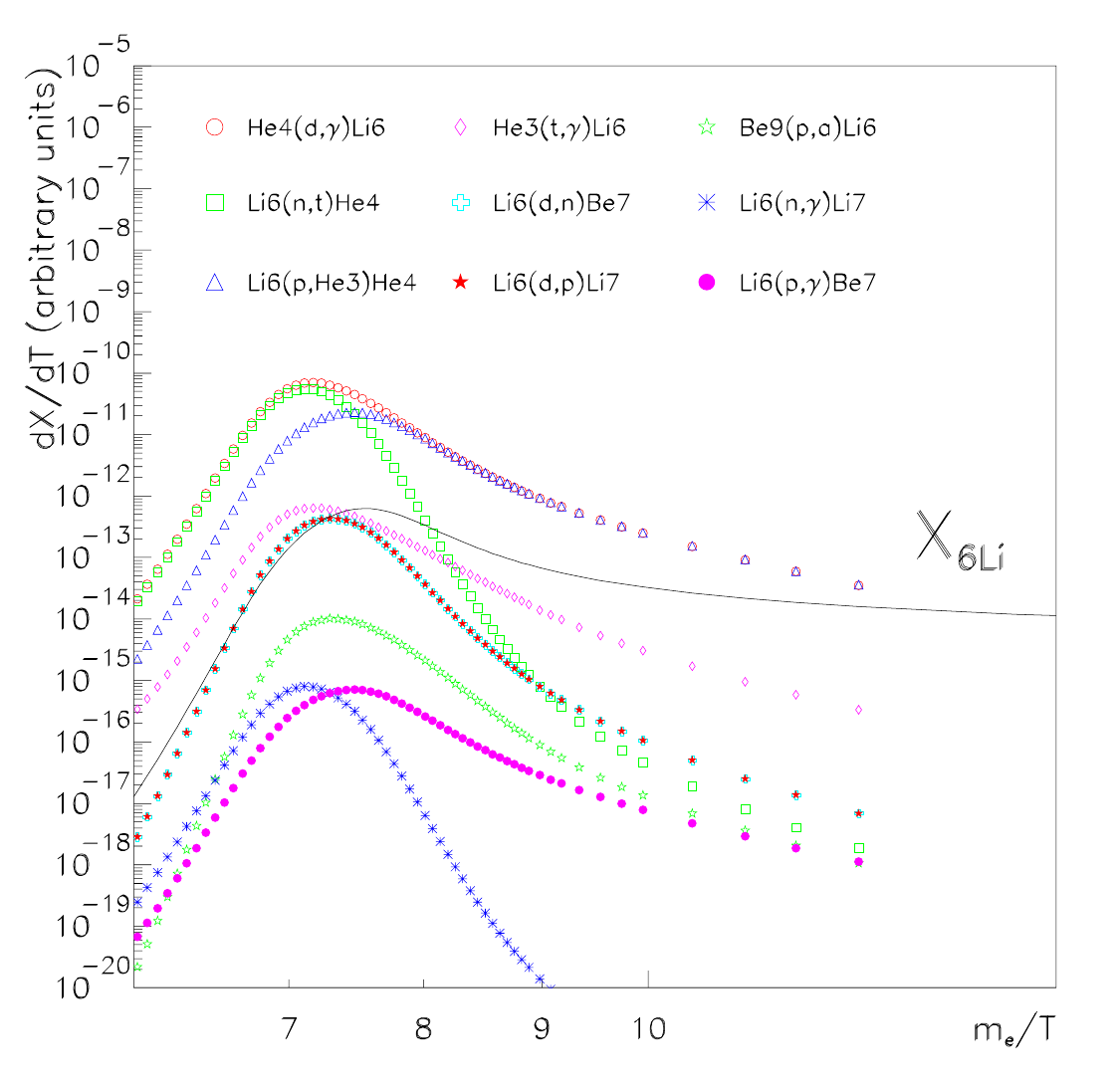}
\end{center}
\caption{Leading processes for production and destruction of
$^6$Li.} \label{synth_li6}
\end{figure}

\begin{figure}[!hbtp]
\begin{center}
\includegraphics[scale=0.5]{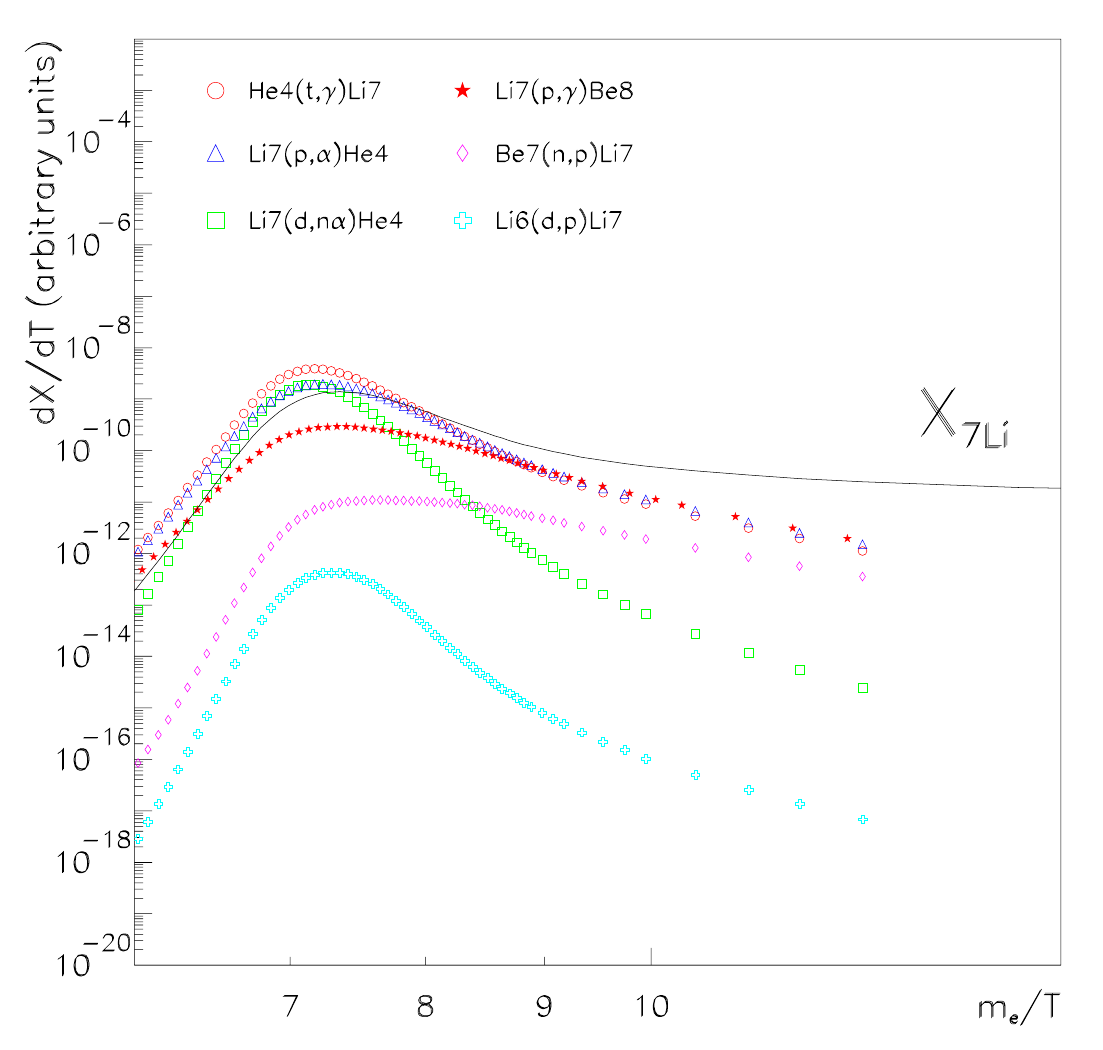}
\end{center}
\caption{Leading processes for production and destruction of
$^7$Li.} \label{synth_li7}
\end{figure}

\begin{figure}[!hbtp]
\begin{center}
\includegraphics[scale=0.5]{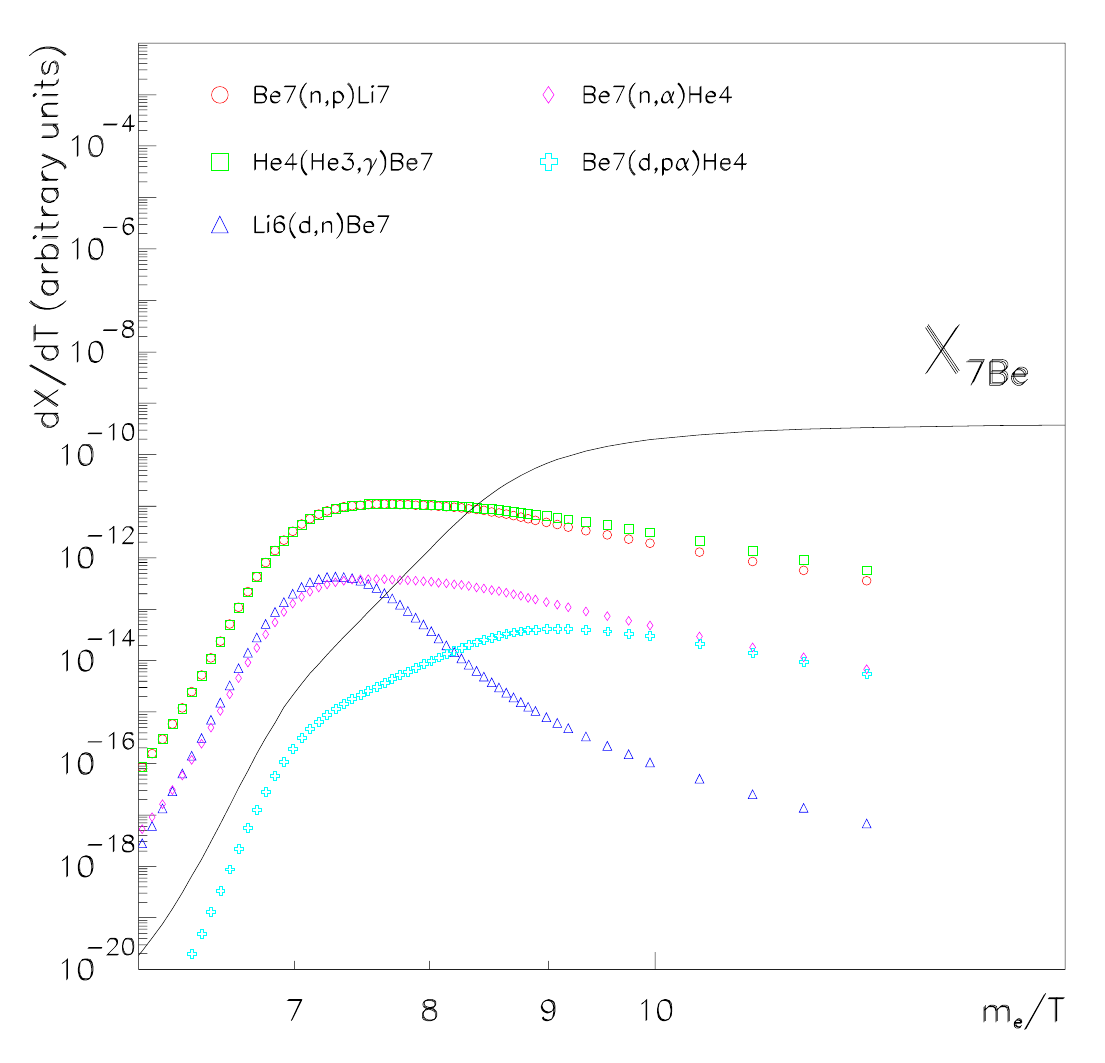}
\end{center}
\caption{Leading processes for production and destruction of
$^7$Be.} \label{synth_be7}
\end{figure}
\newpage

\section{}\label{tabweak}

{\bf Fits of $n \leftrightarrow p$ weak rates}\\

The neutron/proton weak rates have been fitted as function of
$z=m_e/T$ with accuracy better than $0.06\%$ with the following
expressions

\bea \fl \omega(n \rightarrow p) &=& \frac{1}{\tau_n^{ex}} \exp
\left( -q_{np} /z\right) \,\sum_{l=0}^{13} a_l
~z^{-l},\,\,\,\,\,\,\,\,\,\,0.01\leq T/{\rm MeV} \leq 10 \label{e:fitnp} \\
\fl \omega(p \rightarrow n) &=& \left\{ \begin{array} {cc}
\frac{1}{\tau_n^{ex}} \exp \left( -q_{pn} z\right) \,
\sum_{l=1}^{10} b_l ~z^{-l} & 0.1 \leq T/{\rm MeV}  \leq 10 \\
\fl 0 & 0.01\leq T/{\rm MeV}  < 0.1  \end{array} \right.
\label{e:fitpn} \eea with \bea \fl \begin{array}{lll}
a_0= 1 & a_1= 0.15735 & a_2= 4.6172 \\
a_3= -0.40520 {\cdot} 10^2 & a_4= 0.13875 {\cdot} 10^3 & a_5=
-0.59898 {\cdot} 10^2 \\
a_6= 0.66752 {\cdot} 10^2 & a_7= -0.16705 {\cdot} 10^2 & a_8=
3.8071 \\
a_9= -0.39140 & a_{10}= 0.023590 & a_{11}= -0.83696 {\cdot}
10^{-4} \\
a_{12}= -0.42095 {\cdot} 10^{-4} & a_{13}= 0.17675 {\cdot} 10^{-5}
&
q_{np}= 0.33979~~~, \\ \\
b_0= -0.62173 & b_1= 0.22211 {\cdot} 10^2 & b_2=
-0.72798 {\cdot} 10^2 \\
b_3= 0.11571 {\cdot} 10^{3} & b_4= -0.11763 {\cdot} 10^2 & b_5=
0.45521 {\cdot} 10^2
\\  b_6=
-3.7973 & b_7= 0.41266 & b_8= -0.026210 \\
b_9= 0.87934 {\cdot} 10^{-3} & b_{10}= -0.12016 {\cdot} 10^{-4} &
 q_{pn}= 2.8602 ~~~.
\end{array}
\label{e:coeffnp} \\
\label{e:coeffpn} \eea

\section{}
\label{S_Rfactors}
{\bf Fits of the S/R factors for the leading reactions}\\

In this Appendix we report our fits of the S or R factor for each
leading reaction and the largest energy $E_{Max}$ for which the fit is valid.

\begin{enumerate}

\item {\bf Reaction \emph{pn$\gamma$} }: $p+n\lrt
\gamma+{^2H}$
 \begin{eqnarray} \fl R(E)&=& ( 7.31638 {\cdot} 10^{-20} + 2.35455
 {\cdot} 10^{-20}\, E^\frac{1}{2} - 1.55683 {\cdot} 10^{-18}\, E  \nonumber\\
\fl &+& 5.93351 {\cdot} 10^{-18}\, E^\frac{3}{2} - 9.25443 {\cdot}
10^{-18}\, E^2 + 6.6732 {\cdot} 10^{-18}\, E^\frac{5}{2} \nonumber\\
\fl &-& 1.82393 {\cdot} 10^{-18}\, E^3 )~ cm^3 s^{-1}\\
\fl [E_{Max} &=&  1\, MeV] \nonumber
\end{eqnarray}

\bigskip

\item {\bf Reaction  \emph{dp$\gamma$} }: ${^2H}+p\lrt
\gamma+{^3He}$
\begin{eqnarray} \fl S(E)&=& ( 0.214 {\cdot} 10^{-6} + 0.556 {\cdot} 10^{-5}\, E
  + 0.551 {\cdot} 10^{-5}\, E^2 \nonumber\\
\fl &-& 0.157 {\cdot} 10^{-5}\, E^3 )~ MeV~ b\\
\fl [E_{Max} &=&  2\, MeV] \nonumber
\end{eqnarray}

\bigskip

\item {\bf Reaction  \emph{ddn} }: ${^2H} + {^2H}\lrt  n
+ {^3He}$
\begin{eqnarray} \fl
S(E)&=&  ( 0.0522 + 0.370\, E - 0.196\, E^2 + 0.0672\, E^3 -
 0.00885\, E^4 )~ MeV~ b\\
\fl [E_{Max} &=&  3.1\, MeV] \nonumber
\end{eqnarray}

\bigskip

\item {\bf Reaction \emph{ddp} }: ${^2H} + {^2H}\lrt
p+ {^3H}$
\begin{eqnarray} \fl
S(E)&=& ( 0.0542 + 0.205\, E - 0.0240\, E^2 )~ MeV~ b\\
\fl [E_{Max} &=&  3.1\, MeV] \nonumber
\end{eqnarray}

\bigskip

\item {\bf Reaction  \emph{tdn} }:  ${^3H} + {^2H}\lrt
n + {^4He}$
\begin{eqnarray} \fl
S(E)&=& \frac{26 - 0.361\, E + 248\, E^2}{1 +\left( \frac{E -
0.0479}{0.0392} \right)^2}~~ MeV~ b\\
\fl [E_{Max} &=&  2.4\, MeV] \nonumber
\end{eqnarray}

\bigskip

\item {\bf Reaction  \emph{he3dp} }: ${^3He} + {^2H}\lrt
p+ {^4He}$
\begin{eqnarray} \fl
S(E)&=& \frac{19.5 - 22.7\, E + 61.8\, E^2 - 19.5\, E^3 + 4.05\,
E^4}{1 + \left( \frac{E - 0.201}{.132} \right)^2}~~ MeV~ b\\
\fl [E_{Max} &=&  3\, MeV] \nonumber
\end{eqnarray}

\bigskip

\item {\bf Reaction   \emph{he3np} }: ${^3He} + n\lrt  p
+ {^3H}$
\begin{eqnarray} \fl
R(E)\,N_A&=& ( 0.706 {\cdot} 10^9 - 0.149 {\cdot} 10^{10}\, E^\frac{1}{2} +
0.521 {\cdot} 10^{10}\, E - 0.239 {\cdot} 10^{11}\, E^\frac{3}{2}
\nonumber\\
\fl &+& 0.617 {\cdot} 10^{11}\, E^2 - 0.449 {\cdot} 10^{11}\,
E^\frac{5}{2} - 0.540 {\cdot} 10^{11}\, E^3 + 0.951 {\cdot} 10^{11}\,
E^\frac{7}{2} \nonumber\\
\fl &-& 0.375 {\cdot} 10^{11}\, E^4 )~ cm^3 s^{-1} mol^{-1}\\
\fl [E_{Max} &=& 1\, MeV] \nonumber
\end{eqnarray}

\bigskip

\item {\bf Reaction  \emph{be7np} }: ${^7Be} + n\lrt  p+
{^7Li}$
\begin{eqnarray} \fl
R(E)\,N_A&=& \biggl( 0.470 {\cdot} 10^{10} - 0.202 {\cdot} 10^{11}\,
E^\frac{1}{2} + 0.349 {\cdot} 10^{11}\, E - 0.253 {\cdot} 10^{11}\,
E^\frac{3}{2} \nonumber\\
\fl &+& \left. 0.660 {\cdot} 10^{10}\, E^2 + \frac{0.109 {\cdot}
10^{10}}{1 + \left( \frac{E-.317}{0.114} \right)^2} \right)~ cm^3
s^{-1} mol^{-1}\\
\fl [E_{Max} &=& 2\, MeV] \nonumber
\end{eqnarray}

\bigskip

\item {\bf Reaction  \emph{$\alpha$he3$\gamma$} }: ${^4He}+{^3He}
\lrt \gamma + {^7Be}$
\begin{eqnarray} \fl
S(E)&=& 0.107 {\cdot} 10^{-2} + e^{-0.552 E}\, (-0.582 {\cdot} 10^{-3}
- 0.606 {\cdot} 10^{-3}\, E \nonumber\\
\fl &-& 0.154 {\cdot} 10^{-3}\, E^2)~ MeV~ b\\
\fl [E_{Max} &=& 2.5\, MeV] \nonumber
\end{eqnarray}

\bigskip

\item {\bf Reaction  \emph{li7p$\alpha$} }: ${^7Li} + p
\lrt {^4He} + {^4He}$
\begin{eqnarray} \fl
S(E) &=& ( 0.0609 + 0.173\, E - 0.319\, E^2 + 0.217\, E^3 )~ MeV~
b\\
\fl [E_{Max} &=& 1.1\, MeV] \nonumber
\end{eqnarray}

\bigskip

\item {\bf Reaction   \emph{li6phe3} }: ${^6Li}+p\lrt
{^3He} + {^4He}$
\begin{eqnarray} \fl
S(E) &=& ( 3.158 - 1.534\, E )~ MeV~ b\\
\fl [E_{Max} &=& 2.5\, MeV] \nonumber
\end{eqnarray}

\end{enumerate}

\section{}\label{tabrat}

{\bf Fits of reaction rates}\\

In this Appendix we collect our fits for leading and sub-leading reaction
rates and corresponding uncertainties discussed in the paper and adopted in
the numerical solution of BBN equation set.

We recall that reaction rates are usually found in nuclear data catalogues
in the form $f=\langle \sigma v \rangle N_A^{N_i-1}$ where $N_i$ is the
number of ingoing nuclides (note that for $N_i\neq 2$, $\langle \sigma v
\rangle$ has only a formal meaning). This means that $f$ has dimension
$(cm^3 mol^{-1})^{N_i-1}s^{-1}$. The rates $\Gamma_i$ (in $s^{-1}$) are
then obtained by multiplying the $\langle \sigma v \rangle$'s by all
ingoing particle number densities. The reverse process rates can be
obtained from the forward ones through the following equations, which can
be obtained from the detailed balance principle
\begin{enumerate}
\item $a+b \lrt c+d$\\
\be \frac{\langle \sigma v \rangle_{cd}}{\langle \sigma v
\rangle_{ab}} = {\bigg(\frac{\mu_{ab}}{\mu_{cd}}\bigg)}^{3/2}
\frac{g_ag_b}{g_cg_d} e^{-Q/T}\label{reltrarateab1} \ee where $Q$
is the Q-value of the forward reaction.
\item $a+b \lrt c+\gamma $\\
\be \frac{n_\gamma\langle \sigma v \rangle_{c\gamma}}{\langle
\sigma v \rangle_{ab}} \simeq{\bigg(\frac{M_u
T}{2\pi}\bigg)}^{3/2}
{\bigg(\frac{\mu_{ab}}{M_u}\bigg)}^{3/2}\frac{g_ag_b}{g_c}e^{-Q/T}\label{reltraratecg2}
\ee where $[n_\gamma\langle \sigma v \rangle_{c\gamma}]^{-1}$ is
the lifetime of the nucleus $c$ for the photodisintegration process, and
$M_u$ is the nuclear mass unit. \item $a+b\lrt c+d+e$ \be \frac{\langle
\sigma v
\rangle_{cde}}{\langle \sigma v
\rangle_{ab}}\simeq{\bigg(\frac{2\pi}{M_u T}\bigg)}^{3/2}
{\bigg(\frac{M_aM_bM_u}{M_cM_dM_e}\bigg)}^{3/2}\frac{g_ag_b}{g_cg_dg_e}e^{-Q/T}
\ee
\end{enumerate}
In these expressions $g_a$ denotes the statistical factor $2 J_a+1$, $J_a$
being the angular momentum of nuclide $a$. Generalization to other cases is
straightfoward. These expressions also apply to all reactions involving
identical particles, \be N_a(a)+N_b(b)+\ldots \lrt N_c(c)+N_d(d)+\ldots \vv
\ee apart from the replacement \be g_a
\longrightarrow\frac{{g_a}^{N_a}}{N_a!} \pp \ee

\begin{enumerate}

\item {\bf Reaction \emph{pn$\gamma$} }: $p+n\lrt
\gamma+H2$
 \begin{eqnarray} \fl f_{pn \gamma}&=& 44060.\,\left( 1. + 0.106597\,
     {\sqrt{T_9}} -
    2.75037\,T_9 +
    4.62949\,{T_9}^{\frac{3}{2}}
    \right.\nonumber\\
\fl &-& \left. 3.52204\,{T_9}^2
    +1.34596\,{T_9}^{\frac{5}{2}}
    -0.209351\,{T_9}^3 \right)
\end{eqnarray}
\begin{eqnarray} \fl
\delta f_{pn \gamma}&=& 67.9521\,\left[1.25536 -
      25.8289\,{\sqrt{T_9}} +
      618.897\,T_9 -
      5043.28\,{T_9}^{\frac{3}{2}} \right. \nonumber\\ \fl
      &+& \left.
      21793.6\,{T_9}^2 -
      57307.2\,{T_9}^{\frac{5}{2}} +
      97573.4\,{T_9}^3 -
      111072.\,{T_9}^{\frac{7}{2}} \right. \nonumber\\ \fl
      &+& \left.
      85539.6\,{T_9}^4 -
      44154.6\,{T_9}^{\frac{9}{2}} +
      14687.1\,{T_9}^5 -
      2855.93\,{T_9}^{\frac{11}{2}} \right. \nonumber\\ \fl
      &+& \left.
      247.578\,{T_9}^6 \right]^{1/2}
\end{eqnarray}

\bigskip

\item {\bf Reaction  \emph{dp$\gamma$} }: $H2+p\lrt
\gamma+He3$
\begin{eqnarray} \fl
f_{dp\gamma}&=& {T_9}^{-\frac{2}{3}}
\exp\left\{\frac{1.29043}{{T_9}^{\frac{1}{3}}}\right\}
\left[-15.7097 + 126.821\,
     {T_9}^{\frac{1}{3}} -
    206.509\,{T_9}^{\frac{2}{3}} \right. \nonumber\\ \fl
      &-& \left.
    721.914\,T_9 +
    2120.73\,{T_9}^{\frac{4}{3}} -
    369.613\,{T_9}^{\frac{5}{3}} +
    173.239\,{T_9}^2 \right. \nonumber\\ \fl
      &+& \left.
    127.838\,{T_9}^{\frac{7}{3}} +
    100.688\,{T_9}^{\frac{8}{3}} -
    77.3717\,{T_9}^3\right]
\end{eqnarray}
\begin{eqnarray}\fl
\delta f_{dp\gamma}&=& \frac{1}{2} \, {T_9}^{-\frac{2}{3}}
\left[\exp\left\{-\frac{0.822587}
       {{T_9}^{\frac{1}{3}}}\right\} \left(-20.6078 + 134.277\,
     {T_9}^{\frac{1}{3}} \right. \right. \nonumber\\ \fl
     &-& \left. \left.
    148.863\,{T_9}^{\frac{2}{3}} -
    651.425\,T_9 +
    1513.56\,{T_9}^{\frac{4}{3}} -
    668.149\,{T_9}^{\frac{5}{3}} \right. \right. \nonumber\\ \fl
     &+& \left. \left.
    690.01\,{T_9}^2 -
    11.832\,{T_9}^{\frac{7}{3}} -
    1.7648\,{T_9}^{\frac{8}{3}} -
    28.9337\,{T_9}^3\right) \right. \nonumber\\ \fl
    &-& \left.
\exp\left\{-\frac{2.44793}{{T_9}^{\frac{1}{3}}}\right\}
\left(-10.783 + 727.882\,
     {T_9}^{\frac{1}{3}} -
    7736.03\,{T_9}^{\frac{2}{3}} \right. \right. \nonumber\\ \fl
     &+& \left. \left.
    32828.3\,T_9 -
    64848.5\,{T_9}^{\frac{4}{3}} +
    84984.\,{T_9}^{\frac{5}{3}} -
    65943.\,{T_9}^2 \right. \right. \nonumber\\ \fl
     &+& \left. \left.
    30784.9\,{T_9}^{\frac{7}{3}} -
    7555.12\,{T_9}^{\frac{8}{3}} +
    679.86\,{T_9}^3 \right)
    \right]
\end{eqnarray}

\bigskip

\item {\bf Reaction  \emph{ddn} }: ${^2H} + {^2H}\lrt  n
+ {^3He}$
\begin{eqnarray} \fl
f_{ddn}&=&  {{T_9}^{-\frac{2}{3}}}
\exp\left\{-{{T_9}^{-\frac{1}{3}}}\right\}
\left[-1.84664{\cdot}{10}^6 +
    1.22986{\cdot}{10}^7\,
     {T_9}^{\frac{1}{3}} \right. \nonumber\\ \fl
     &-& \left.
    1.3761{\cdot}{10}^7\,{T_9}^{\frac{2}{3}} -
    6.11628{\cdot}{10}^7\,T_9 +
    1.3329{\cdot}{10}^8\,{T_9}^{\frac{4}{3}} \right. \nonumber\\ \fl
     &-& \left.
    1.24333{\cdot}{10}^7\,
     {T_9}^{\frac{5}{3}} -
    2.72404{\cdot}{10}^7\,{T_9}^2 +
    8.52947{\cdot}{10}^6\,
     {T_9}^{\frac{7}{3}} \right. \nonumber\\ \fl
     &+& \left.
    2.2519{\cdot}{10}^6\,{T_9}^{\frac{8}{3}} -
    2.31204{\cdot}{10}^6\,{T_9}^3 -
    294342.\,{T_9}^{\frac{10}{3}} \right. \nonumber\\ \fl
     &+& \left.
    911550.\,{T_9}^{\frac{11}{3}} -
    252211.\,{T_9}^4\right]
\end{eqnarray}
\begin{eqnarray} \fl
\delta f_{ddn}&=& {{T_9}^{-\frac{2}{3}}}
\exp\left\{-{T_9}^{-\frac{1}{3}}\right\} \left[56996.9 - 433312.\,
     {T_9}^{\frac{1}{3}} +
    952341.\,{T_9}^{\frac{2}{3}} \right. \nonumber\\ \fl
     &+& \left.
    451314.\,T_9 -
    4.57126\,{10}^6\,
     {T_9}^{\frac{4}{3}} +
    5.85118\,{10}^6\,
     {T_9}^{\frac{5}{3}} \right. \nonumber\\ \fl
     &-& \left.
    1.30553\,{10}^6\,{T_9}^2 -
    1.5064\,{10}^6\,{T_9}^{\frac{7}{3}} +
    320431.\,{T_9}^{\frac{8}{3}} \right. \nonumber\\ \fl
     &+& \left.
    426550.\,{T_9}^3 -
    16565.1\,{T_9}^{\frac{10}{3}} -
    131474.\,{T_9}^{\frac{11}{3}} \right. \nonumber\\ \fl
     &+& \left.
    35632.\,{T_9}^4 \right]
\end{eqnarray}

\bigskip

\item {\bf Reaction \emph{ddp} }: ${^2H} + {^2H}\lrt
p+ {^3H}$
\begin{eqnarray} \fl
f_{ddp}&=& {{T_9}^{-\frac{2}{3}}} \exp\left\{-1.06765 \,
{T_9}^{-\frac{1}{3}} \right\} \left[-5.85032{\cdot}{10}^6 +
    5.23171\,{10}^7\,
     {T_9}^{\frac{1}{3}} \right. \nonumber\\ \fl
     &-& \left.
    1.70199{\cdot}{10}^8\,
     {T_9}^{\frac{2}{3}} +
    2.32242{\cdot}{10}^8\,T_9 -
    1.18812{\cdot}{10}^8\,
     {T_9}^{\frac{4}{3}} \right. \nonumber\\ \fl
     &+& \left.
    5.28874{\cdot}{10}^7\,
     {T_9}^{\frac{5}{3}} -
    9.85542{\cdot}{10}^6\,{T_9}^2 \right]
\end{eqnarray}
\begin{eqnarray} \fl
\delta f_{ddp}&=& \frac{{T_9}^{-\frac{2}{3}}}{2}\left[
 \exp\left\{-1.04657 \, {T_9}^{-\frac{1}{3}} \right\}
\left(-5.74135{\cdot}{10}^6 +
    5.10713{\cdot}{10}^7\,
     {T_9}^{\frac{1}{3}} \right. \right. \nonumber\\ \fl
     &-& \left. \left.
    1.64992{\cdot}{10}^8\,
     {T_9}^{\frac{2}{3}} +
    2.22795{\cdot}{10}^8\,T_9 -
    1.11871{\cdot}{10}^8\,
     {T_9}^{\frac{4}{3}} \right. \right. \nonumber\\ \fl
     &+& \left. \left.
    5.02777{\cdot}{10}^7\,
     {T_9}^{\frac{5}{3}} -
    9.45483{\cdot}{10}^6\,{T_9}^2\right) \right.\nonumber\\ \fl
     &-& \left.
\exp\left\{-1.06576 \, {T_9}^{-\frac{1}{3}} \right\}
\left(-5.8649\,{10}^6 +
    5.24033\,{10}^7\,
     {T_9}^{\frac{1}{3}} \right. \right. \nonumber\\ \fl
     &-& \left. \left.
    1.70365\,{10}^8\,
     {T_9}^{\frac{2}{3}} +
    2.32446\,{10}^8\,T_9 -
    1.19167\,{10}^8\,
     {T_9}^{\frac{4}{3}} \right. \right. \nonumber\\ \fl
     &+& \left. \left.
    5.30638\,{10}^7\,
     {T_9}^{\frac{5}{3}} -
    9.88923\,{10}^6\,{T_9}^2\right)
\right]
\end{eqnarray}

\bigskip

\item {\bf Reaction  \emph{tdn} }:  ${^3H} + {^2H}\lrt
n + {^4He}$
\begin{eqnarray} \fl
f_{tdn}&=& {{T_9}^{-\frac{2}{3}}} \exp\left\{-1.34274 \,
{T_9}^{-\frac{1}{3}} - 0.233098\,{T_9}^2\right\} \,
    \left( -8.11449{\cdot}{10}^7 \right. \nonumber\\  \fl &+& \left.
      2.23153{\cdot}{10}^9\,T_9 -
      2.94397{\cdot}{10}^9\,{T_9}^2 +
      1.87645{\cdot}{10}^9\,{T_9}^3 \right. \nonumber\\  \fl &-& \left.
      6.05116{\cdot}{10}^8\,{T_9}^4 +
      9.51966{\cdot}{10}^7\,{T_9}^5 -
      5.29011{\cdot}{10}^6\,{T_9}^6 \right)\nonumber\\ \fl
      &+& 6.22657{\cdot}{10}^8 \, {T_9}^{-0.567854} \,
  \exp\left\{-\frac{0.497116}{T_9}\right\}
\end{eqnarray}
\begin{eqnarray} \fl
\delta f_{tdn}&=&\frac 12 \left[ {{T_9}^{-\frac{2}{3}}}
\exp\left\{-1.37848 \, {T_9}^{-\frac{1}{3}} -
       0.237971\,{T_9}^2\right\}
    \left( -8.70182{\cdot}{10}^7 \right. \right. \nonumber\\  \fl &+&
    \left. \left.
      2.41143{\cdot}{10}^9\,T_9 -
      3.22272{\cdot}{10}^9\,{T_9}^2 +
      2.07799{\cdot}{10}^9\,{T_9}^3 \right. \right. \nonumber\\  \fl &-&
    \left. \left.
      6.77396{\cdot}{10}^8\,{T_9}^4 +
      1.07624{\cdot}{10}^8\,{T_9}^5 -
      6.03483{\cdot}{10}^6\,{T_9}^6 \right)\right. \nonumber\\  \fl &+&
\left. 6.20059{\cdot}{10}^8
  \exp\left\{-\frac{0.49496}{T_9}\right\}\,
    {T_9}^{-0.560781} \right. \nonumber\\ \fl
&-& \left. {{T_9}^{-\frac{2}{3}}} \exp\left\{-1.05088
{T_9}^{-\frac{1}{3}} -0.332736\,{T_9}^2\right\}
 \left( -4.09641{\cdot}{10}^7 \right. \right.
\nonumber\\  \fl &+&
    \left. \left.
      1.0649{\cdot}{10}^9\,T_9 -
      7.15272{\cdot}{10}^8\,{T_9}^2 -
      1.41552{\cdot}{10}^8\,{T_9}^3 \right. \right. \nonumber\\  \fl &+&
    \left. \left.
      3.92762{\cdot}{10}^8\,{T_9}^4 -
      1.58174{\cdot}{10}^8\,{T_9}^5 +
      2.12803{\cdot}{10}^7\,{T_9}^6
      \right) \right. \nonumber\\ \fl &-& \left.
      6.37982{\cdot}{10}^8 \,
  \exp\left\{-\frac{0.495982}{T_9}\right\}\,{T_9}^{-0.584609}
    \right]
\end{eqnarray}

\bigskip

\item {\bf Reaction  \emph{he3dp} }: ${^3He} + {^2H}\lrt
p+ {^4He}$
\begin{eqnarray} \fl
f_{he3dp}&=& {T_9}^{-\frac{2}{3}} \exp\left\{-1.45406 \,
       T_9^{-\frac{1}{3}} -
      0.00623408\,{T_9}^2\right\}
      \left[-3.13359{\cdot}{10}^7
      \right. \nonumber\\ \fl
      &+&
      \left.
 1.14185{\cdot}{10}^8\,{T_9}^{\frac{1}{3}} +
 1.75002{\cdot}{10}^8\,{T_9}^{\frac{2}{3}} -
 6.20511{\cdot}{10}^8\,{T_9}       \right. \nonumber\\ \fl
      &-&
      \left.
 1.75134{\cdot}{10}^9\,{T_9}^{\frac{4}{3}} +
 5.27922{\cdot}{10}^9\,{T_9}^{\frac{5}{3}} -
 1.87822{\cdot}{10}^9\,{T_9}^2       \right. \nonumber\\ \fl
      &-&
      \left.
 3.32382{\cdot}{10}^9\,{T_9}^{\frac{7}{3}} +
 2.03463{\cdot}{10}^9\,{T_9}^{\frac{8}{3}} +
 6.56428{\cdot}{10}^8\,{T_9}^3       \right. \nonumber\\ \fl
      &-&
      \left.
 4.95421{\cdot}{10}^8\,{T_9}^{\frac{10}{3}} -
 1.77029{\cdot}{10}^8\,{T_9}^{\frac{11}{3}} +
 1.53089{\cdot}{10}^8\,{T_9}^4       \right. \nonumber\\ \fl
      &-&
      \left.
  2.51653{\cdot}{10}^7\,{T_9}^{\frac{13}{3}} \right]
 \nonumber\\ \fl
      &+&
3.10384{\cdot}{10}^8\,
  \exp\left\{-\frac{1.6191}{T_9}\right\}\,
    {T_9}^{-0.121595}
\end{eqnarray}
\begin{eqnarray} \fl
\delta f_{he3dp}&=& \frac 12 \left[ {{T_9}^{-\frac{2}{3}}}
\exp\left\{-1.76473 \,
       {{T_9}^{-\frac{1}{3}}} -
      0.0115845\,{T_9}^2\right\}
\left(-4.05392{\cdot}{10}^7 \right. \right. \nonumber\\ \fl &+&
\left. \left.  1.39903{\cdot}{10}^8\,
   {T_9}^{\frac{1}{3}} +
  2.41212{\cdot}{10}^8\,{T_9}^{\frac{2}{3}} -
  6.80688{\cdot}{10}^8\,T_9 \right. \right. \nonumber\\ \fl
&-& \left. \left.
  2.39601{\cdot}{10}^9\,{T_9}^{\frac{4}{3}} +
  5.33313{\cdot}{10}^9\,{T_9}^{\frac{5}{3}} +
  1.67555{\cdot}{10}^9\,{T_9}^2 \right. \right. \nonumber\\ \fl
&-& \left. \left.
  7.79969{\cdot}{10}^9\,{T_9}^{\frac{7}{3}} +
  3.34874{\cdot}{10}^9\,{T_9}^{\frac{8}{3}} +
  1.33272{\cdot}{10}^9\,{T_9}^3 \right. \right. \nonumber\\ \fl
&-& \left. \left.
  8.00725{\cdot}{10}^8\,{T_9}^{\frac{10}{3}} -
  3.23328{\cdot}{10}^8\,{T_9}^{\frac{11}{3}} +
  2.52841{\cdot}{10}^8\,{T_9}^4 \right. \right. \nonumber\\ \fl
&-& \left. \left.
  4.01975{\cdot}{10}^7\,{T_9}^{\frac{13}{3}}\right)
\right. \nonumber\\ \fl &+& \left. {T_9}^{-0.42536}
2.75408{\cdot}{10}^8
  \exp\left\{-\frac{1.78958}{T_9}\right\}
\right. \nonumber\\ \fl &-& \left. {{T_9}^{-\frac{2}{3}}}
\exp\left\{-2.04847 \,
       {{T_9}^{-\frac{1}{3}}} -
      0.00678199\,{T_9}^2\right\}
\left( -4.63896{\cdot}{10}^6 \right. \right. \nonumber\\ \fl &-&
\left. \left. 2.34671{\cdot}{10}^8\,
   {T_9}^{\frac{1}{3}} +
  1.6138{\cdot}{10}^9\,{T_9}^{\frac{2}{3}} -
  3.22641{\cdot}{10}^9\,T_9 \right. \right. \nonumber\\ \fl
&+& \left. \left.
  5.50235{\cdot}{10}^9\,{T_9}^{\frac{4}{3}} -
  2.96688{\cdot}{10}^{10}\,
   {T_9}^{\frac{5}{3}} +
  8.37688{\cdot}{10}^{10}\,{T_9}^2 \right. \right. \nonumber\\ \fl
&-& \left. \left.
  9.56773{\cdot}{10}^{10}\,
   {T_9}^{\frac{7}{3}} +
  2.44982{\cdot}{10}^{10}\,
   {T_9}^{\frac{8}{3}} +
  4.65934{\cdot}{10}^{10}\,{T_9}^3 \right. \right. \nonumber\\ \fl
&-& \left. \left.
  5.17303{\cdot}{10}^{10}\,
   {T_9}^{\frac{10}{3}} +
  2.36306{\cdot}{10}^{10}\,
   {T_9}^{\frac{11}{3}} -
  5.30274{\cdot}{10}^9\,{T_9}^4 \right. \right. \nonumber\\ \fl
&+& \left. \left.
  4.80234{\cdot}{10}^8\,{T_9}^{\frac{13}{3}}
\right) \right. \nonumber\\ \fl &-& \left. 2.75528\,{10}^8 \,
{T_9}^{-0.00704741} \,
  \exp\left\{-\frac{1.59705}{T_9}\right\}
\right]
\end{eqnarray}

\bigskip

\item {\bf Reaction   \emph{he3np} }: ${^3He} + n\lrt  p
+ {^3H}$
\begin{eqnarray} \fl
f_{he3np}&=&\left[7.06494 - 4.92737\,
   {\sqrt{T_9}} +
  6.73321\,T_9 -
  13.6597\,{T_9}^{\frac{3}{2}} +
  17.1812\,{T_9}^2 \right. \nonumber\\ \fl
      &-& \left.
  6.62993\,{T_9}^{\frac{5}{2}} -
  4.53677\,{T_9}^3 +
  4.83495\,{T_9}^{\frac{7}{2}} -
  1.22167\,{T_9}^4\right] {10}^8
\end{eqnarray}
\begin{eqnarray} \fl
\delta f_{he3np}&=& \left[35.1 -
     50.7448\,{\sqrt{T_9}} +
     31.121\,T_9 -
     13.2721\,
      {T_9}^{\frac{3}{2}} +
     7.90158\,{T_9}^2\right. \nonumber\\ \fl
      &-& \left.
     3.9825\,
      {T_9}^{\frac{5}{2}} +
     2.04542\,{T_9}^3 -
     1.4832\,
      {T_9}^{\frac{7}{2}} +
     0.511089\,{T_9}^4 \right. \nonumber\\ \fl
      &-& \left.
     0.390935\,
      {T_9}^{\frac{9}{2}} +
     0.390167\,{T_9}^5 -
     0.109195\,
      {T_9}^{\frac{11}{2}} +
     0.121065\,{T_9}^6 \right. \nonumber\\ \fl
      &-& \left.
     0.0907356\,
      {T_9}^{\frac{13}{2}} +
     0.0284269\,{T_9}^7 -
     0.0493578\,
      {T_9}^{\frac{15}{2}} \right. \nonumber\\ \fl
      &+& \left.
     0.025025\,{T_9}^8\right]^{1/2} 10^5
\end{eqnarray}

\bigskip

\item {\bf Reaction  \emph{be7np} }: ${^7Be} + n\lrt  p+
{^7Li}$
\begin{eqnarray} \fl
f_{be7np}&=& \left[6.8423 {\cdot} {10}^9  -
  1.4988 {\cdot} {10}^{10}\,{\sqrt{T_9}} +
  1.76749 {\cdot} {10}^{10}\,T_9 \right. \nonumber\\ \fl
      &-& \left.
  1.05769 {\cdot} {10}^{10}\,
   {T_9}^{\frac{3}{2}}
  + 2.6622 {\cdot} {10}^9\,{T_9}^2 +
  2.74476 {\cdot} {10}^8\,{T_9}^{\frac{5}{2}} \right. \nonumber\\ \fl
      &-& \left.
  3.35616 {\cdot} {10}^8\,{T_9}^3 +
  7.64252 {\cdot} {10}^7\,{T_9}^{\frac{7}{2}} -
  5.93091 {\cdot} {10}^6\,{T_9}^4 \right. \nonumber\\ \fl
      &-& \left. 2.28294 {\cdot} {10}^7 \,
\exp\left\{-\frac{0.0503518}{T_9}\right\}\, {T_9}^{-\frac{3}{2}}
\right]
\end{eqnarray}
\begin{eqnarray} \fl
\delta f_{be7np}&=& \frac{1}{2}\left[-6.01026{\cdot}{10}^7 \,
\exp\left\{-\frac{0.276138}{T_9}\right\}\,
       {T_9}^{-\frac{3}{2}} \right. \nonumber\\ \fl
      &-& \left.
    2.29462{\cdot}{10}^7 \, \exp\left\{- \frac{0.0504213}{T_9}\right\}\,
       {T_9}^{-\frac{3}{2}} \right. \nonumber\\ \fl
      &-& \left.
    2.8789\,{10}^6\,\left( -3.42532 +
       {\sqrt{T_9}} \right) \,
     \left( 3.09635 + {\sqrt{T_9}} \right)
     \right. \nonumber\\ \fl
      &{\times}& \left.
       \,\left( 8.12909 -
       5.62253\,{\sqrt{T_9}} +
       T_9 \right) \,
     \left( 4.24739 -
       3.80751\,{\sqrt{T_9}} +
       T_9 \right) \,\right. \nonumber\\ \fl
      &{\times}& \left.
     \left( 1.44946 -
       1.86774\,{\sqrt{T_9}} +
       T_9 \right)\right]
\end{eqnarray}

\bigskip

\item {\bf Reaction  \emph{$\alpha$he3$\gamma$} }: ${^4He}+{^3He}
\lrt \gamma + {^7Be}$
\begin{eqnarray} \fl
f_{\alpha he3 \gamma}&=&\frac{1}{\sqrt{T_9}}\,
\frac{\exp\left\{-0.481029\,T_9\right\}}
    {\left( 1 + 1.17918\,T_9 \right)^3}
\left[0.0000461656 - 0.000460361\,T_9 \right. \nonumber\\ \fl &-&
\left.
    0.0216009\,{T_9}^2 +
    0.0696278\,{T_9}^3 +
    7.34661\,{T_9}^4 -
    95.1232\,{T_9}^5 \right. \nonumber\\ \fl
&+& \left.
    391.131\,{T_9}^6 -
    187.237\,{T_9}^7 +
    86.1115\,{T_9}^8 -
    21.6302\,{T_9}^9 \right. \nonumber\\ \fl
&+& \left.
    3.60069\,{T_9}^{10} -
    0.343228\,{T_9}^{11} +
    0.0181067\,{T_9}^{12} \right. \nonumber\\ \fl
&-& \left.
    0.000356815\,{T_9}^{13}\right]
\end{eqnarray}
\begin{eqnarray} \fl
\delta f_{\alpha he3 \gamma}&=&\frac{1}{2 \sqrt{T_9}}\,\left[
    \frac{\exp\left\{-0.228239\,T_9\right\}}
    {\left( 1 + 1.31654\,T_9 \right)^3}
\left(0.0000507127 - 0.000482028\,T_9 \right. \right. \nonumber\\
\fl &-& \left. \left.
    0.0238316\,{T_9}^2 +
    0.0560337\,{T_9}^3 +
    8.40897\,{T_9}^4 -
    106.227\,{T_9}^5 \right. \right. \nonumber\\ \fl
&+& \left. \left.
    434.79\,{T_9}^6 -
    238.48\,{T_9}^7 +
    94.7573\,{T_9}^8 -
    23.7058\,{T_9}^9 +
    3.80071\,{T_9}^{10} \right. \right. \nonumber\\ \fl
&-& \left. \left.
    0.370295\,{T_9}^{11} +
    0.0199336\,{T_9}^{12} -
    0.000452817\,{T_9}^{13}\right) \right. \nonumber\\ \fl
    &-& \left.
\frac{\exp\left\{0.253254\,T_9\right\}}
    {\left( 1 + 1.2857\,T_9 \right)^3}
    \left(0.0000497987 - 0.000478013\,T_9 \right. \right. \nonumber\\ \fl
&-& \left. \left.
    0.0233624\,{T_9}^2 +
    0.0617906\,{T_9}^3 +
    8.05896\,{T_9}^4 -
    102.196\,{T_9}^5 \right. \right. \nonumber\\ \fl
&+& \left. \left.
    418.237\,{T_9}^6 -
    229.349\,{T_9}^7 +
    92.6387\,{T_9}^8 -
    23.3706\,{T_9}^9 \right. \right. \nonumber\\ \fl
&+& \left. \left.
    3.76443\,{T_9}^{10} -
    0.367266\,{T_9}^{11} +
    0.0197845\,{T_9}^{12} \right. \right. \nonumber\\ \fl
&-& \left. \left.
    0.000449519\,{T_9}^{13}\right)
    \right]
\end{eqnarray}

\bigskip

\item {\bf Reaction  \emph{li7p$\alpha$} }: ${^7Li} + p
\lrt {^4He} + {^4He}$
\begin{eqnarray} \fl
f_{li7pa} &=& {T_9}^{-\frac{2}{3}} \exp\left\{-\frac{7.7339}
       {{T_9}^{\frac{1}{3}}}\right\}\, \left[-8.96541{\cdot}{10}^7 +
    3.86917{\cdot}{10}^8\,
     {T_9}^{\frac{1}{3}} \right. \nonumber\\ \fl
     &+& \left.
    4.97213{\cdot}{10}^8\,
     {T_9}^{\frac{2}{3}} -
    2.58516{\cdot}{10}^8\,T_9 +
    2.64448{\cdot}{10}^7\,
     {T_9}^{\frac{4}{3}} \right. \nonumber\\ \fl
     &-& \left.
    1.29464{\cdot}{10}^6\,
     {T_9}^{\frac{5}{3}} -
    2.68313{\cdot}{10}^7\,{T_9}^2 -
    1.09411{\cdot}{10}^8\,
     {T_9}^{\frac{7}{3}} \right. \nonumber\\ \fl
     &+& \left.
    9.98996{\cdot}{10}^7\,{T_9}^{\frac{8}{3}}\right]
\end{eqnarray}
\begin{eqnarray} \fl
\delta f_{li7pa} &=& \frac 12 \, {T_9}^{-\frac{2}{3}} \left[
\exp\left\{-\frac{6.34173}
       {{T_9}^{\frac{1}{3}}}\right\}\,
\left(1.64256{\cdot}{10}^7 -
    1.19144{\cdot}{10}^8\,
     {T_9}^{\frac{1}{3}} \right. \right. \nonumber\\ \fl
     &+& \left.\left.
    3.36593{\cdot}{10}^8\,
     {T_9}^{\frac{2}{3}} -
    7.68266{\cdot}{10}^8\,T_9 +
    1.82342{\cdot}{10}^9\,
     {T_9}^{\frac{4}{3}} \right. \right. \nonumber\\ \fl
     &-& \left.\left.
    1.99627{\cdot}{10}^9\,
     {T_9}^{\frac{5}{3}} +
    1.24618{\cdot}{10}^9\,{T_9}^2 -
    5.49787{\cdot}{10}^8\,
     {T_9}^{\frac{7}{3}} \right. \right. \nonumber\\ \fl
     &+& \left.\left.
    1.42145{\cdot}{10}^8\,{T_9}^{\frac{8}{3}}\right) -
\exp\left\{-\frac{5.35733}
       {{T_9}^{\frac{1}{3}}}\right\}\,
    \left(-2.99794{\cdot}{10}^6 \right. \right. \nonumber\\ \fl
     &+& \left.\left.
    5.70066{\cdot}{10}^6\,
     {T_9}^{\frac{1}{3}} +
    1.33079{\cdot}{10}^8\,
     {T_9}^{\frac{2}{3}} -
    7.81101{\cdot}{10}^8\,T_9 \right. \right. \nonumber\\ \fl
     &+& \left.\left.
    1.73392{\cdot}{10}^9\,
     {T_9}^{\frac{4}{3}} -
    1.82843{\cdot}{10}^9\,
     {T_9}^{\frac{5}{3}} +
    1.18162{\cdot}{10}^9\,{T_9}^2 \right. \right. \nonumber\\ \fl
     &-& \left.\left.
    5.03632{\cdot}{10}^8\,
     {T_9}^{\frac{7}{3}} +
    1.10262{\cdot}{10}^8\,{T_9}^{\frac{8}{3}}\right)
    \right]
\end{eqnarray}

\bigskip

\item {\bf Reaction   \emph{$\alpha$d$\gamma$} }:
${^4He}+{^2H}\lrt \gamma + {^6Li}$
\begin{eqnarray} \fl
f_{\alpha d \gamma} &=& 14.82 \, {T_9}^{-\frac 23} \,
\exp\left\{-7.435 \, {T_9}^{-\frac 13}\right\} \left(1.+6.572 \,
T_9 \right. \nonumber\\ \fl & + & \left. 7.6 {\cdot} 10^{-2} \,
{T_9}^2+2.48 {\cdot} 10^{-2} \, {T_9}^3\right) \nonumber\\ \fl &+&
82.8 \, {T_9}^{-\frac 32}\, \exp\left\{-\frac{7.904}{T_9}\right\}
\end{eqnarray}
For $T_9 \leq 10.$ the lower and upper bounds result
\begin{eqnarray} \fl
\delta f_{\alpha d \gamma} = f_{\alpha d \gamma} \left\{
\begin{tabular}{l}
$-.9813 \, +.355 \sqrt{T_9}-.0411 \, T_9$ \\  \\ $+.289 \,
+5.612d0 \, e^{- 3 \, T_9} - 2.63 \, e^{- 2 \, T_9} +.773\, e^{-
T_9}$
\end{tabular}
\right.
\end{eqnarray}

\bigskip

\item {\bf Reaction   \emph{li6phe3} }: ${^6Li}+p\lrt
{^3He} + {^4He}$
\begin{eqnarray} \fl
f_{li6phe3}&=& {T_9}^{-\frac{2}{3}} \, \exp\left\{-\frac{4.62619}
       {{T_9}^{\frac{1}{3}}}\right\}\, \left[-7.49662{\cdot}{10}^7 +
    2.05335{\cdot}{10}^7\,
     {T_9}^{\frac{1}{3}} \right. \nonumber\\ \fl
     &+& \left.
    3.95475{\cdot}{10}^9\,
     {T_9}^{\frac{2}{3}} -
    1.94116{\cdot}{10}^{10}\,T_9 +
    3.79074{\cdot}{10}^{10}\,
     {T_9}^{\frac{4}{3}} \right. \nonumber\\ \fl
     &-& \left.
    3.43138{\cdot}{10}^{10}\,
     {T_9}^{\frac{5}{3}} +
    1.62629{\cdot}{10}^{10}\,{T_9}^2 -
    3.99652{\cdot}{10}^9\,
     {T_9}^{\frac{7}{3}} \right. \nonumber\\ \fl
     &+& \left.
    4.03339{\cdot}{10}^8\,{T_9}^{\frac{8}{3}}\right]
\end{eqnarray}
\begin{eqnarray} \fl
\delta f_{li6phe3}&=&\frac 12 \, {T_9}^{-\frac{2}{3}} \left[
\exp\left\{-\frac{2.47111}
       {{T_9}^{\frac{1}{3}}}\right\}\,
\left(4.67941{\cdot}{10}^7 -
    4.18832{\cdot}{10}^8\,
     {T_9}^{\frac{1}{3}} \right. \right. \nonumber\\ \fl
     &+& \left. \left.
    1.34221{\cdot}{10}^9\,
     {T_9}^{\frac{2}{3}} -
    1.50774{\cdot}{10}^9\,T_9 -
    9.45974{\cdot}{10}^8\,
     {T_9}^{\frac{4}{3}} \right. \right. \nonumber\\ \fl
     &+& \left. \left.
    3.60732{\cdot}{10}^9\,
     {T_9}^{\frac{5}{3}} -
    2.86104{\cdot}{10}^9\,{T_9}^2 +
    9.40736{\cdot}{10}^8\,
     {T_9}^{\frac{7}{3}} \right. \right. \nonumber\\ \fl
     &-& \left.\left.
    1.15472{\cdot}{10}^8\,{T_9}^{\frac{8}{3}}\right) -
\exp\left\{-\frac{4.53459}
       {{T_9}^{\frac{1}{3}}}\right\}\,
\left(-2.15414{\cdot}{10}^7 \right. \right. \nonumber\\ \fl
     &-& \left.\left.
    4.59395{\cdot}{10}^8\,
     {T_9}^{\frac{1}{3}} +
    5.52086{\cdot}{10}^9\,
     {T_9}^{\frac{2}{3}} -
    2.14539{\cdot}{10}^{10}\,T_9 \right. \right. \nonumber\\ \fl
     &+& \left.\left.
    3.82668{\cdot}{10}^{10}\,
     {T_9}^{\frac{4}{3}} -
    3.30682{\cdot}{10}^{10}\,
     {T_9}^{\frac{5}{3}} +
    1.51652{\cdot}{10}^{10}\,{T_9}^2 \right. \right. \nonumber\\ \fl
     &-& \left.\left.
    3.6301{\cdot}{10}^9\,{T_9}^{\frac{7}{3}} +
    3.58414{\cdot}{10}^8\,{T_9}^{\frac{8}{3}}\right)
    \right]
\end{eqnarray}

\bigskip

\item {\bf Reaction   \emph{$\alpha$t$\gamma$} }: ${^4He}+{^3H}
\lrt \gamma + {^7Li}$
\begin{eqnarray} \fl
f_{\alpha t \gamma}&=&\frac{1}{\sqrt{T_9}} \,
\frac{\exp\left\{-8.4{\cdot}{10}^{-7}\,T_9\right\}}
    {\left( 1 + 1.78617\,T_9 \right)^3}
    \left(0.0946142 - 4.92731\,T_9 \right. \nonumber\\ \fl
    &+& \left.
    99.359\,{T_9}^2 -
    989.812\,{T_9}^3 +
    4368.45\,{T_9}^4 +
    931.936\,{T_9}^5 \right. \nonumber\\ \fl
    &-& \left.
    391.079\,{T_9}^6 +
    159.231\,{T_9}^7 -
    34.4076\,{T_9}^8 +
    3.3919\,{T_9}^9 \right. \nonumber\\ \fl
    &+& \left.
    0.0175562\,{T_9}^{10} -
    0.0362534\,{T_9}^{11} +
    0.00311188\,{T_9}^{12} \right. \nonumber\\ \fl
    &-& \left.
    0.0000871447\,{T_9}^{13}\right)
\end{eqnarray}
\begin{eqnarray} \fl
\delta f_{\alpha t \gamma}&=& \frac{1}{2\sqrt{T_9}}
\left[\frac{\exp\left\{-9.3{\cdot}{10}^{-7}\,T_9\right\}}
    {\left( 1 + 1.60171\,T_9 \right)^3}
\left(0.083877 - 4.54089\,T_9 \right. \right. \nonumber\\ \fl
     &+& \left.\left.
    96.3161\,{T_9}^2 -
    1016.55\,{T_9}^3 +
    4809.48\,{T_9}^4 -
    168.102\,{T_9}^5 \right. \right. \nonumber\\ \fl
     &+& \left.\left.
    208.818\,{T_9}^6 -
    64.6182\,{T_9}^7 +
    10.4789\,{T_9}^8 -
    0.417824\,{T_9}^9 \right. \right. \nonumber\\ \fl
     &-& \left.\left.
    0.0645353\,{T_9}^{10} +
    0.00477763\,{T_9}^{11} +
    0.000200272\,{T_9}^{12} \right. \right. \nonumber\\ \fl
     &-& \left.\left.
    0.0000178642\,{T_9}^{13}\right) -
        \frac{\exp\left\{-0.0111133\,T_9\right\}}
    {\left( 1 + 1.63278\,T_9 \right)^3}
\left(0.0660966 \right. \right. \nonumber\\ \fl
     &-& \left.\left. 3.56229\,T_9 +
    75.1382\,{T_9}^2 -
    788.241\,{T_9}^3 +
    3705.89\,{T_9}^4 \right. \right. \nonumber\\ \fl
     &-& \left.\left.
    106.986\,{T_9}^5 +
    139.556\,{T_9}^6 -
    7.89845\,{T_9}^7 -
    1.60357\,{T_9}^8 \right. \right. \nonumber\\ \fl
     &-& \left.\left.
    0.175089\,{T_9}^9 +
    0.0464259\,{T_9}^{10} +
    0.00302332\,{T_9}^{11} \right. \right. \nonumber\\ \fl
     &-& \left.\left.
    0.000816826\,{T_9}^{12} +
    0.0000345452\,{T_9}^{13}\right)
    \right]
\end{eqnarray}

\bigskip

\item {\bf Reaction  \emph{tp$\gamma$} }: ${^3H} +p \lrt
\gamma+{^4He}$
\begin{eqnarray} \fl
f_{tp\gamma}&=& 2.2 {\cdot} 10^4 \, {T_9}^{-\frac 23} \,
\exp\left\{-3.869 \, {T_9}^{-\frac 13} \right\} \left(1.\, +.108
\, {T_9}^{\frac 13} +1.68d0 \, {T_9}^{\frac 23} \right.
\nonumber\\ \fl &+& \left. 1.26 \, T_9 + .551 \, {T_9}^{\frac
43}+1.06 \, {T_9}^{\frac 53}\right)
\end{eqnarray}
\begin{eqnarray} \fl
\delta f_{tp\gamma} \approx 0.2 \, f_{tp\gamma}
\end{eqnarray}

\bigskip

\item {\bf Reaction  \emph{li7p$\gamma$} }: ${^7Li} +p
\lrt \gamma+{^8Be}$
\begin{eqnarray} \fl
f_{li7p\gamma}&=& {T_9}^{-\frac{2}{3}} \, \exp\left\{-8.62567 \,
{T_9 }^{-\frac{1}{3}} -
    1.13752\,{T_9}^2\right\}
\left[3.00142{\cdot}{10}^7 \right. \nonumber\\ \fl &-& \left.
1.83661{\cdot}{10}^8\,T_9 +
  1.76881{\cdot}{10}^9\,{T_9}^2 -
  8.47723{\cdot}{10}^9\,{T_9}^3 \right. \nonumber\\ \fl &+& \left.
  2.02374{\cdot}{10}^{10}\,{T_9}^4 -
  1.96501{\cdot}{10}^{10}\,{T_9}^5 +
  7.94528{\cdot}{10}^8\,{T_9}^6 \right. \nonumber\\ \fl &+& \left.
  1.31325{\cdot}{10}^{10}\,{T_9}^7 -
  8.20935{\cdot}{10}^9\,{T_9}^8 -
  9.10992{\cdot}{10}^8\,{T_9}^9 \right. \nonumber\\ \fl &+& \left.
  2.78141{\cdot}{10}^9\,{T_9}^{10} -
  1.07853{\cdot}{10}^9\,{T_9}^{11} \right. \nonumber\\ \fl &+& \left.
  1.39934{\cdot}{10}^8\,{T_9}^{12}\right]
\end{eqnarray}
\begin{eqnarray} \fl
\delta f_{li7p\gamma}&=& \frac 12 {T_9}^{-\frac{2}{3}} \left[
\exp\left\{-5.55707 \, {T_9}^{-\frac{1}{3}} -
    1.04184\,{T_9}^2\right\}
\left(-25145.5 \right. \right. \nonumber\\ \fl &+& \left. \left.
 1.07873{\cdot}{10}^6\,T_9 -
  1.58997{\cdot}{10}^7\,{T_9}^2 +
  1.71826{\cdot}{10}^8\,{T_9}^3 \right. \right. \nonumber\\ \fl &-& \left. \left.
  8.31031{\cdot}{10}^8\,{T_9}^4 +
  2.12435{\cdot}{10}^9\,{T_9}^5 -
  2.87231{\cdot}{10}^9\,{T_9}^6 \right. \right. \nonumber\\ \fl &+& \left. \left.
  2.0104{\cdot}{10}^9\,{T_9}^7 -
  4.38596{\cdot}{10}^8\,{T_9}^8 -
  3.52934{\cdot}{10}^8\,{T_9}^9 \right. \right. \nonumber\\ \fl &+& \left. \left.
  2.98156{\cdot}{10}^8\,{T_9}^{10} -
  8.89207{\cdot}{10}^7\,{T_9}^{11} +
  9.98509{\cdot}{10}^6\,{T_9}^{12}\right) \right. \nonumber\\ \fl
&-& \left. \exp\left\{-5.20925 \, {T_9}^{-\frac{1}{3}} -
    1.00686\,{T_9}^2\right\}
\left(-14997.5 + 665017.\,T_9 \right. \right. \nonumber\\ \fl &-&
\left. \left.
  1.08801{\cdot}{10}^7\,{T_9}^2 +
  1.12999{\cdot}{10}^8\,{T_9}^3 -
  5.30972{\cdot}{10}^8\,{T_9}^4 \right. \right. \nonumber\\ \fl &+& \left. \left.
  1.32888{\cdot}{10}^9\,{T_9}^5 -
  1.7653{\cdot}{10}^9\,{T_9}^6 +
  1.21966{\cdot}{10}^9\,{T_9}^7 \right. \right. \nonumber\\ \fl &-& \left. \left.
  2.68716{\cdot}{10}^8\,{T_9}^8 -
  2.01198{\cdot}{10}^8\,{T_9}^9 +
  1.70323{\cdot}{10}^8\,{T_9}^{10} \right. \right. \nonumber\\ \fl &-& \left. \left.
  5.04165{\cdot}{10}^7\,{T_9}^{11} +
  5.61882{\cdot}{10}^6\,{T_9}^{12}\right)
\right]
\end{eqnarray}

\bigskip

\item {\bf Reaction  \emph{be7n$\alpha$} }: ${^7Be} + n
\lrt {^4He} + {^4He}$\\
\begin{eqnarray} \fl
f_{be7n\alpha}&=& 2.05 {\cdot} 10^4  \, \left(1. \, +3760 \,
T_9\right)
\end{eqnarray}
\begin{eqnarray} \fl
\delta f_{be7n\alpha}&\approx& 0.9 \, f_{be7n\alpha}
\end{eqnarray}

\bigskip

\item {\bf Reaction  \emph{li7dn$\alpha$} }: ${^7Li} + {^2H}
\lrt n+{^4He} + {^4He}$\\
\begin{eqnarray} \fl
f_{li7dn\alpha}&=& 1.71{\cdot}{10}^6 \, {T_9}^{-\frac{3}{2}} \,
\exp\left\{-\frac{3.246}{T_9}\right\}\, +
1.49{\cdot}{10}^{10}\,{T_9}^{-\frac{3}{2}} \,
\exp\left\{-\frac{4.0894}{T_9}\right\} \nonumber\\ \fl &{\times} &
\left( -2.1241 +
       \frac{0.0257}{T_9} +
       \frac{2.6314}
        {{T_9}^{\frac{2}{3}}} -
       \frac{4.1929}
        {{T_9}^{\frac{1}{3}}} +
       4.1136\,{T_9}^{\frac{1}{3}} \right)\nonumber \\ \fl
       &+&
1.66{\cdot}{10}^{11} \, {T_9}^{-\frac{2}{3}} \,
\exp\left\{-\frac{10.254}
        {{T_9}^{\frac{1}{3}}}\right\}\,
\end{eqnarray}
\begin{equation}
\delta f_{li7dn\alpha} \approx 0.5 \, f_{li7dn\alpha}
\end{equation}

\bigskip

\item {\bf Reaction  \emph{be7dp$\alpha$} }: ${^7Be} + {^2H}
\lrt p+{^4He} + {^4He}$\\
\begin{eqnarray} \fl
f_{be7dp\alpha}&=& 1.07 {\cdot} 10^{12} \, {T_9}^{-\frac 23} \,
\exp\left\{-12.428  \, {T_9}^{-\frac 13} \right\}
\end{eqnarray}
\begin{equation}
\delta f_{be7dp\alpha} \approx 0.9 \, f_{be7dp\alpha}
\end{equation}

\end{enumerate}

\section{}
\label{fitrate}

{\bf Fit of nuclide abundances}\\

We report a fit of the main nuclide abundances as functions of
$\eta$ and the number of effective extra degrees of freedom
$\Delta N$. The fits hold for $\eta$ corresponding to the 3
$\sigma$ range of $\omega_b$ as found by WMAP, {\it i.e.} $(5.48
\div 7.12){\cdot} 10^{-10}$, while $\Delta N$ varies in the interval
$-3 \div 3$. The fitting function is chosen for all nuclei as
follows \bea && \left( \sum_{n=1}^{8} a_n x^{n-1}+\sum_{n=1}^{8}
b_n x^{n-1}\Delta N+\sum_{n=1}^{8} c_n x^{n-1}(\Delta N)^2
\nonumber \right. \\  && \left. + \sum_{n=1}^{8} d_n
x^{n-1}(\Delta N)^3 \right) \exp \left( \sum_{n=1}^{6} e_n x^{n}
\right) \vv \label{formulafit} \eea where $x\equiv \log_{10}
\left( \eta {\cdot} 10^{10} \right)$ and the values of the
coefficients are reported in Tables
\ref{tablecoeffa}-\ref{tablecoeffe}. The fit accuracy is better
than 0.05\% for all nuclides but for $^7{\rm Li}$ (0.2\%).

We also report the squared error and correlations (see Equations
(\ref{sisi}) and (\ref{rhoij})) for $^2$H, $^4$He and $^7$Li, as
function of $\eta$ and $\Delta N$, valid in the same ranges for
$\eta$ and $\Delta N$. The accuracy of these fitting expressions
is better than 10 \%. \bea  \sigma^2_{^2{\rm H} ^2{\rm H}}  {\cdot}
10^{10} &=& 0.0151-0.0271 \ x +
0.0126 \ x^2 + 1.277 {\cdot} 10^{-3}\  \Delta N \nonumber \\
 &-&  1.288 {\cdot} 10^{-3} \ x \ \Delta N + 0.81 {\cdot} 10^{-5} \ (\Delta N)^2\\
 \sigma^2_{^4{\rm He} ^4{\rm He}} {\cdot} 10^{8} &=& 2.74 \\
 \sigma^2_{^7{\rm Li} ^7{\rm Li}} {\cdot} 10^{20}&=& 3.013 -9.015 \ x +6.901 \ x^2 + 0.116 \ \Delta N
 \nonumber \\ &-& 0.190 \ x \ \Delta N + 2.04 {\cdot} 10^{-3} \
(\Delta
 N)^2\\
\rho_{^2{\rm H} ^4{\rm He}} &=& -0.108 \\
\rho_{^2{\rm H} ^7{\rm Li}} &=& -0.255 +0.243 \ x -0.252 \ x^2 +
0.018 \ \Delta N
\nonumber \\
& - &  0.015 \ x \ \Delta N - 1.1
{\cdot} 10^{-4} \ (\Delta N)^2 \\
 \rho_{^4{\rm He} ^7{\rm Li}} &=& 0.035 \eea
\begin{table*}
\begin{center}
\begin{tabular}{|c|c|c|c|c|}
\hline & $X_{^2{\rm H}}/X_p$ & $X_{^3{\rm He}}/X_p$ & $Y_p$ &
$X_{^7{\rm Li}}/X_p$ \\
\hline $a_1$ & 1.2507${\cdot} 10^{-2}$ & 0.27737 & 1.8737 & -1.5477${\cdot} 10^{-2}$ \\
\hline $a_2$ & 4.9638${\cdot} 10^{-3}$ & 9.7400${\cdot} 10^{-2}$ & 0.59507 & 5.2207${\cdot} 10^{-2}$\\
\hline $a_3$ &-2.9574${\cdot} 10^{-3}$ & -4.3242${\cdot} 10^{-3}$ & -0.38065 & 9.6345${\cdot} 10^{-2}$\\
\hline $a_4$ & -9.7061${\cdot} 10^{-3}$ & -2.5863${\cdot} 10^{-2}$ & -0.81999 & 0.15855\\
\hline $a_5$ & -1.2869${\cdot} 10^{-2}$& 5.8552${\cdot} 10^{-2}$ & -0.53340 & 0.21032\\
\hline $a_6$ & -8.9681${\cdot} 10^{-3}$ & 0.26452& 0.56298 & 0.24617\\
\hline $a_7$ & 6.7369${\cdot} 10^{-3}$ & 0.58868 &2.3498 & 0.26478\\
\hline $a_8$ & 4.0429${\cdot} 10^{-2}$ & 0.99729 & 4.3744 &
0.27337\\ \hline
\end{tabular}
\end{center}
\caption{Coefficients $a_i$ of the nuclei abundance fit of
Equation (\ref{formulafit}).} \label{tablecoeffa}
\end{table*}
\begin{table*}
\begin{center}
\begin{tabular}{|c|c|c|c|c|}
\hline & $X_{^2{\rm H}}/X_p$ & $X_{^3{\rm He}}/X_p$ & $Y_p$ &
$X_{^7{\rm Li}}/X_p$ \\
\hline $b_1$ & 1.5633${\cdot} 10^{-3}$ & 1.1181${\cdot} 10^{-2}$ & 7.6314${\cdot} 10^{-2}$ & 1.7068${\cdot} 10^{-2}$\\
\hline $b_2$ & 7.5528${\cdot} 10^{-4}$ & 8.9021${\cdot} 10^{-3}$ & 5.7666${\cdot} 10^{-2}$ & -6.9421${\cdot} 10^{-3}$\\
\hline $b_3$ & -2.5357${\cdot} 10^{-4}$ & 5.6831${\cdot} 10^{-3}$ & 2.1875${\cdot} 10^{-2}$ & -3.4891${\cdot} 10^{-2}$\\
\hline $b_4$ & -1.2834${\cdot} 10^{-3}$ & 2.4859${\cdot} 10^{-3}$ & -2.3930${\cdot} 10^{-2}$ & -4.2916${\cdot} 10^{-2}$\\
\hline $b_5$ & -1.9679${\cdot} 10^{-3}$& 1.0731${\cdot} 10^{-3}$ & -6.1388${\cdot} 10^{-2}$ & -4.4713${\cdot} 10^{-2}$\\
\hline $b_6$ & -1.6636${\cdot} 10^{-3}$ & 4.2430${\cdot} 10^{-3}$ & -5.5100${\cdot} 10^{-2}$ & -2.8193${\cdot} 10^{-2}$\\
\hline $b_7$ &  6.6829${\cdot} 10^{-4}$ & 1.6082${\cdot} 10^{-2}$ & 5.4745${\cdot} 10^{-2}$ & 1.1372${\cdot} 10^{-2}$\\
\hline $b_8$ & 6.6106${\cdot} 10^{-3}$ &4.2236${\cdot} 10^{-2}$& 0.36162 & 7.6082${\cdot} 10^{-2}$\\
\hline
\end{tabular}
\end{center}
\caption{Coefficients $b_i$ of the nuclei abundance fit of
Equation (\ref{formulafit}).} \label{tablecoeffb}
\end{table*}
\begin{table*}
\begin{center}
\begin{tabular}{|c|c|c|c|c|}
\hline & $X_{^2{\rm H}}/X_p$ & $X_{^3{\rm He}}/X_p$ & $Y_p$ &
$X_{^7{\rm Li}}/X_p$ \\
\hline $c_1$ & 6.3841${\cdot} 10^{-5}$ & 3.8084${\cdot} 10^{-3}$ & 2.8799${\cdot} 10^{-2}$ & -5.3604${\cdot} 10^{-5}$\\
\hline $c_2$ & -5.2844${\cdot} 10^{-5}$ & -4.2935${\cdot} 10^{-3}$ & -3.6642${\cdot} 10^{-2}$ & 2.3056${\cdot} 10^{-3}$\\
\hline $c_3$ & -1.0118${\cdot} 10^{-4}$ & -6.7032${\cdot} 10^{-3}$ & -5.6964${\cdot} 10^{-2}$ & 3.2785${\cdot} 10^{-3}$ \\
\hline $c_4$ & -6.2060${\cdot} 10^{-5}$ & -3.0083${\cdot} 10^{-3}$ & -2.6008${\cdot} 10^{-2}$& 2.3728${\cdot} 10^{-3}$\\
\hline $c_5$ & 5.5453${\cdot} 10^{-5}$ & 5.1127${\cdot} 10^{-3}$& 4.4481${\cdot} 10^{-2}$ & -4.3057${\cdot} 10^{-4}$\\
\hline $c_6$ &1.8655${\cdot} 10^{-4}$ & 1.2626${\cdot} 10^{-2}$ & 0.11194 & -4.1902${\cdot} 10^{-3}$\\
\hline $c_7$ & 1.7096${\cdot} 10^{-4}$ & 9.5251${\cdot} 10^{-3}$ & 8.5888${\cdot} 10^{-2}$ & -6.3226${\cdot} 10^{-3}$\\
\hline $c_8$ & -3.0158${\cdot} 10^{-4}$ & -2.1193${\cdot} 10^{-2}$& -0.19410 & -1.6925${\cdot} 10^{-3}$\\
\hline
\end{tabular}
\end{center}
\caption{Coefficients $c_i$ of the nuclei abundance fit of
Equation (\ref{formulafit}).} \label{tablecoeffc}
\end{table*}
\begin{table*}
\begin{center}
\begin{tabular}{|c|c|c|c|c|}
\hline & $X_{^2{\rm H}}/X_p$ & $X_{^3{\rm He}}/X_p$ & $Y_p$ &
$X_{^7{\rm Li}}/X_p$ \\
\hline $d_1$ &8.1121${\cdot} 10^{-6}$ & 6.1008${\cdot} 10^{-6}$ & -7.8725${\cdot} 10^{-4}$& 1.1108${\cdot} 10^{-4}$\\
\hline $d_2$ & 7.4834${\cdot} 10^{-6}$ & 5.4409${\cdot} 10^{-5}$ &1.6205${\cdot} 10^{-3}$ & -4.7543${\cdot} 10^{-4}$\\
\hline $d_3$ & 1.3879${\cdot} 10^{-5}$ & 4.1976${\cdot} 10^{-5}$ & 2.2747${\cdot} 10^{-3}$ & -5.6928${\cdot} 10^{-4}$\\
\hline $d_4$ &  8.4894${\cdot} 10^{-6}$& -1.3248${\cdot} 10^{-5}$& 9.7351${\cdot} 10^{-4}$ & -8.0121${\cdot} 10^{-5}$ \\
\hline $d_5$ & -7.5232${\cdot} 10^{-6}$ & -7.4104${\cdot} 10^{-5}$ & -1.7800${\cdot} 10^{-3}$ & 8.7257${\cdot} 10^{-4}$\\
\hline $d_6$ & -2.5569${\cdot} 10^{-5}$& -8.4875${\cdot} 10^{-5}$& -4.2767${\cdot} 10^{-3}$ & 1.7785${\cdot} 10^{-3}$\\
\hline $d_7$ & -2.4336${\cdot} 10^{-5}$& 2.4265${\cdot} 10^{-5}$& -2.9394${\cdot} 10^{-3}$ & 1.4867${\cdot} 10^{-3}$\\
\hline $d_8$ & 3.7481${\cdot} 10^{-5}$&  3.2913${\cdot} 10^{-4}$ & 8.5262${\cdot} 10^{-3}$& -2.1277${\cdot} 10^{-3}$\\
\hline
\end{tabular}
\end{center}
\caption{Coefficients $d_i$ of the nuclei abundance fit of
Equation (\ref{formulafit}).} \label{tablecoeffd}
\end{table*}
\begin{table*}
\begin{center}
\begin{tabular}{|c|c|c|c|c|}
\hline & $X_{^2{\rm H}}/X_p$ & $X_{^3{\rm He}}/X_p$ & $Y_p$ &
$X_{^7{\rm Li}}/X_p$ \\
\hline $e_1$ & 4.8889 & 6.8625 & -1.8375 & 1.6107\\
\hline $e_2$ & -2.7519 & -5.9003 & 5.9213 & -3.5500\\
\hline $e_3$ & -2.4691 & -5.7748 & -0.69772 & 3.1784\\
\hline $e_4$ & -0.23165 & 2.1014 & -4.5914 & -3.7802\\
\hline $e_5$ & 1.7868 & 2.4221 &  -5.3678 & 5.2812\\
\hline $e_6$ & -2.4201 & -0.75970 & 5.4446 & -2.9140
\\\hline
\end{tabular}
\end{center}
\caption{Coefficients $e_i$ of the nuclei abundance fit of
Equation (\ref{formulafit}).} \label{tablecoeffe}
\end{table*}

\end{document}